\documentclass[reprint,superscriptaddress,tightenlines,bibnotes, amsmath,amssymb,aps,prl]{revtex4-2}

\usepackage{graphicx}
\usepackage{dcolumn}
\usepackage{bm}

\usepackage{amsmath,amsfonts,amssymb}
\usepackage{subfigure}
\usepackage{graphics,graphicx}
\usepackage{hyperref}

\usepackage{multirow}
\usepackage{array}
\usepackage{pdfrender}
\newcommand*{\boldcheckmark}{%
  \textpdfrender{
    TextRenderingMode=FillStroke,
    LineWidth=.5pt, 
  }{\checkmark}%
}

\usepackage{xcolor}

\begin{document}

\title{Distinguishing simple and complex contagion processes on networks}

\author{Giulia Cencetti}
\affiliation{Fondazione Bruno Kessler, Trento, Italy}
\author{Diego Andr\'es Contreras}
\affiliation{Aix-Marseille Univ, Université de Toulon, CNRS, Centre de Physique Théorique, Turing Center for Living Systems, Marseille, France}
\author{Marco Mancastroppa}
\affiliation{Aix-Marseille Univ, Université de Toulon, CNRS, Centre de Physique Théorique, Turing Center for Living Systems, Marseille, France}
\author{Alain Barrat}
\affiliation{Aix-Marseille Univ, Université de Toulon, CNRS, Centre de Physique Théorique, Turing Center for Living Systems, Marseille, France}

\begin{abstract}
Contagion processes on networks, including disease spreading, information diffusion, or social behaviors propagation, can be modeled as simple contagion, i.e. involving one connection at a time, or as complex contagion, in which multiple interactions are needed for a contagion event. Empirical data on spreading processes however, even when available, do not easily allow to uncover which of these underlying contagion mechanisms is at work. We propose a strategy to discriminate between these mechanisms upon the observation of a single instance of a spreading process. The strategy is based on the observation of the order in which network nodes are infected, and on its correlations with their local topology: these 
correlations differ between processes of simple contagion, processes involving threshold mechanisms and processes driven by group interactions
(i.e., by “higher-order” mechanisms). Our results 
improve our understanding of contagion processes and provide a method using
 only limited information to distinguish between several possible contagion mechanisms.
\end{abstract}

\maketitle

Many phenomena can be described as contagions, such as disease spreading, information diffusion, or propagation of social behaviors
\cite{anderson1992infectious,keeling2011modeling,centola2007complex,barrat2008dynamical,castellano2009statistical,pastor2015epidemic}. Modelling contagion processes in a population typically includes two main steps. First, one describes how the state of the hosts
(individuals who receive and propagate the disease/information/behavior) can evolve. For instance, one often assumes that they can
only be in one of few possible states, such as susceptible (healthy), infectious (having the disease/information and able to transmit it), or 
recovered (cured and immunized). Second, the
propagation is described along the structure of interactions between hosts, often encoded through a network in which nodes represent hosts and links represent their interactions.
The resulting network epidemiology framework has been applied to the spread of human and computer viruses
\cite{pastor2001epidemic,barrat2008dynamical,pastor2015epidemic},  rumors
\cite{daley1964epidemics,moreno2004dynamics}, 
innovations \cite{valente1996network,rogers2003diffusion,watts2002simple,watts2007influentials,iacopini2018network} or  behavior \cite{centola2010spread}. 

Depending on the phenomenon, the fundamental propagation mechanisms are different. To describe the spread of infectious diseases,  
models of \textit{simple contagion}, in which a single interaction between a susceptible and an infectious can lead to a transmission event, are adequate \cite{anderson1992infectious,pastor2015epidemic}. 
In social contagion of behaviors, peer influence and reinforcement mechanisms can play an important role, 
and empirical evidence 
indicate that single interactions are not sufficient to cause transmission \cite{centola2010spread,weng2012competition,ugander2012structural,karsai2014complex,monsted2017evidence,notarmuzi2022universality}. 
These cases are hence better described by so-called \textit{complex contagion} models, in which each transmission event requires interactions with multiple infectious hosts \cite{watts2002simple,karsai2014complex,iacopini2019simplicial,ferrazdearruda2021phase,st2022influential}.

For both simple and complex contagions, most studies start from a propagation mechanism and design models to represent it and study how the structure of the interaction network impacts the spread \cite{pastor2015epidemic}. 
In general, these investigations focus on averages over realizations of the process, and compare the  phenomenology of processes and how they depend on the network structure. However, when empirical data related to a spreading process is observed, it concerns a single instance and one cannot average over multiple instances to obtain overall statistics.   
Therefore, here we address the issue of determining, from the observation of a single
instance of a contagion process on a network, whether it is governed by a simple or complex contagion, and whether threshold or group 
effects are involved. 
Previous works have tried to identify the footprint of different contagion models on real or simulated processes.
Evidence of complex contagion has been found in real data, observing the temporal evolution of the number of infectious \cite{karsai2014complex,fink2016investigating,sprague2017evidence,notarmuzi2022universality}, or by investigating how the contagion probability of a node depends on infectious neighbors
\cite{aiello2012link,ugander2012structural,monsted2017evidence,karsai2014complex}. Other rather data demanding techniques involve using Deep Learning~\cite{murphy2021deep} or comparing spreading processes on different network structures \cite{centola2010spread,horsevad2022transition}. However, we still lack a clear identification of the main features distinguishing simple and complex processes.

\begin{figure*}[thb]
\includegraphics[width=\textwidth]{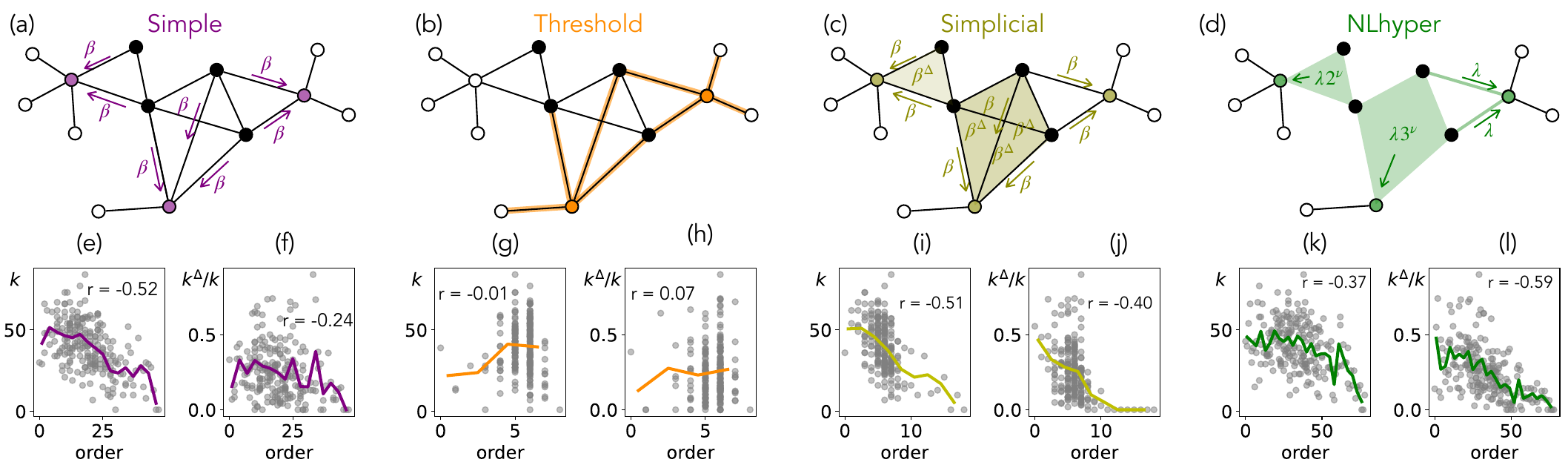}
\caption{
The first row reports a toy network at an intermediate stage of the process (in which contagion events have taken place, resulting in 4 infected nodes, shown in black) and how the different models of propagation would imply contagion of further nodes (colored). 
Contagion events occur respectively
(a) in simple contagion, along the network edges, with probability $\beta$ per unit time for each edge;
(b) when a susceptible node sees a fraction of infected neighbors that is above a threshold $\theta$ (here $\theta=0.5$);
(c) both along network edges (rate $\beta$) and 
if a susceptible is part of a simplex of 3 nodes in which the two others are infected (rate $\beta^{\Delta}$);
(d) along hyperedges, a susceptible node sharing a hyperedge with $n$ infected becoming infected at rate $\lambda n^\nu$.
The second row gives 
scatterplots of $k_i$ vs. $o_i$ and $k_i^{\Delta}/k_i$ vs. $o_i$ for single numerical realizations of each model 
on the workplace data set, where $o_i$ is the order in which the node $i$, with degree 
$k_i$ and belonging to $k_i^{\Delta}$ hyperedges of size $3$, has been reached by the propagation in that realization. The values of the corresponding correlation coefficients are given in the plots. Parameters: $\beta=0.005$ in panels (e) and (f), $\theta=0.007$ in (g) and (h), $\beta=0.005$, $\beta^{\Delta}=0.8$ in (i) and (j), $\nu=4$, $\lambda=0.001$ in (k) and (l). The colored curves report the mean of all $k_i$ or $k_i^{\Delta}/k_i$ for each $o_i$.
}\label{fig1}
\end{figure*}

Here, we put forward a new method based on the correlations between the order in which successive nodes in the network are reached by the spread and their basic local properties. 
We show that paradigmatic models of simple and complex contagion lead to different correlation patterns, and how to exploit them to build a classification tool able to determine whether a given instance of a spread is due to a simple contagion, a threshold model or a higher-order contagion model. We investigate the robustness of our classifier with respect to incomplete data, and the possibility to apply it to a process taking place on an unknown network, i.e., different from the one(s) on which the classifier has been calibrated, as knowing the detailed structure of the network on which a contagion occurs is often challenging \cite{eames2015challenges,newman2018network,nuria2020mobile}. 

\paragraph*{Models of contagion.}
We consider four contagion processes on networks. For simplicity, we use SI models, i.e., each node can only be in two states, susceptible (S) or infected (I), and infected nodes do not recover. We consider processes in discrete time,  differing in the mechanism determining how a node can switch from the S to the I state. 

We first consider a \textit{simple contagion} process
(Fig.~\ref{fig1}(a)): every susceptible node can be infected independently by each of its infected neighbors with a probability per unit time $\beta$. The disease spreads thus along the pairwise links among nodes.

The second process (Fig.~\ref{fig1}(b)) is the deterministic 
\textit{threshold model}~\cite{watts2002simple}: 
 a susceptible node becomes infected when the fraction of its neighbors that are infected reaches a threshold $\theta$, to mimic the fact that an individual may adopt an innovation only if enough friends are already adopters.  

We also consider two models of complex contagion that involve higher-order contagion mechanisms, i.e., interactions among groups of nodes \cite{battiston2020networks}. First,
the  \textit{simplicial model} takes place on simplicial complexes and the higher-order contagion is regulated by a parameter $\beta^{\Delta}$ (Fig.~\ref{fig1}(c)). Second, the \textit{non-linear hypergraph (NL-hyper) model} takes place on hypergraphs and is regulated by parameters $\lambda$ and $\nu$ (Fig.~\ref{fig1}(d)). See Appendix and Fig.~\ref{fig1} for details.

Given an observed single realization of one of these models on a network, our goal is to devise a method to determine which model it corresponds to. To this aim, we consider several empirical data sets as the substrates on which the processes unfold. We use data representing physical or online interactions between individuals in several contexts: a workplace \cite{Genois2018}, 
educational contexts
\cite{stehle2011high,sapiezynski2019interaction,mastrandrea2015contact,toth2015role},
a scientific conference~\cite{Genois2018}, a hospital~\cite{vanhems2013estimating}, and an email dataset~\cite{klimt2004enron} (see Supplementary Material -SM). 
These data are temporally resolved but we consider the aggregated network $G_{\cal D}$, the aggregated hypergraph $H_{\cal D}$ and its projection on hyperedges of size at most 3 $H^3_{\cal D}$ (defined in Appendix). The degree $k_i$ of an individual is the number of links involving $i$ in $G_{\cal D}$, while we denote by $k_i^{\Delta}$ the number of hyperedges of size 3 to which $i$ belongs in $H^3_{\cal D}$.
We consider here for simplicity unweighted networks and hypergraphs, but each link or hyperedge can be weighted by the number of times that the corresponding interaction has been observed during the data collection. We discuss the case of weighted networks and hypergraphs 
in the SM.
In the main text, we give mostly results obtained with the workplace data set and refer to the SM for the other data sets.

\begin{figure}
\includegraphics[width=0.5\textwidth]{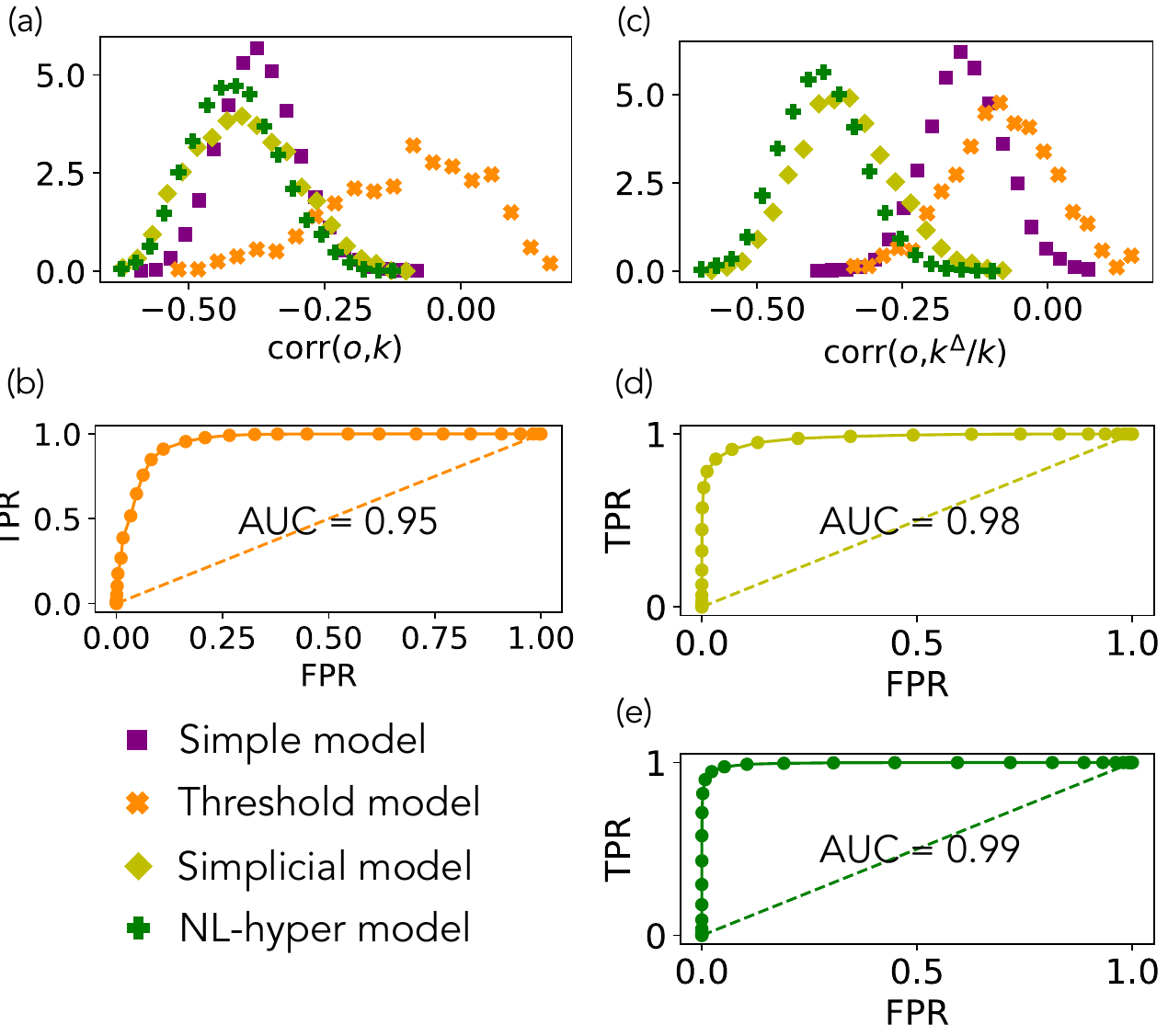}
\caption{Results for the workplace data set.
(a): Distributions of $C_1$ for the four contagion models. 
(b): ROC curve when using $C_1$ to classify threshold model processes against the other three. 
(c): Distributions of $C_2$ for the four contagion models. 
(d): ROC curve when using $C_2$ to classify simplicial against simple and threshold models. 
(e): ROC curve when using $C_2$ to classify NL-hyper against simple and threshold models. 
For the stochastic models (simple, simplicial and NL-hyper) 1000 realizations are implemented for each parameter setting, always starting with one random infected node. For the deterministic threshold model we use only one realization for each different initial condition, i.e. one for each network node, for each parameter value.
Parameters: $\beta \in \{0.005, 0.008, 0.014, 0.023, 0.039\}$ for both simple and simplicial models, $\beta^{\Delta} = 0.8$, $\lambda \in \{0.0001, 0.001, 0.006, 0.011, 0.015\}$, $\nu=4$,
$\theta \in \{0.01, 0.02, 0.03, 0.04, 0.05, 0.06, 0.07, 0.08, 0.09, 0.10\}$.}
\label{fig2}
\end{figure}

\paragraph*{Results.} 
Simple contagion processes on networks are
characterized by hierarchical dynamics: large degree nodes are reached early, and a  cascade follows towards small degree nodes 
\cite{barthelemy2004velocity}. In general, the order in which nodes are infected can be influenced by their degrees, as illustrated in Fig.~\ref{fig1} for single instances of each process.
This is explored further in Fig.~\ref{fig2}(a), where we show the distribution of the Spearman correlation $C_1 = \text{corr}(o,k)$ between the order in which nodes are infected and their degree $k$ \cite{spearman1987proof}, computed for each numerical realization of each model and for a wide range of parameter values. The distributions are similar for the simple and higher-order contagion processes, with similar ranges of (only negative) correlation values, while the distribution of $C_1$ obtained for the threshold process has a broader support including  positive values. We thus consider the possibility to use the value of $C_1$ to identify whether a given realization results from the threshold model or from another model. We use the parametric Receiver Operating Characteristic (ROC) curve (see Appendix)
to summarize the quality of the classifier as the area under the ROC curve (AUC), which is $0.5$ for a random classification and $1$ for a perfect one. The  ROC curve for the workplace data set is shown in Fig.~\ref{fig2}(b), with a very high AUC of $0.95$.

To identify processes involving higher-order mechanisms, we need to 
take into account the participation of nodes to higher-order structures.
We thus consider the correlation between the order in which nodes are infected and the ratio $k^{\Delta}/k$ between the number of hyperedges of size $3$ to which they belong and their degree: $C_2 = \text{corr}(o,k^{\Delta}/k)$
\footnote{It is in principle possible to consider the number of hyperedges of larger sizes to which each node participates; for the sake of simplicity, we limit here to size $3$, and, if a node belongs to a hyperedge of size $m > 3$, we decompose the hyperedge into triangles to compute $k^{\Delta}$.}.
Figure \ref{fig1} (f), (h), (j), (l) illustrate the correlation $C_2$ on specific instances of each process, and Fig. \ref{fig2}(c) shows its distribution over multiple instances. The distributions are similar for the 
simplicial and NL-hyper contagion cases on the one hand, and for the simple and threshold models on the other hand. Very good classification performances are attained, as quantified by the ROC curves obtained when using $C_2$ to classify instances of the simplicial model (figure \ref{fig2}(d)) or of the NL-hyper model (figure \ref{fig2}(e)) against instances of simple and threshold processes. We study in the SM 
how this performance depends on the model parameters.

\begin{figure}[thb]
\includegraphics[width=\columnwidth]{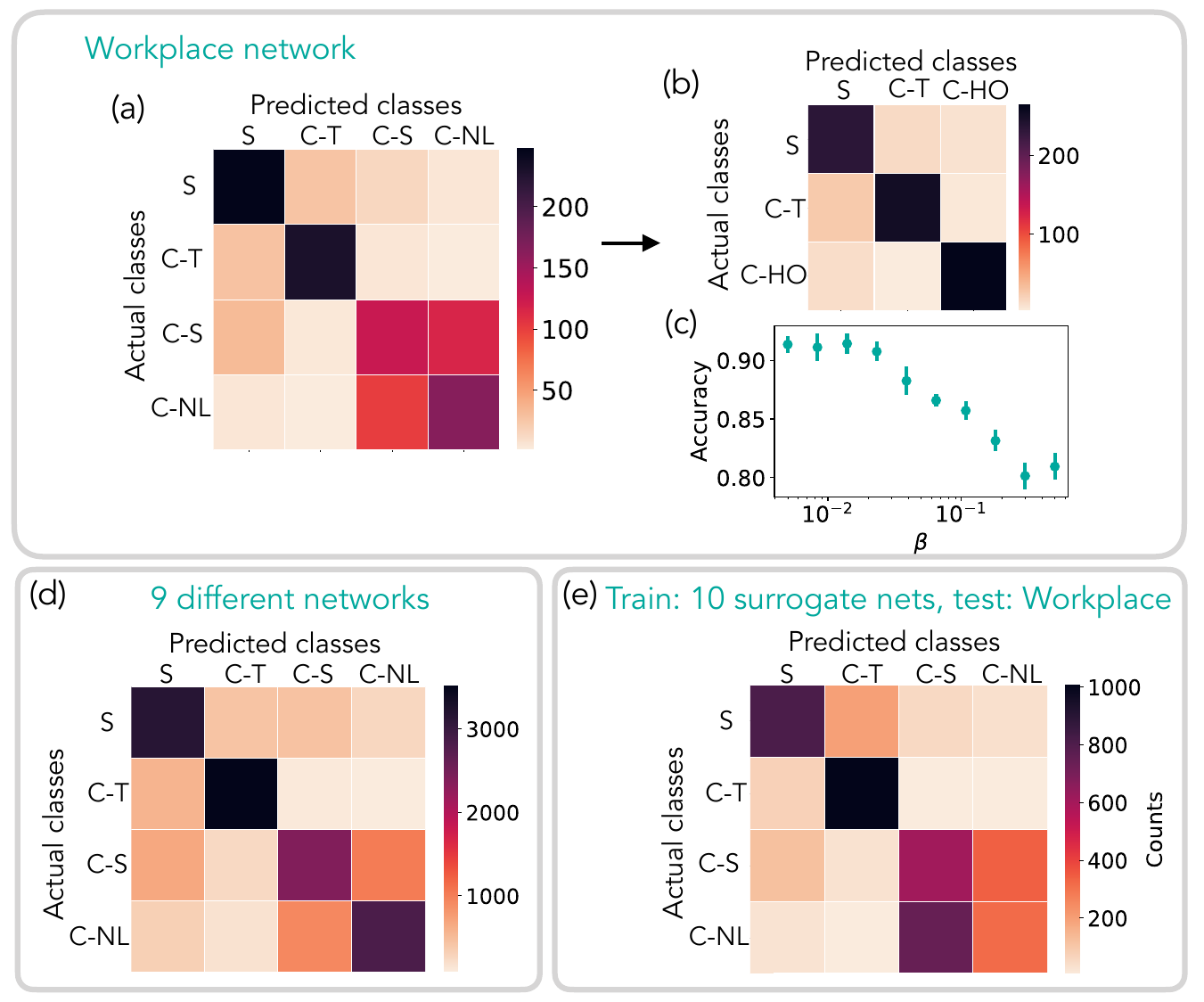}
\caption{(a): Confusion matrix with 4 classes: S (simple), C-S (complex simplicial), C-NL (complex NL-hyper), and C-T (complex threshold). 3288 instances used for training and 1096 for testing (with approximately the same number of instances for each model).
(b): Confusion matrix merging C-S and C-NL into C-HO (complex higher-order). 2466 instances used for training and 822 for testing.
(c): Accuracy of RF classification with classes S, C-T, C-HO at varying $\beta$ in simple and simplicial. 2247 instances used for training and 750 for testing for each value of $\beta$.
(d): Confusion matrix obtained by combining results on 9 different networks. The relative accuracy assembling in classes S, C-T, C-HO is 0.84. 51609 instances used for training and 17203 for testing.
(e): Confusion matrix obtained by training the classifier on 10 surrogate networks (obtained with SDC method, see \cite{Note1}) with statistical properties similar to the workplace data set and testing using processes run on the real data. The accuracy when considering three classes S, C-T, C-HO is 0.85. 33220 instances used for training and 4384 for testing.
The parameters are set as in Fig.~\ref{fig2}. 
}
\label{fig3}
\end{figure}

We now combine the previous results using $C_1$ and $C_2$ to build a global classifier. 
We consider in addition the correlations between the order $o$ of infection of nodes with $k^{\Delta}$ and with $k^|$, their number of purely pairwise links (excluding connections part of higher-order interactions): respectively $C_3 = \text{corr}(o,k^{\Delta})$ and $C_4 = \text{corr}(o,k^|)$.
We use a Random Forest (RF) classifier~\cite{ho1995random}, a standard Machine Learning method, to perform the overall classification of instances of the four models.
The performance of this classification task can be  assessed by the confusion matrix depicted in Fig.~\ref{fig3}(a): it gives in row $x$ and column $y$ the number of instances of a model $x$ that are classified as resulting from model $y$. Simple and threshold model instances are identified almost perfectly, while simplicial and NL-hyper model instances are confused more easily. Merging the higher-order model realizations as one unique class of Complex Higer-Order (C-HO) processes results in a very high accuracy (Fig.~\ref{fig3}(b-c)).
The classifier yields similar results for simulations implemented on each of the nine interaction networks we consider \cite{Note1}.

Up to now, we have considered that the classifier is trained using data coming from one network, and applied on processes run on the same network, and  that the order of contagion and the local properties of all nodes are known. 
We now examine less idealized conditions. In particular, we start by relaxing the hypothesis of a full knowledge of the network structure, as measuring the full detailed structure of networks on which spreading processes can occur, such as contact networks, is a much more challenging task \cite{eames2015challenges,nuria2020mobile} than getting information on purely local properties \cite{mossong2008social}. 
First, we examine the case in which instances of the contagion processes occurring on different data sets are mixed: one can thus consider that the process to be classified has taken place on an unknown network, but that network has been used among others to train the classifier. The resulting accuracy is still very high
(Fig.~\ref{fig3}(d)) for the distinction between simple, threshold-based and higher-order based contagion processes.
To understand further the generalizability of the classifier, we also consider the case of a process observed on a completely ``new'' network, while the classifier has been trained using processes run on other network(s).
A first case consists in using one or several of the available data sets for training, and another for testing. The resulting accuracy depends on the data sets chosen for training and testing \cite{Note1}, and remains high in many cases, which indicates a certain generalizability. These results also have limitations: the unfolding of a spread depends on the network structure, so that a classifier trained using one network cannot be blindly applied to a completely different one.
However, recent works have also shown that  statistical properties describing a spreading process can be obtained even from limited information on the network it unfolds on \cite{machens2013infectious,contreras2022impact}. 
We thus assume that the detailed structure of the data set $H_{\cal D}$ on which the spread to be classified has occurred is unknown. However, using the information on the degrees in $H^3_{\cal D}$ 
(assumed known anyway, as they represent the minimum information needed to compute the correlations fed into the classifier), we can generate surrogate data, i.e. synthetic networks having similar degree distributions as 
$H^3_{\cal D}$. In addition, we envision the case in which the group structure of the data is known (e.g., classes in a school), as it is known to be relevant to spreading processes \cite{machens2013infectious,contreras2022impact}
and also build surrogate data reproducing this structure. We consider in the SM three possible ways to built such surrogate data sets. In each case we train the classifier using processes run on the surrogate data and classify processes run on the original data. We show that
large values of the accuracy are recovered, as long as the algorithm for creating the surrogate data performs sufficiently well (see 
Fig.~\ref{fig3}(e) for the workplace data set, and \cite{Note1} for a more extensive
analysis concerning all data sets and surrogate data types).

We finally report in the SM results obtained when relaxing the assumption of a complete observation of the spreading instance. First, we assume that only a fraction of the nodes (chosen arbitrary at random) can be observed. The values of the correlations are thus computed using only the order of infection and degrees of the observed nodes (both for training and testing). In this case, the accuracy of the classifier remains high, with values above $0.7$ even when only $20\%$ of the nodes are observed. This can be understood by the robustness of the correlations when randomly removing a fraction of the data points.
If instead only the order of infection and the degrees of the first $h$ infected are known, the performances are more impacted. To observe e.g. the occurrence of a cascade from large to small degrees, the first data points might indeed not be sufficient, and having information beyond the initial phase brings more accurate results.

\paragraph*{Discussion.}
We have  developed a framework for classifying contagion processes on networks through the observation of correlations between the order in which nodes are reached and their local structural properties. The classification task  
(i) uses only local information, without the need to access the whole network structure,
(ii) does not use any information on the infection status of the nodes' neighbours, (iii) is applied on single instances of a process, and 
(iv) can distinguish between a simple contagion process, a process driven by a threshold mechanism and a process involving contagion on higher-order interactions. The proposed classifier yields a very good accuracy on several real-world networks, remaining robust against partial observation of the process. 
Moreover, although it cannot be trained with and applied to processes run on networks 
with very different properties, it can be applied 
on a process occurring on an unknown network as long as it is possible to generate
surrogate data with similar statistical properties to produce the training set.

Our work has several limitations worth discussing. 
First, we have assumed to have access to the precise values of the degrees for all observed nodes, as well as the precise ordering of infection. To be more realistic, one could simulate the use of estimated values by inserting noise in the degree values. As the classifier is based on the measure of correlations, we expect that its accuracy should not be strongly impacted. It could however be affected by e.g. extreme errors such as hubs classified mistakenly as low degree nodes or vice-versa.
Second, we have considered SI models, where nodes do not recover, and all nodes are finally reached if the network is connected. More realistic models consider that nodes do not remain infectious at all times.A preliminary study using the SIR framework yields similar results \cite{Note1}, but we leave further investigations of the role of the recovery parameters and of more complex models to future works (e.g. including a latency period and/or asymptomatic state).

Finally, we have considered a limited series of data sets as substrate, and the classifier performance depends on the network characteristics and on the networks used to produce the training set. 
Other data sets might have different properties, such as e.g. geometric embeddings, which could impact the spreading properties; additional features might then  be added to the classifier.
Overall, it is expected that the classifier cannot be fully general (trained using a randomly chosen network and tested on another one), since the properties of a spread depend on the network's structure. As obtaining detailed knowledge of the structure of networks on which spreading processes occur is challenging \cite{eames2015challenges,nuria2020mobile},
we have started to explore the generalizability of the classifier to a new data set, showing 
the possibility to train it using surrogate data sets which respect the known statistical properties of the new data. 
These considerations open the door to future studies investigating how different network structures affect the classifier's performance, which properties are the most important for building surrogate data, and to find new algorithms to this aim.

\vspace{1cm}

\paragraph*{Appendix on higher-order models of contagion.}
Higher-order models of contagion can be defined on hypergraphs or simplicial complexes, in which a hyperedge of size $m$ represents an interaction among a group of $m$ nodes (simplicial complexes are hypergraphs $\mathcal{H}$ such that, for each hyperedge --simplex-- $e=\{i_1,...,i_m\}$, all subsets of $e$ are also hyperedges of $\mathcal{H}$).
We use the \textit{simplicial contagion}~\cite{iacopini2019simplicial} model, considering simplices up to the second order, i.e. three nodes interactions, and neglecting structures of higher orders (which thus appear only as decomposed into second order simplices). Each susceptible node can be infected by an infectious neighbour with rate $\beta$ (as in the simple contagion), 
but also if it belongs to a simplex of $3$ nodes in which the two others are infectious. This second process happens with rate $\beta^{\Delta}$ (Fig.~\ref{fig1}(c)) and is thus specific to the
existence of hyperedges (no such process takes place on "empty" triangles which are cliques in the projected networks but not hyperedges).
In addition, we consider the \textit{non-linear hypergraph (NL-hyper) model}~\cite{st2022influential}, which includes contagion processes in interactions of arbitrary sizes: if in a hyperedge of size $m$ there are $n$ infected individuals, each of the remaining $m - n$ susceptible nodes is independently infected with probability $\lambda n^{\nu}$ at each time step (Fig.\ref{fig1}(d)), with $\lambda$ and $\nu$ free parameters. The case $\nu=1$  reduces to a simple contagion, while the non-linearity for $\nu \neq 1$ leads to reinforcement (for $\nu > 1$) or
inhibition (for $\nu < 1$) effects and thus to a
complex contagion phenomenology, as explored in 
 \cite{st2022influential}.

\vspace{1cm}

\paragraph*{Appendix on aggregated networks, hypergraphs and degrees.}
All the data that we use are temporally resolved, giving the specific time of each interaction and, for each data set ${\cal D}$, we consider the aggregated network $G_{\cal D}$: each link in $G_{\cal D}$
represents the fact that the two corresponding individuals have been in contact at least once during the data collection. The degree $k_i$ of an individual is then the number of links involving $i$ in $G_{\cal D}$.
Similarly, we define the aggregated hypergraph $H_{\cal D}$:
a hyperedge of size $m$ represents a simultaneous group of $m$ nodes observed at least once: the availability of temporally resolved data allows thus to distinguish in the aggregated network between hyperedges and cliques. 
We also consider the hypergraph $H^3_{\cal D}$ restricted to hyperedges of size at most $3$: each hyperedge of larger size in $H_{\cal D}$ is simply decomposed into all its groups of nodes of size $3$, and we denote by $k^\Delta_i$ the number of hyperedges of size $3$ to which $i$ belongs in $H^3_{\cal D}$. 

\vspace{1cm}

\paragraph*{Appendix on ROC curve.}
The Receiver Operating Characteristic curve is a parametric method to assess the possibility to classify data. 
Let us consider the case of classification between threshold and non-threshold models observing instances of correlation $C_1$. The curve is built as follows:
given the parameter $c \in [-1,1]$, we classify each instance
having $C_1 \ge c$ as resulting from a threshold model.
If the instance was really produced by the threshold model, it is a true positive (TP), and else a false positive (FP). If the instance instead has a correlation $C_1 < c$, it is classified as resulting from one of the simple, simplicial or NL-hyper processes: if this is correct, it is a true negative (TN), while if it was a threshold process it is a false negative (FN). The ROC curve presents, as $c$ varies, the true positive ratio $TPR = TP/(TP+FN)$, i.e. the fraction of instances of threshold model that are correctly classified, versus the false positive ratio $FPR= FP/(FP+TN)$, i.e. the fraction of instances of the other models that are wrongly classified. 
This example of classification leads to the resulting curves reported in Fig.~\ref{fig2}(b),(d),(e).

\bibliography{bib}
\clearpage
\newpage
\onecolumngrid

\appendix
\setcounter{figure}{0}
\setcounter{equation}{0}
\setcounter{table}{0}
\renewcommand{\thefigure}{S\arabic{figure}}
\renewcommand{\thesection}{S\arabic{section}}
\renewcommand{\thesubsection}{S\arabic{section}.\arabic{subsection}}
\renewcommand{\thetable}{S\arabic{table}}
\renewcommand{\theequation}{S\arabic{equation}}


\section*{Distinguishing simple and complex contagion processes on networks: Supplementary Material}

\section{Data description and generation of interaction networks and hypergraphs}
\label{SMsec_data}

The data sets we consider describe interactions between individuals, either in terms of physical proximity or online, and are temporally resolved.
We thus preprocess these data sets in order to build the  three types of structures needed for the numerical simulations, for each data set ${\cal D}$: 
the temporally aggregated pairwise network $G_{\cal D}$, 
the aggregated hypergraph with all orders of interactions $H_{\cal D}$, 
and the aggregated hypergraph restricted to second order (i.e., interactions of size $3$ at most) $H^3_{\cal D}$.

First we build the aggregated network $G_{\cal D}$: we consider each interaction between two individuals as a pairwise link of a static network with weight given by the number of times (consecutive or not) that the connection has appeared in the temporal data. The aggregated network is used to simulate the simple and threshold contagion models, and the contagion events of the simplicial contagion model involving pairwise links.

\begin{figure}[thb]
\includegraphics[width=0.75\textwidth]{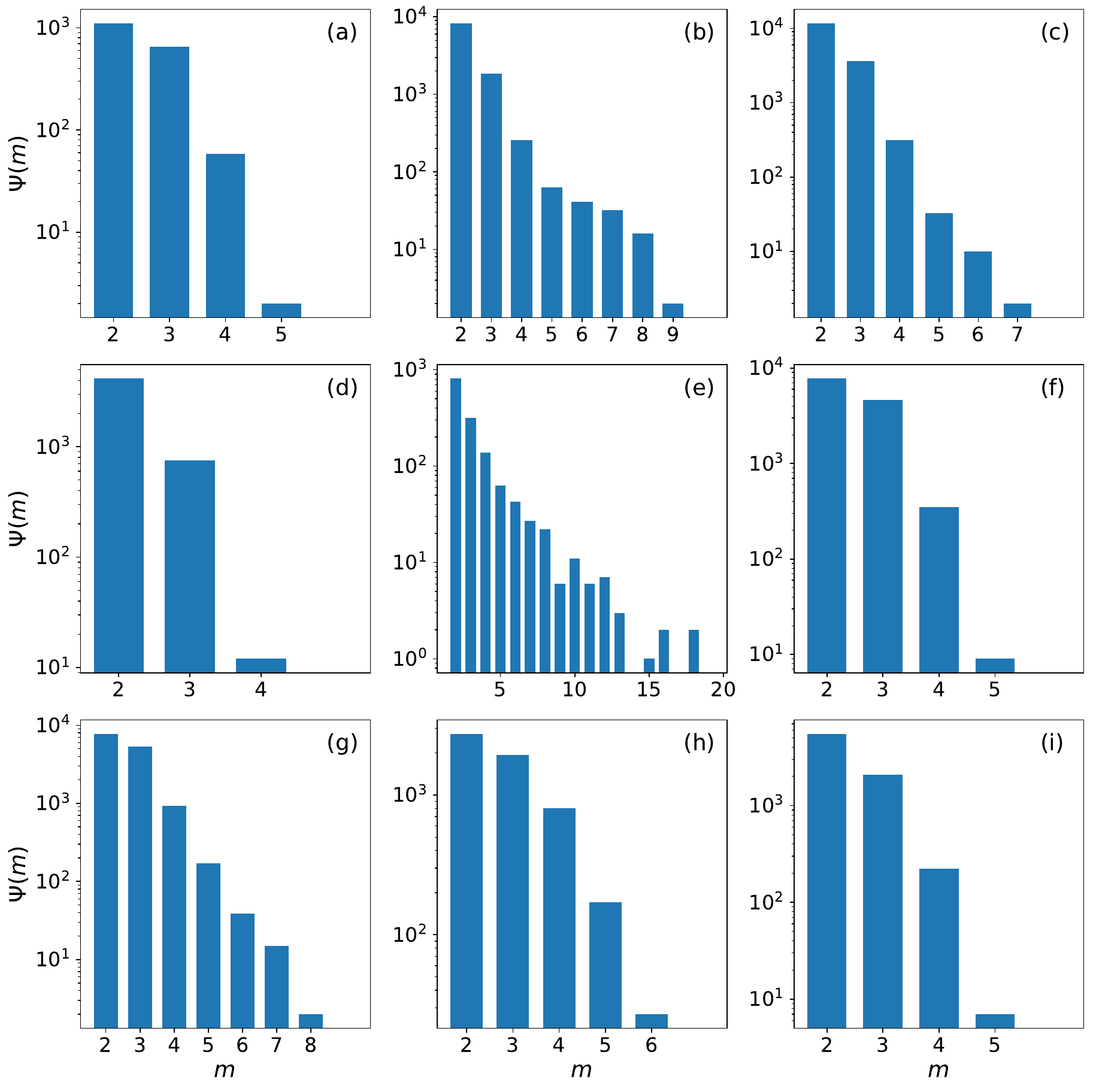}
\caption{\textbf{Hyperedges size distribution.} We show the hyperedges size distribution $\Psi(m)$ for the following data sets: (a) hospital, (b) conference, (c) middle school Utah, (d) workplace, (e) email, (f) primary school France, (g) elementary school Utah, (h) university campus and (i) high-school France.}
\label{fig_Psi_m}
\end{figure}

We then generate the hypergraph $H_{\cal D}$ by considering as hyperedges all the groups of simultaneous all-to-all interactions in the temporal data (see Fig. \ref{fig_Psi_m} for the distributon of the hyperedges size, $\Psi(m)$, of all the data sets). Each hyperedge is stored with its weight given by the number of temporal repetitions. This is used to simulate the NL-hyper contagion process, where each hyperedge is characterized by its own infection probability.

Finally, we consider each hyperedge involving more than $3$ nodes in $H_{\cal D}$ and we decompose it into multiple groups of three nodes (in other words we project these interactions on hyperedges of size 3). The set of the obtained three-nodes interactions, together with the hyperedges of size $2$, forms the aggregated restricted hypergraph
$H^3_{\cal D}$. It is used to simulate the infections taking place in the simplicial contagion model, as it is indeed a simplicial complex (for each hyperedge of size $3$, the three nodes are also linked by $3$ pairwise edges).

The unweighted networks are obtained by setting all the weights equal to $1$.

\begin{table}[hb]
\centering
\begin{tabular}{|c|c|c|c|c|}
	\hline
	data set & Nb. of nodes & Nb. of pairwise & Nb. of size $3$ & Max. hyperedge \\
    &  & connections & hyperedges
    (in $H^3_{\cal D}$) & size\\
	\hline
	Workplace & 217 & 4274 & 772 & 4\\
	Hospital  & 75 & 1139 & 703 & 5\\
	Conference & 403 & 9565 & 2634 & 9\\
	Primary school France & 242 & 8317 & 5138 & 5\\
	High school France & 327 & 5818 & 2370 & 5\\
	University campus (48h) & 495 & 3486 & 2981 & 6\\
	Elementary school Utah (1d) & 339 & 8763 & 6868 & 8\\
	Middle school Utah (1d) & 591 & 13550 & 4453 & 7\\
	Email & 143 & 1800 & 6578 & 18\\
	\hline
\end{tabular}
\caption{\textbf{Size details about the used data sets}. The third column reports the number of links in the aggregated network $G_{\cal D}$. The fourth column reports the number of 
hyperedges of size $3$ in the aggregated hypernetwork restricted to hyperedges of 
size $2$ and $3$ ($H^3_{\cal D}$), i.e. the simplicial complex obtained by projecting larger hyperedges on simplices of 3 nodes. The fifth column reports the number of nodes in the largest hyperedge.}
\label{tab_datasets}
\end{table}

We use the following publicly available data sets of social interactions (see also Table \ref{tab_datasets}):
\begin{itemize}
	\item \textbf{Sociopatterns}: data sets describing pairwise face-to-face interactions among individuals, collected using RFID devices. These data have a temporal resolution of 20 seconds and describe contacts between individuals collected during several successive days. We consider five of these data sets: a workplace~\cite{Genois2018}, a conference~\cite{Genois2018}, a hospital~\cite{vanhems2013estimating}, a high school~\cite{mastrandrea2015contact}, and a primary school~\cite{stehle2011high}.
	\item \textbf{University campus}~\cite{sapiezynski2019interaction}: data set describing pairwise co-location interactions among  students at the campus of the Technical University of Denmark for one month, recorded by the exchange of Bluetooth radio packets between smartphones every 5 minutes for 1 month. We consider a reduced version of the original data set, involving only the first 48 hours, in order to obtain an aggregated network with a moderate density.
	\item \textbf{Contacts among Utah's School-age Population}~\cite{toth2015role}: data set describing pairwise interactions among students in an elementary and a middle school in Utah. The interactions were collected with a temporal resolution of 20 seconds for 2 days via Wireless Ranging Enabled Nodes. We consider a limited version of the original data set, involving only the first day.
	\item  \textbf{Email Enron}~\cite{klimt2004enron}: data set describing group interactions among a company employees via email messages: each node corresponds to an email address and a hyperedge includes the sender and all receivers of an email.
\end{itemize}

\clearpage
\newpage

\section{Models of contagion}
\label{secSM_models}

We describe in details the four different contagion models used in the main text. For the sake of generality we define the models on weighted networks, making use of the weighted adjacency matrix $W$, which reduces to the adjacency matrix for non-weighted networks.
In order to make clear the distinction between models we give for each of them the node infection probability.

In simple contagion processes, infections take place independently on each link of 
$G_{\cal D}$ connecting infected to susceptible nodes at rate $\beta$ multiplied by the weight of the link (normalized such that $\max_{(i,j)}W_{ij} = 1$). The probability that a susceptible node $i$ is infected at a specific time $t$ is hence given by:
\begin{equation}
	p_i(t) = 1 - \prod_{j\in I(t)}(1-\beta W_{ij}),
	\label{eq_simple_W}
\end{equation}
where $I(t)$ is the set of infected nodes at time $t$, so the second term is the product of the independent probabilities that node $i$ is not infected by each of its infected neighbors.

\smallskip

The threshold model also takes place on $G_{\cal D}$. It is a deterministic model where a susceptible node is infected when the fraction of its neighbors that are infected exceeds a  threshold $\theta$. More generally, if the network is weighted, a susceptible node is infected when the weight of its connections with infected nodes divided by the total weight of its connections exceeds the  threshold. The general probability can be described by a Heaviside function:
\begin{equation}
	p(t) = H\biggl(\frac{1}{s_i}\sum_{j\in I(t)}W_{ij}-\theta \biggr) = 
	\begin{cases}
		1 \text{ if } \frac{1}{s_i}\sum_{j\in I(t)}W_{ij} \geq \theta\\
		0 \text{ if } \frac{1}{s_i}\sum_{j\in I(t)}W_{ij} < \theta
	\end{cases}\,. 
	\label{eq_thresh_W}
\end{equation}
where $s_i=\sum_jW_{ij}$ is the total strength of node $i$.

\smallskip

The simplicial contagion model can be considered as a generalization of the simple model on simplicial complexes, where nodes are connected via simplices instead of only pairwise links. We only consider simplices up to the second order, i.e. three nodes interactions, neglecting structures of higher orders (which will only appear decomposed into edges of first and second order, as explained above), i.e., $H^3_{\cal D}$. 
Each susceptible node can be infected via its first-order connections (links), at rate $\beta$ (like in simple contagion), or via its second-order connections (hyperedges of size $3$), if present, at rate $\beta^{\Delta}$. Second order contagions can only happen if the susceptible node is part of one (or more) hyperedges of size $3$ ("closed triangles") where the other two nodes are both infected. Notice that in this case, as the model is defined on a simplicial complex, the two nodes additionally contribute independently to the first-order contagion. All infection probabilities are multiplied by the weight of the involved simplex, via the matrix $W$ at first order and tensor $W^{\Delta}$ at second order, where the element $W^{\Delta}_{ikl}$ is the weight of the closed triangle $(i,k,l)$, normalized by the same factor as the link weights. The probability per unit time that a susceptible node is infected is hence given by the combination of its first and second order interactions with infected nodes: 
\begin{equation}
	p_i(t) = 1 - 
 \prod_{j\in I(t)}(1-\beta W_{ij})
 \prod_{k,l\in I(t)}(1-\beta^{\Delta} W^{\Delta}_{ikl}) \ .
	\label{eq_simplicial_W}
\end{equation}

\smallskip

The non-linear hypergraph contagion model is defined on hypergraphs rather than on simplicial complexes, and  multi-body interactions are modelled through a non-linear infection term. In this case, we consider the SI epidemic model on a hypergraph with hyperedges of arbitrary size (i.e. links, triangles and also higher-order structures up to the maximum hyperedge size $M$): if in a hyperedge $h=(i_1,i_2,...i_m)$ of size $m$ there are $n_h$ infected individuals, each of the remaining $(m-n_h)$ susceptible nodes is independently infected with probability $\lambda n_h^{\nu} W^m_h$. Note that $\nu$ sets the non-linearity, $\lambda \in [0,1]$ and $m \in [2,M]$. All the infection terms are multiplied by the weight of the involved hyperedge, via the tensor $W^m$ at the $(m-1)$th order, whose element $W^m_h=W^m_{i_1,i_2,...,i_m}$ is the weight of the hyperedge $h=(i_1,i_2,...i_m)$. By assuming that each hyperedge produces an independent contagion process, the probability per unit time that a susceptible node $i$ is infected is hence given by:
\begin{equation}
	p_i(t)=1-\prod_{m=2}^M\prod_{h \in \mathcal{E}^m_i} (1-\lambda n_h^{\nu} W^m_h)
\end{equation}
where $\mathcal{E}^m_i$ is the set of all hyperedges of size $m$ containing $i$ and $n_h$ is the number of infected individuals in $h$.


\section{Classification tasks for processes on all the data sets considered}

\subsection{SI models on unweighted networks}

\begin{figure*}[thb]
\includegraphics[width=0.41\textwidth]{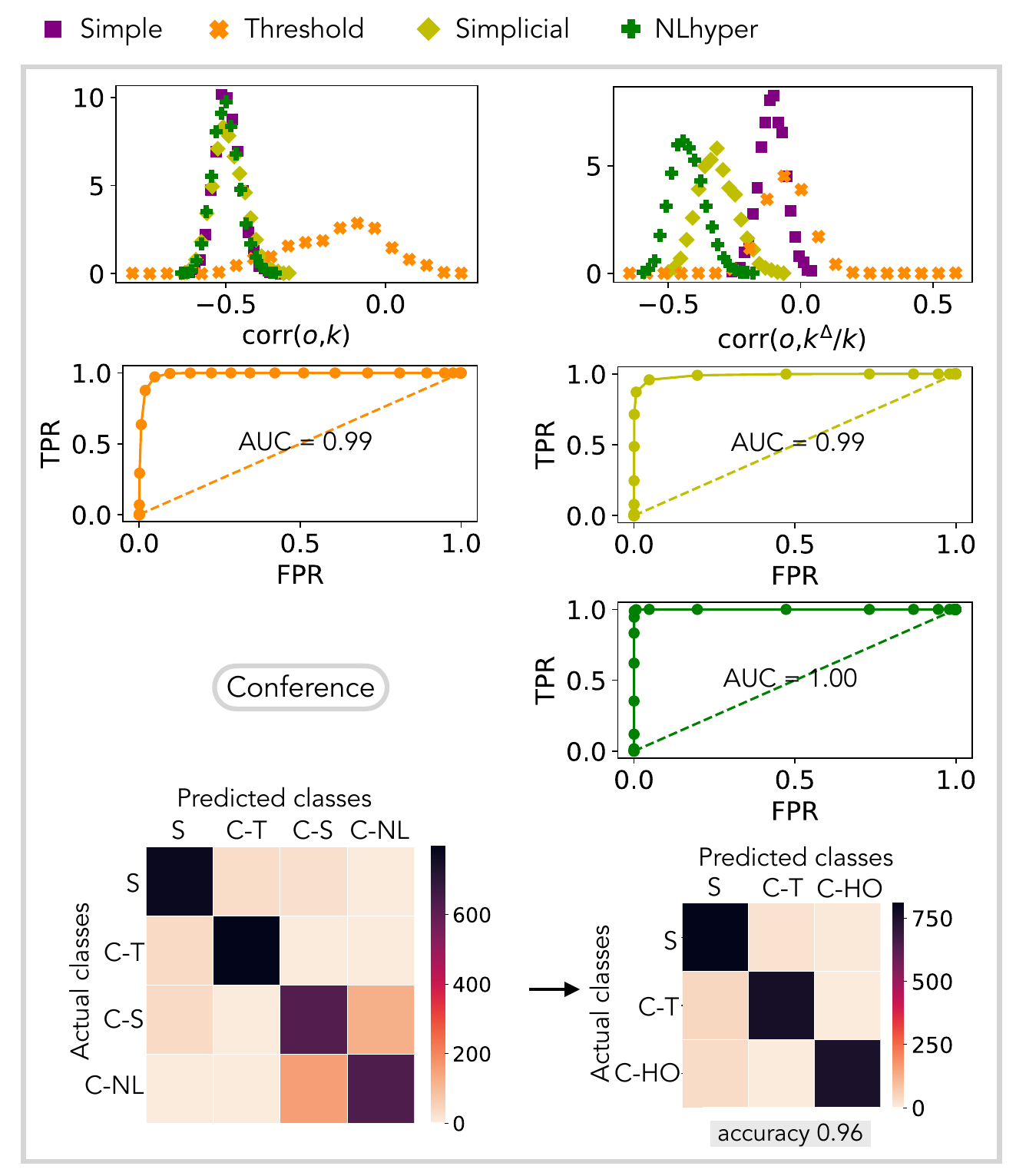}
\includegraphics[width=0.41\textwidth]{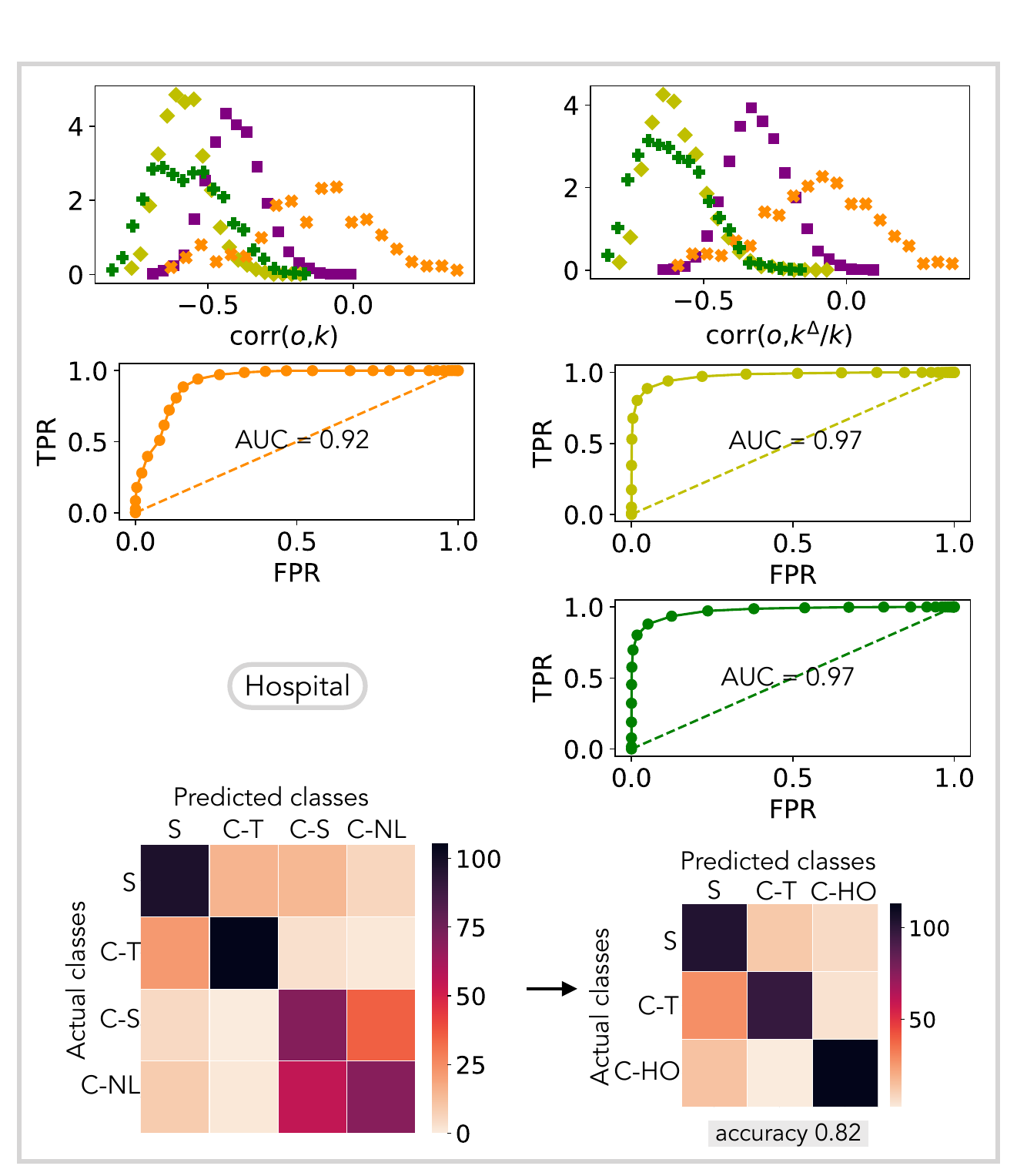}
\includegraphics[width=0.41\textwidth]{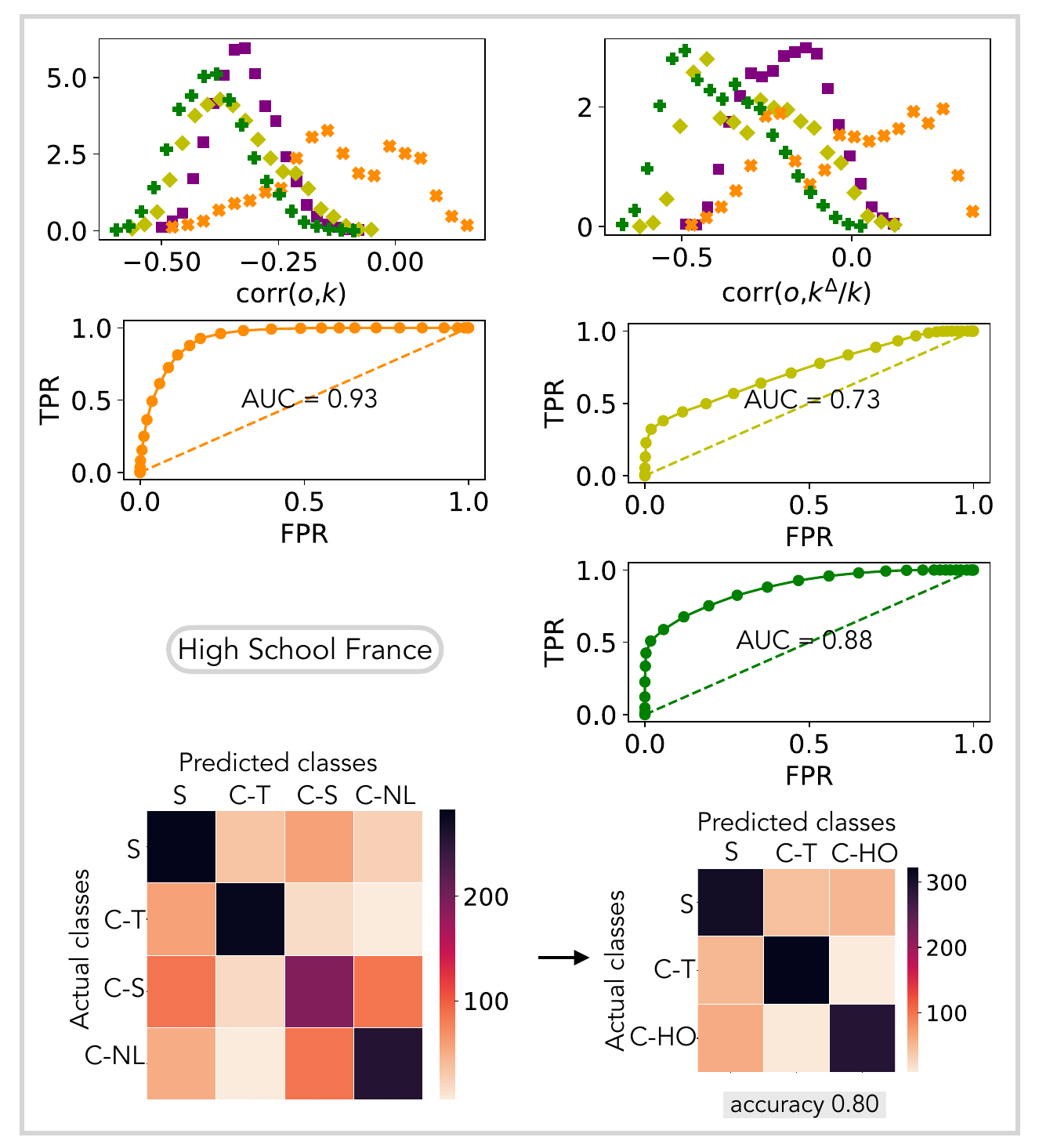}
\includegraphics[width=0.41\textwidth]{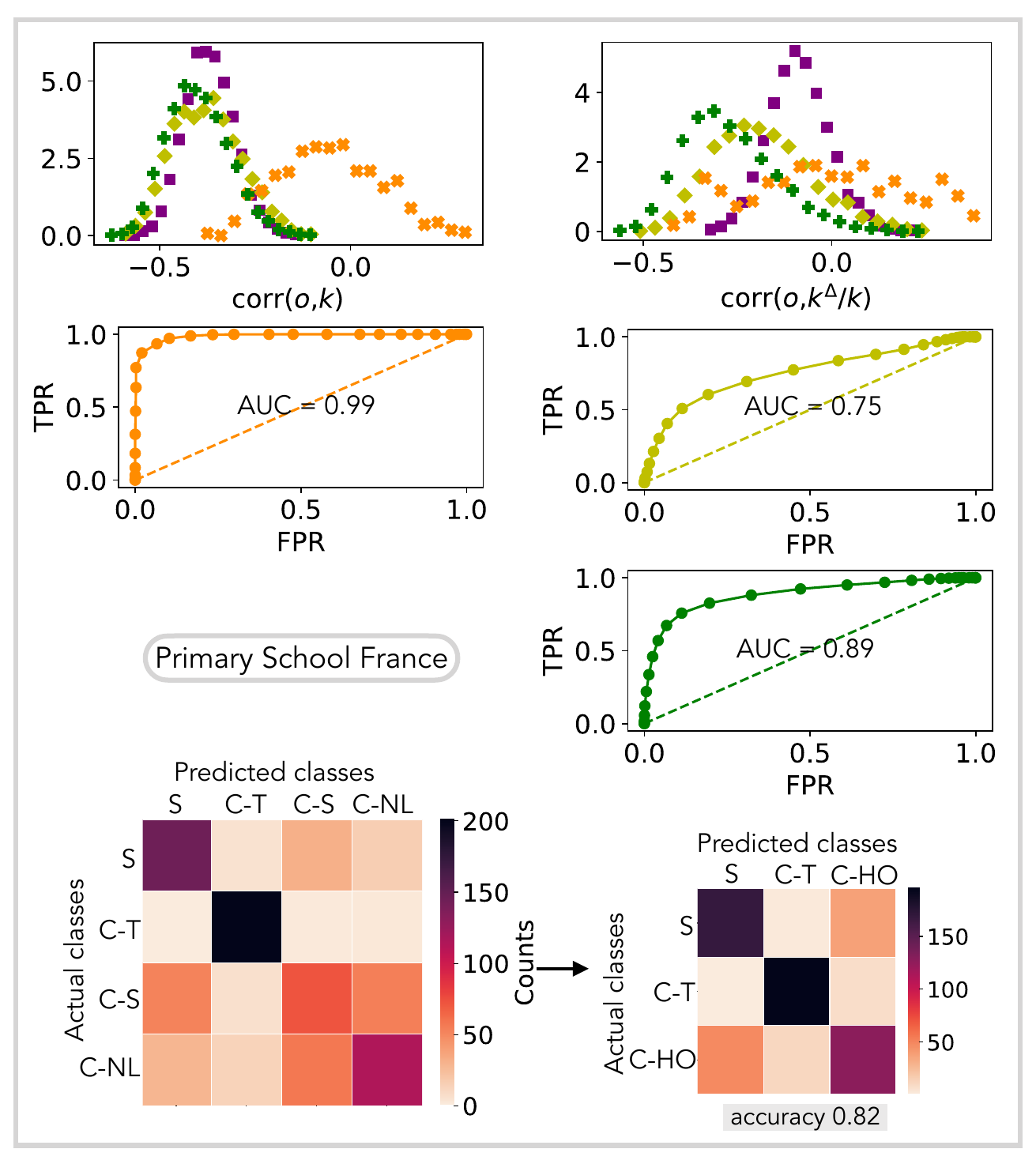}
\caption{\textbf{Unweighted SI model.} 
Analogous of Fig.~2 and Fig.~3(a) of the main text for four other data sets: conference, hospital, high school, and primary school. For each data set, the top panels report the distributions of $C_1$ and $C_2$. The three panels below give: in orange, the ROC curve when using $C_1$ to classify 
threshold model processes against the other three; in light green, the ROC curve when using $C_2$ to classify the simplicial model against simple and threshold; in green, the ROC curve when using $C_2$ to classify the NL-hyper model against simple and threshold. The bottom panels report the confusion matrices with 4 classes (simple, complex threshold, complex simplicial and complex NL-hyper) and with 3 classes (simple, complex threshold and complex higher-order).
For the stochastic models (simple, simplicial and NL-hyper) 1000 realizations are implemented for each parameter setting, always starting with one random infected node. For the deterministic threshold model, we use only one realization for each different initial condition, i.e. one for each network node, for each parameter value. Confusion matrices are obtained using the same number of instances for each model.
Parameter values: $\beta \in \{0.005, 0.008, 0.014, 0.023, 0.039\}$ for both simple and simplicial models, $\beta^{\Delta} = 0.8$, $\lambda \in \{0.0001, 0.001, 0.006, 0.011, 0.015\}$, $\nu=4$,
$\theta \in \{0.01, 0.02, 0.03, 0.04, 0.05, 0.06, 0.07, 0.08, 0.09, 0.10\}$. 
Additional data sets are shown in Fig.~\ref{figSM_distr_unweighted2}.
\label{figSM_distr_unweighted1}}
\end{figure*}

\begin{figure*}[thb]
\includegraphics[width=0.41\textwidth]{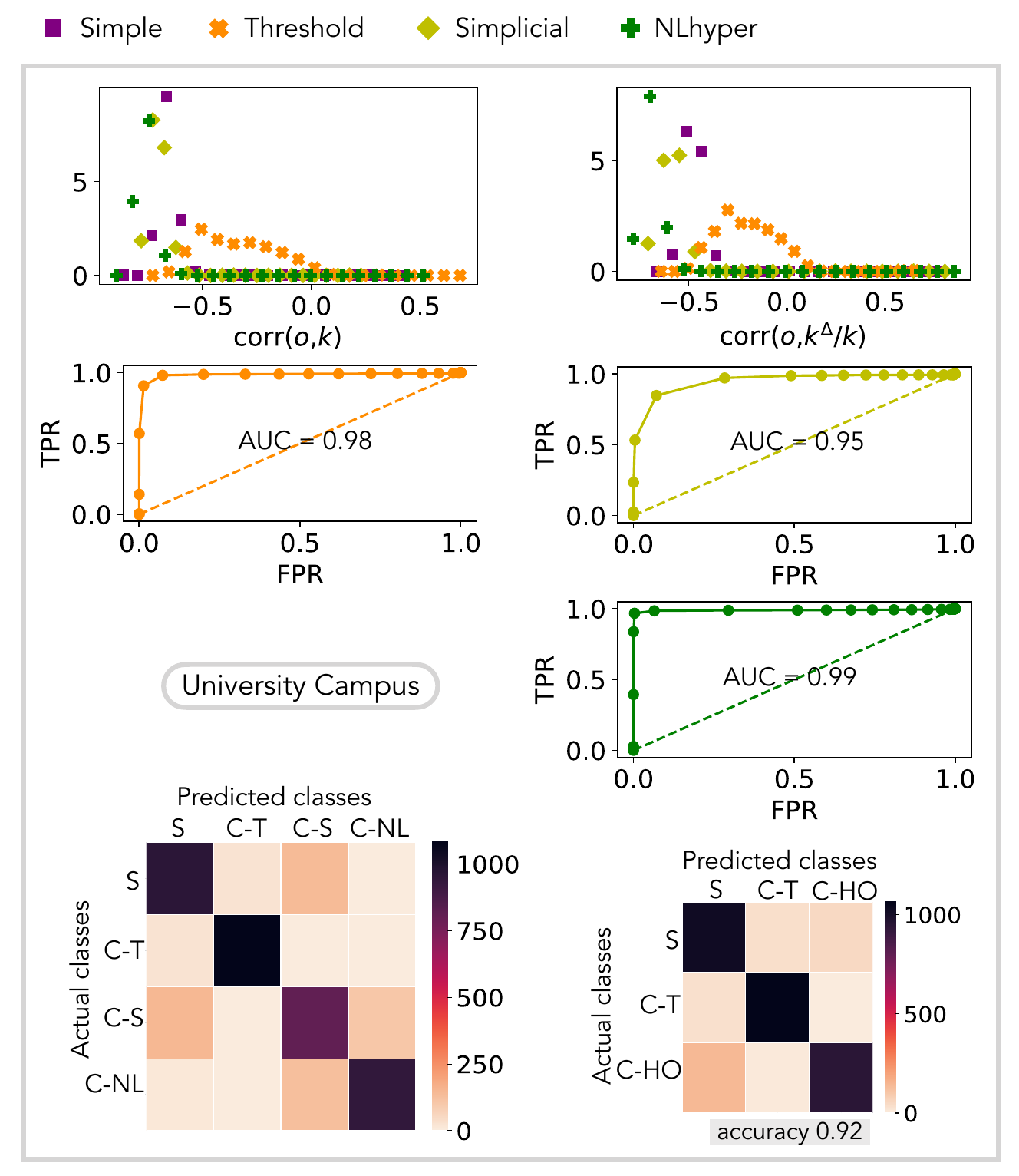}
\includegraphics[width=0.41\textwidth]{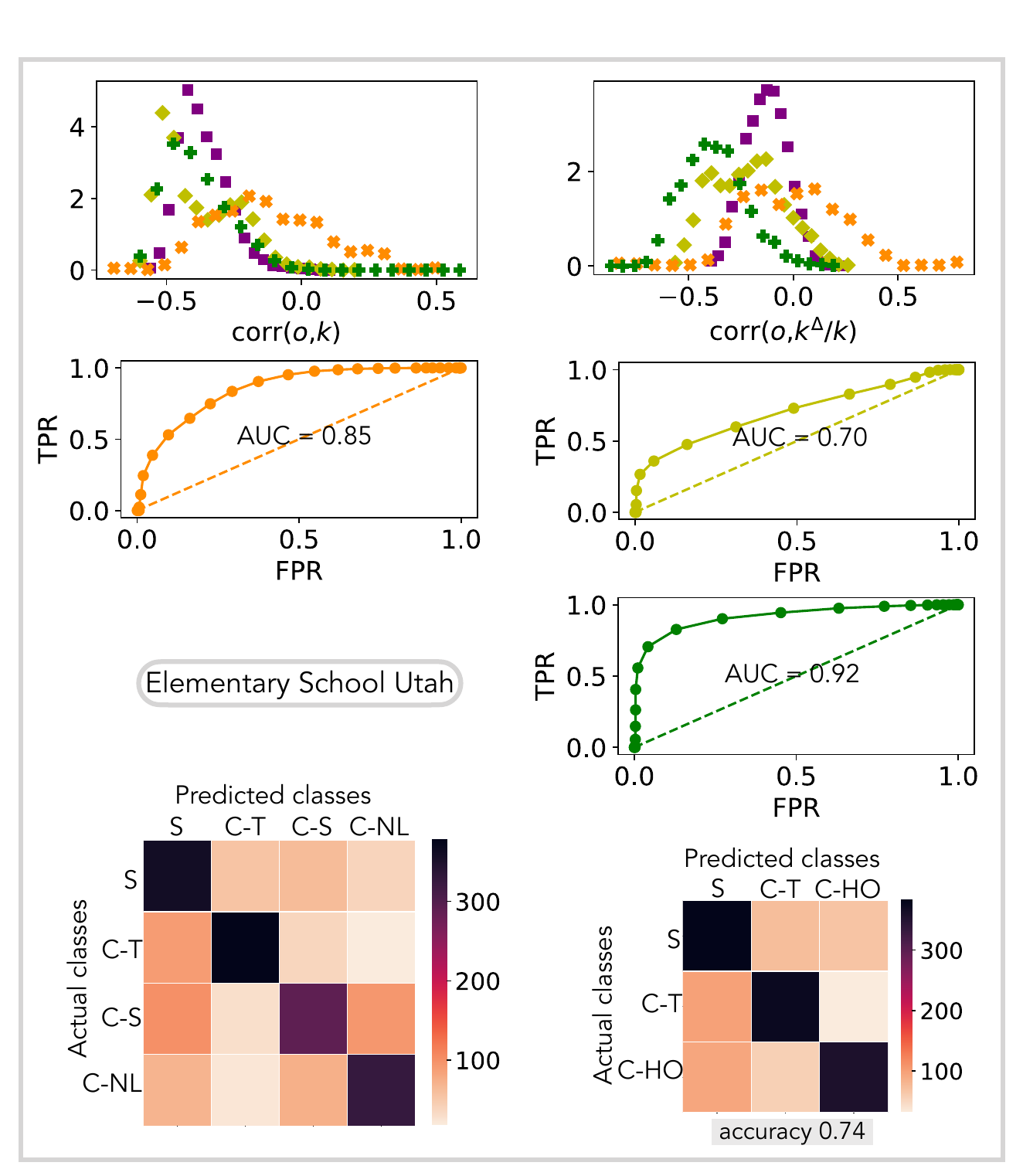}
\includegraphics[width=0.41\textwidth]{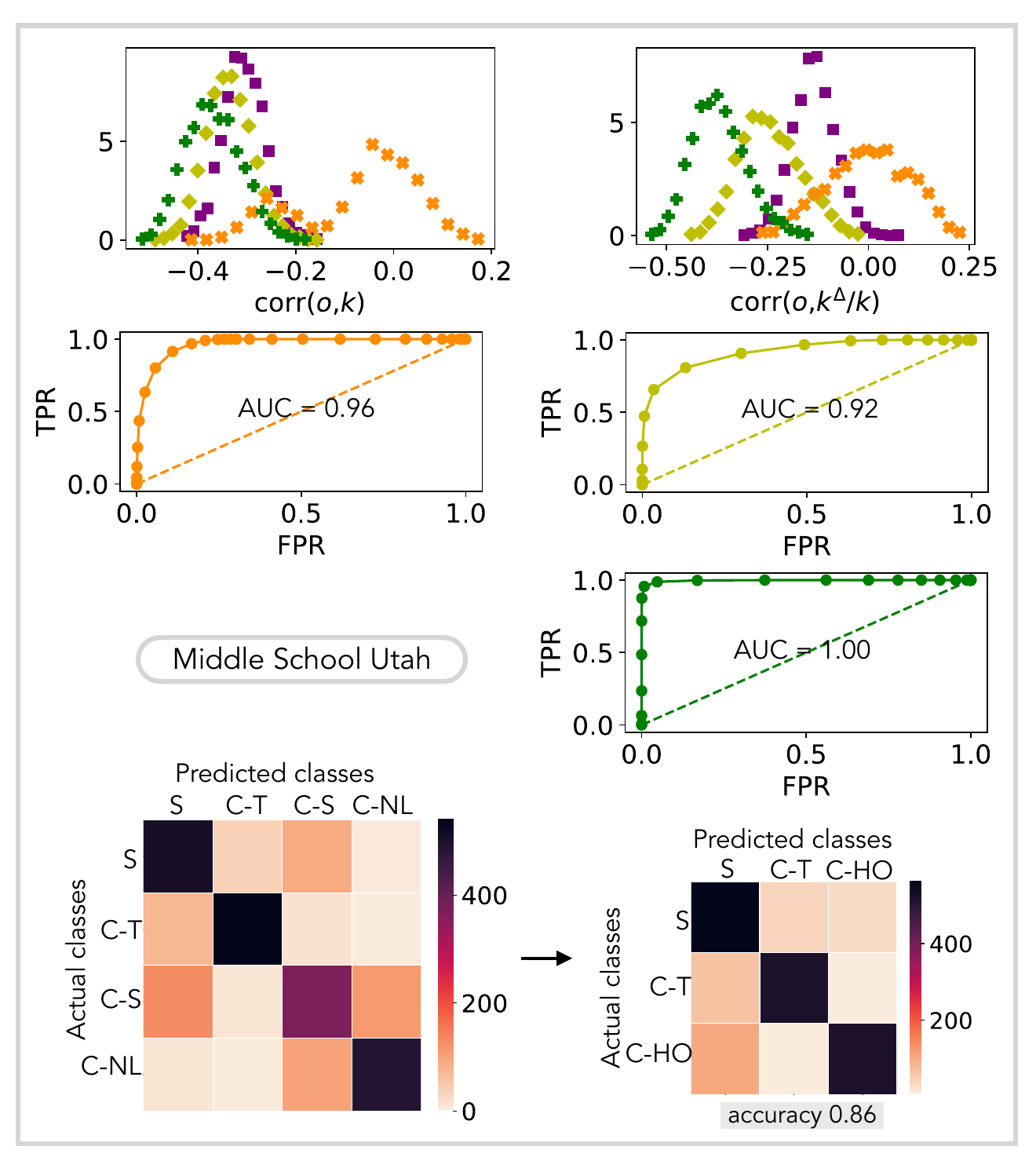}
\includegraphics[width=0.41\textwidth]{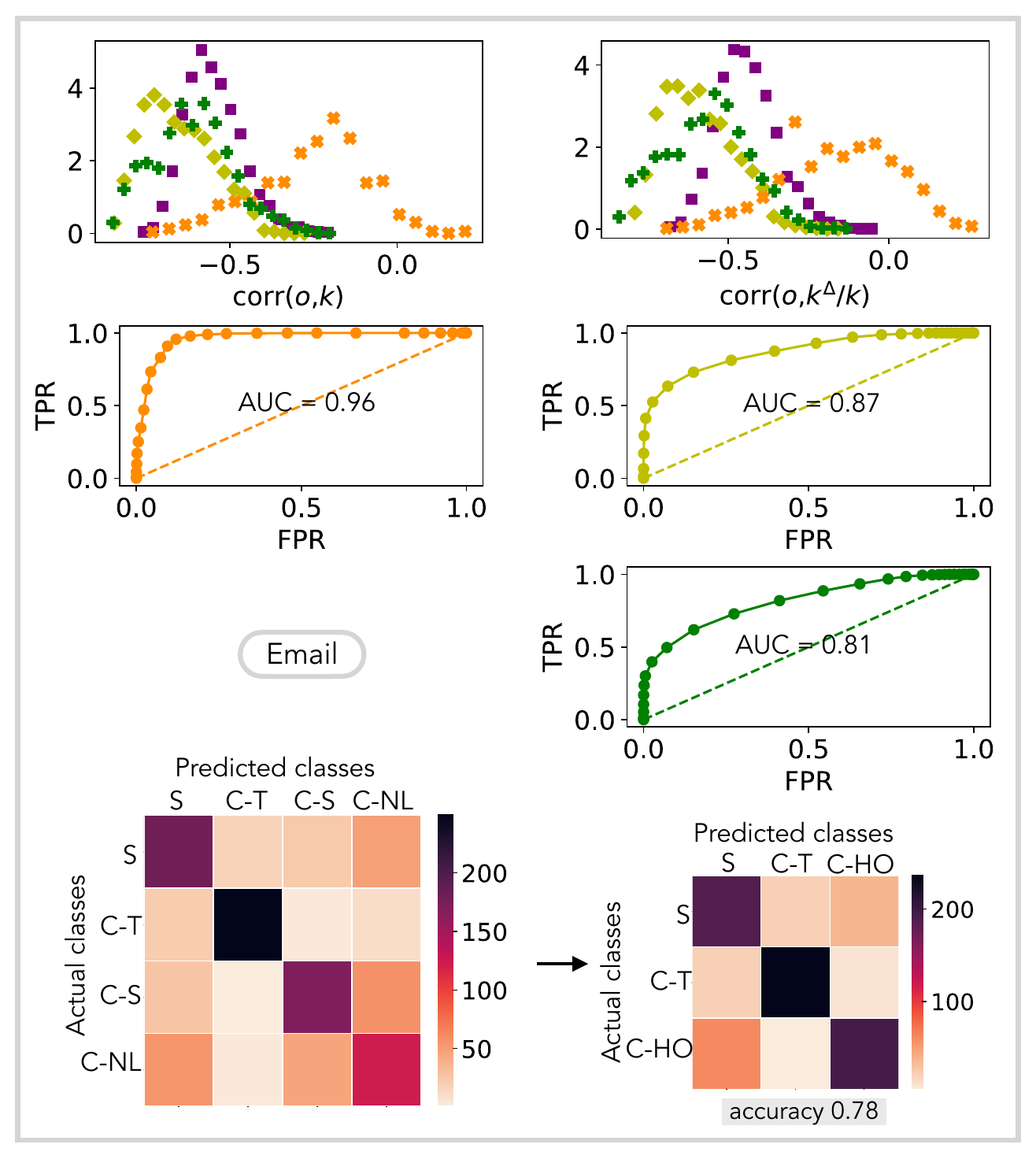}
\caption{\textbf{Unweighted SI model.} 
Same as Fig.~\ref{figSM_distr_unweighted1}
for the data sets: University campus, elementary school in Utah, middle school in Utah, and email.
\label{figSM_distr_unweighted2}}
\end{figure*}

We repeat here the analyses reported in the main text with the data described in section~\ref{SMsec_data}. The results for SI model on unweighted networks are shown in Figs.~\ref{figSM_distr_unweighted1} and \ref{figSM_distr_unweighted2}, which report the distributions of the correlations $C_1$ and $C_2$, the resulting ROC curves and the confusion matrices, similarly to Fig.~2 and Fig.~3(a,b) in the main text. The figures show that the results obtained in the main text can be generalized to other networks, with a classification accuracy always above 0.7.
We however notice that the classification performances are worse for the networks of the two schools in France and the elementary school in Utah, with in particular 
a lower ability to distinguish between simple and simplicial model when using $C_2$ (even if the AUC is still above 0.7). This could be due to the existence of a strong community structure in these networks, but future work is needed to investigate how different network topologies affect the classification based on the measure of correlations.

\clearpage
\newpage
\subsection{Feature importance}
\begin{figure*}[thb]
\includegraphics[width=0.7\textwidth]{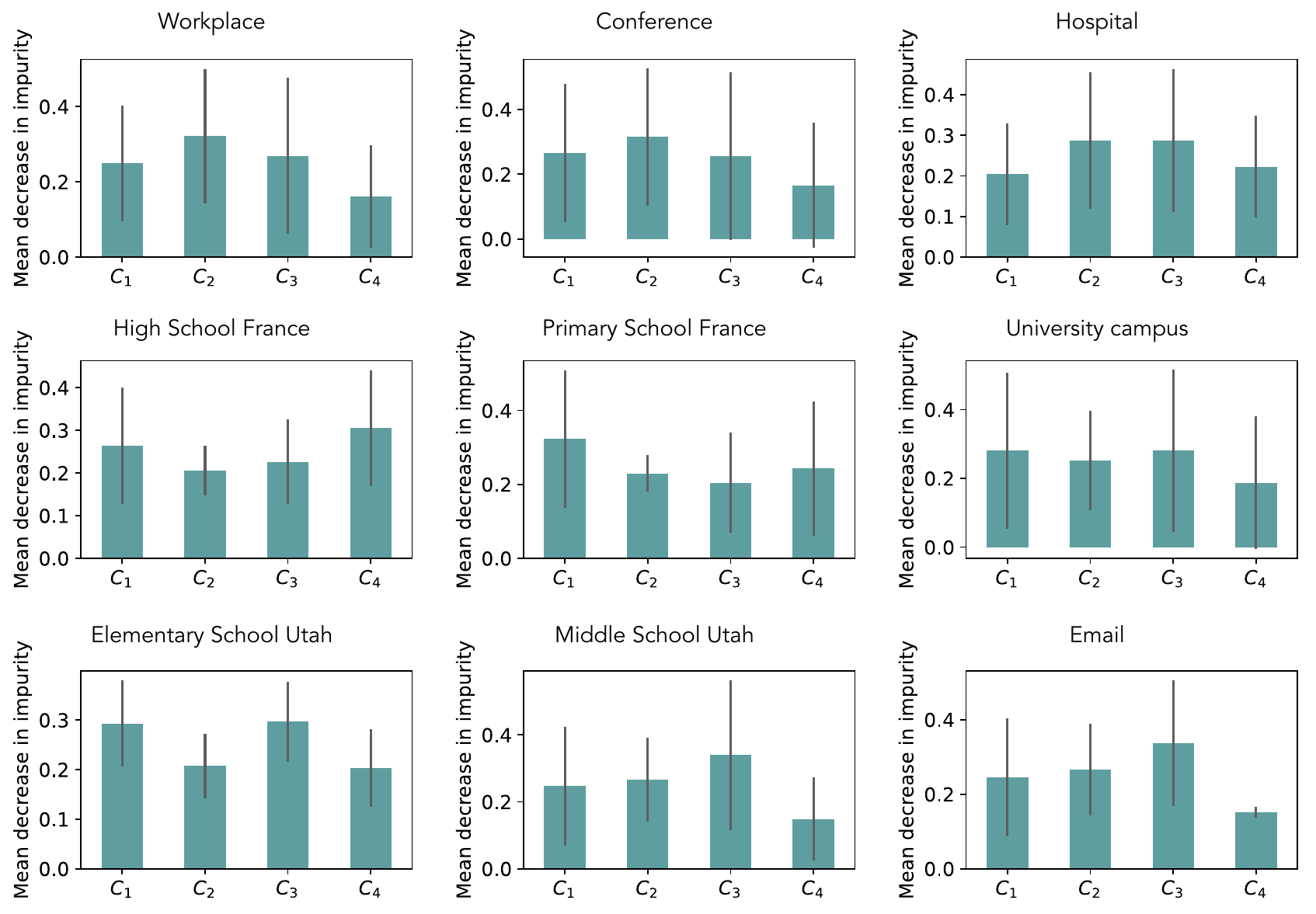}
\caption{\textbf{Feature importance.} 
We report the importance given by Random Forest to each of the four correlation measures, for each classification task 
 relative to figure 3(a) of the main text and to figures \ref{figSM_distr_unweighted1} and \ref{figSM_distr_unweighted2}.}
\label{figSM_feat_imp}
\end{figure*}

The Random Forest classification takes in input, as stated in the main text, four features: $C_1$, $C_2$, $C_3$ and $C_4$. The importance given by the algorithm to these features can be computed using the Mean Decrease in Impurity or Gini method~\cite{breiman2001random}. The results, shown in Fig.~\ref{figSM_feat_imp} for the SI unweighted cases, show that all the features appear important, with no feature that stands out above the rest.

\clearpage
\newpage

\subsection{SI models on weighted networks}
\label{secSM_weight}

The classifier can be applied to spreading processes occurring on weighted networks: the processes spread according to the links' and hyperedges' weights, 
as described in section~\ref{secSM_models}.
The classifier takes as input the correlations between the order of infection of the nodes and their strengths instead of their degrees.

\begin{figure*}[thb]
\includegraphics[width=0.41\textwidth]{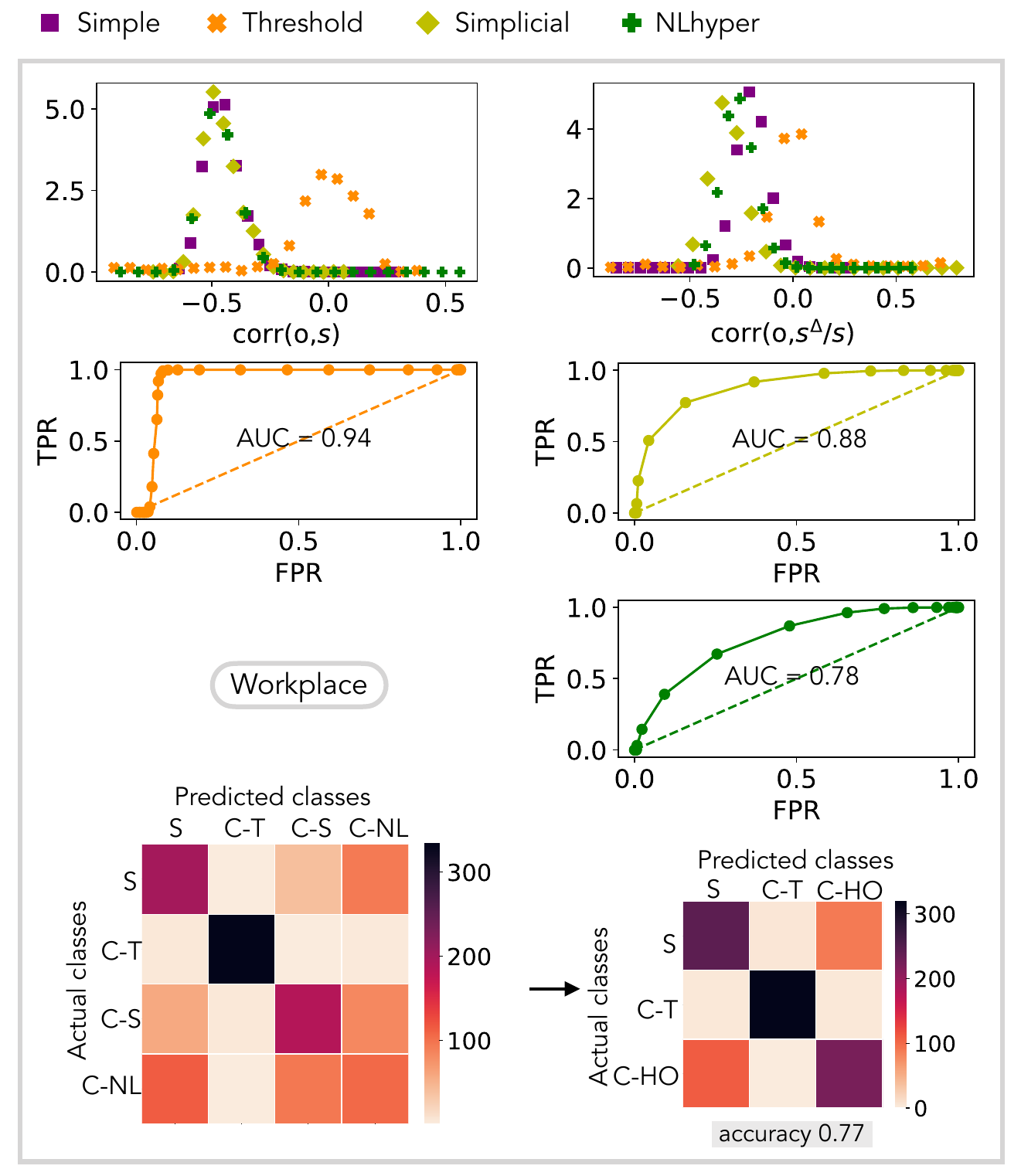}
\caption{\textbf{Weighted SI model.} Analogous to Figs.~2 and 3 of the main text for the workplace data set, using the weighted aggregated networks for the spreading processes. The parameters are chosen as in Fig.~\ref{figSM_distr_unweighted1}, except for $\theta$, which becomes a threshold for nodes strength instead of nodes degree and is consequently adapted, taking the values $\{0.04, 0.06, 0.08, 0.10, 0.12, 0.14, 0.16, 0.18, 0.20, 0.22\}$. 
\label{figSM_distr_weighted0}}
\end{figure*}

\begin{figure*}[thb]
\includegraphics[width=0.41\textwidth]{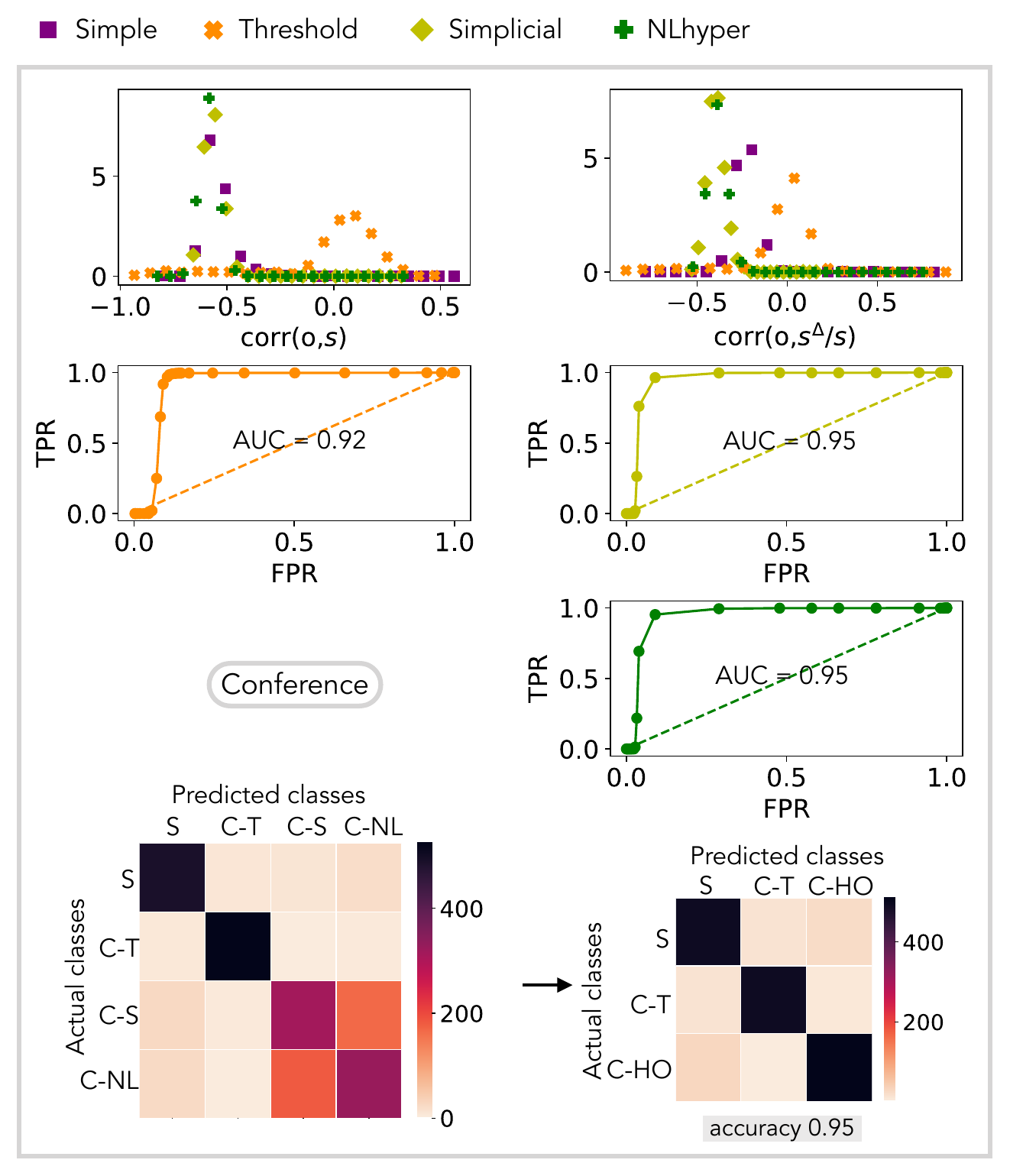}
\includegraphics[width=0.41\textwidth]{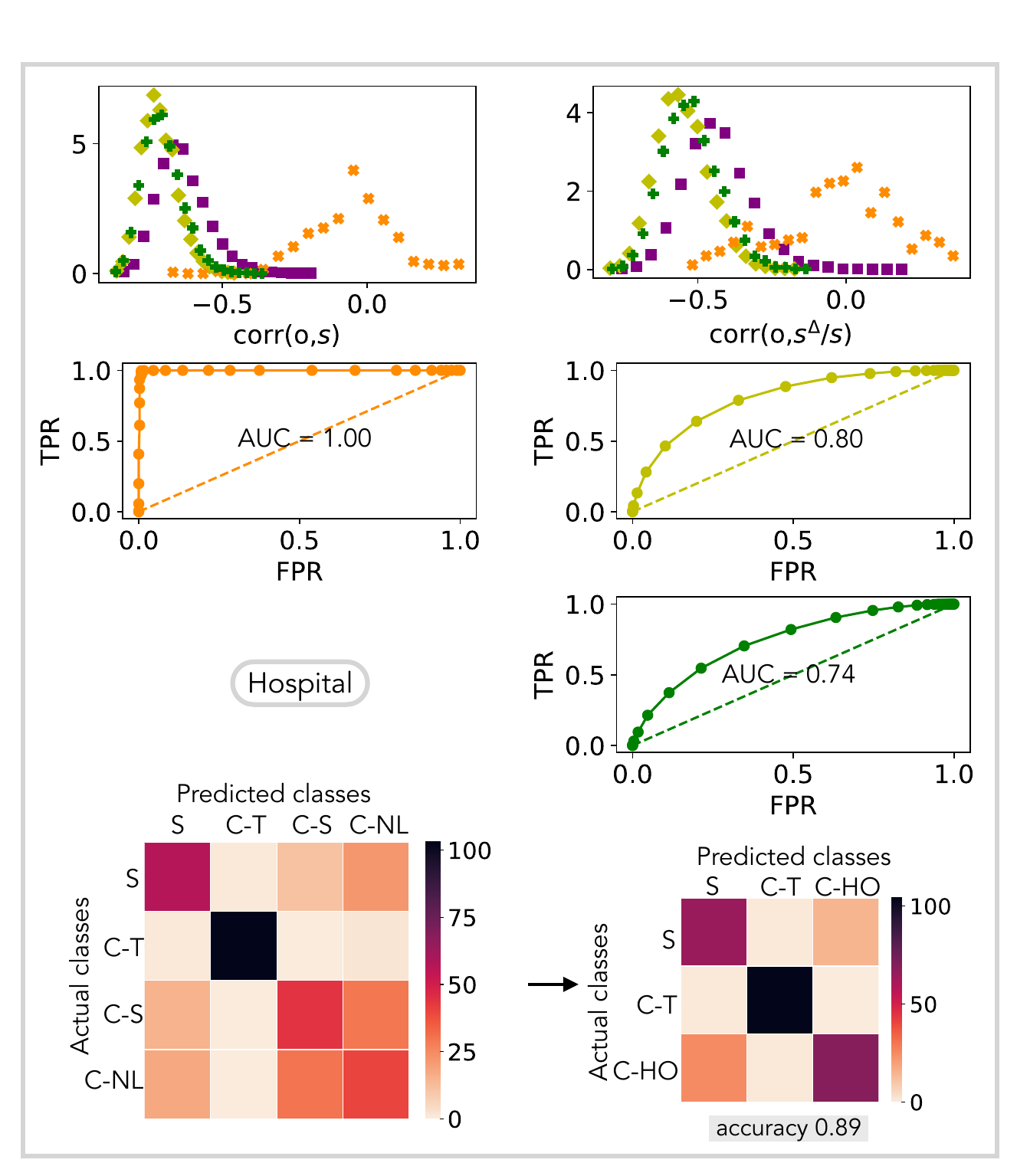}
\includegraphics[width=0.41\textwidth]{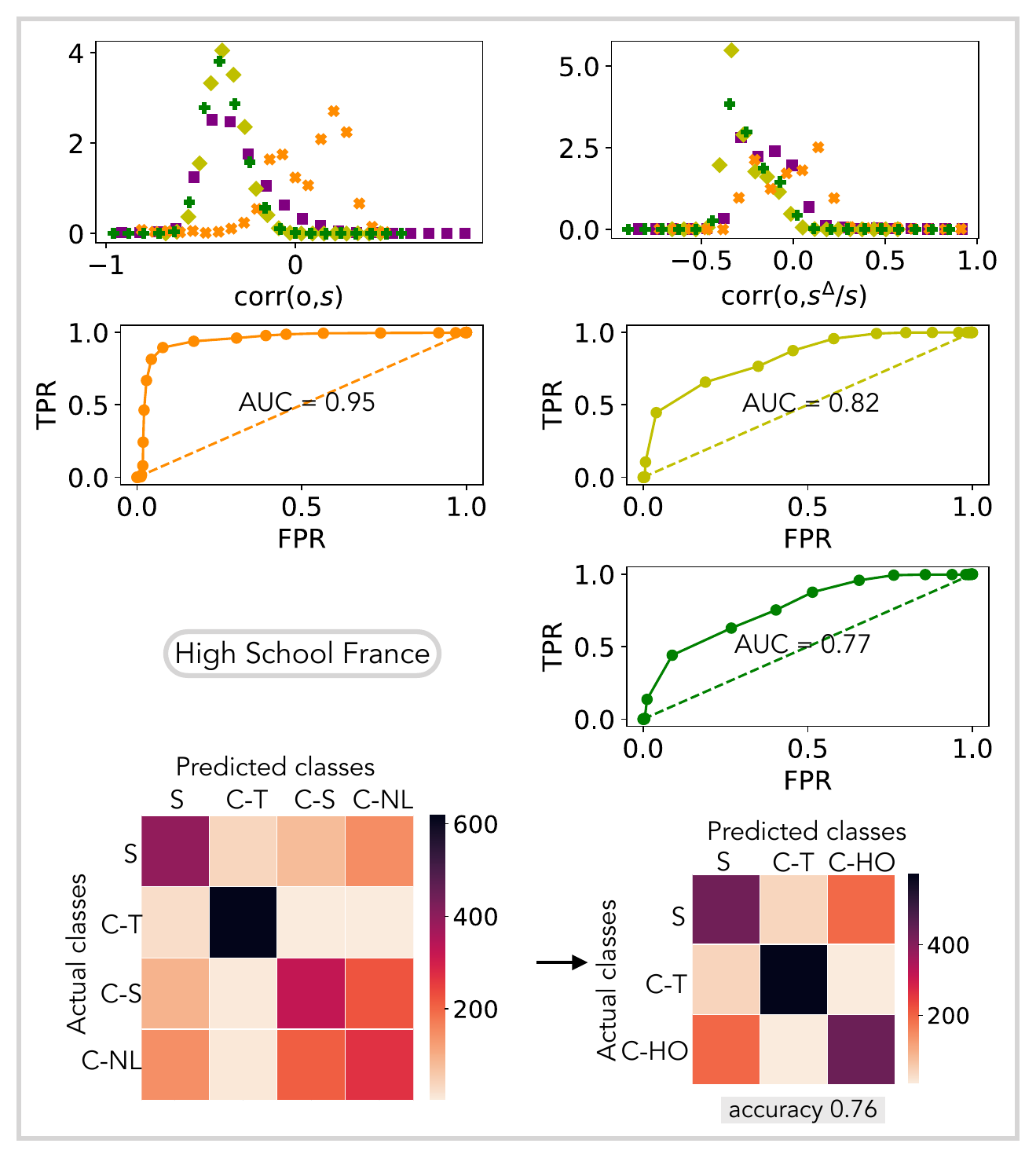}
\includegraphics[width=0.41\textwidth]{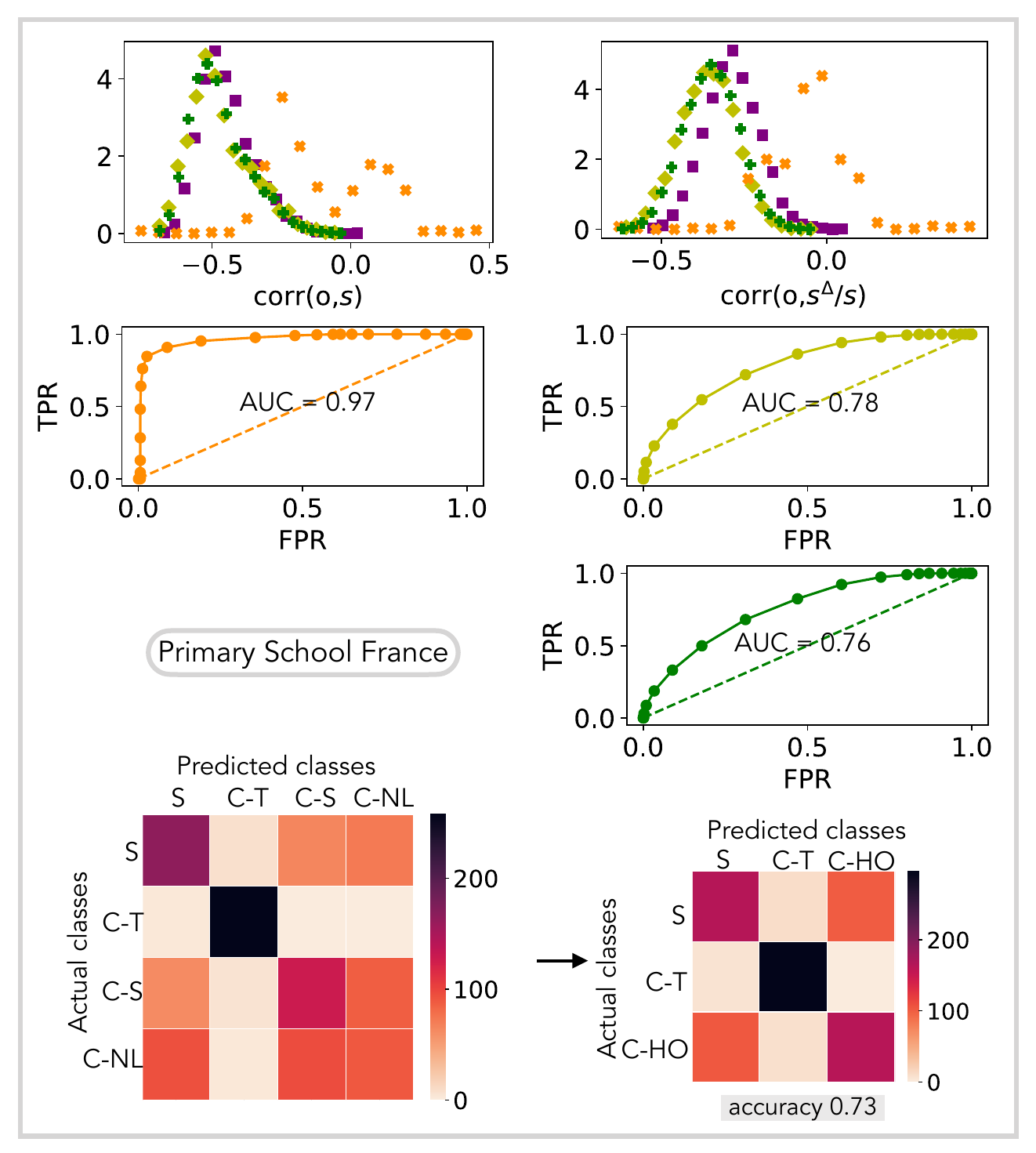}
\caption{\textbf{Weighted SI model.} Analogous of Fig.~\ref{figSM_distr_weighted0} for other networks: conference, hospital, high school in France, and primary school in France. Additional data sets are shown in Fig.~\ref{figSM_distr_weighted2}. 
\label{figSM_distr_weighted1}}
\end{figure*}

\begin{figure*}[thb]
\includegraphics[width=0.41\textwidth]{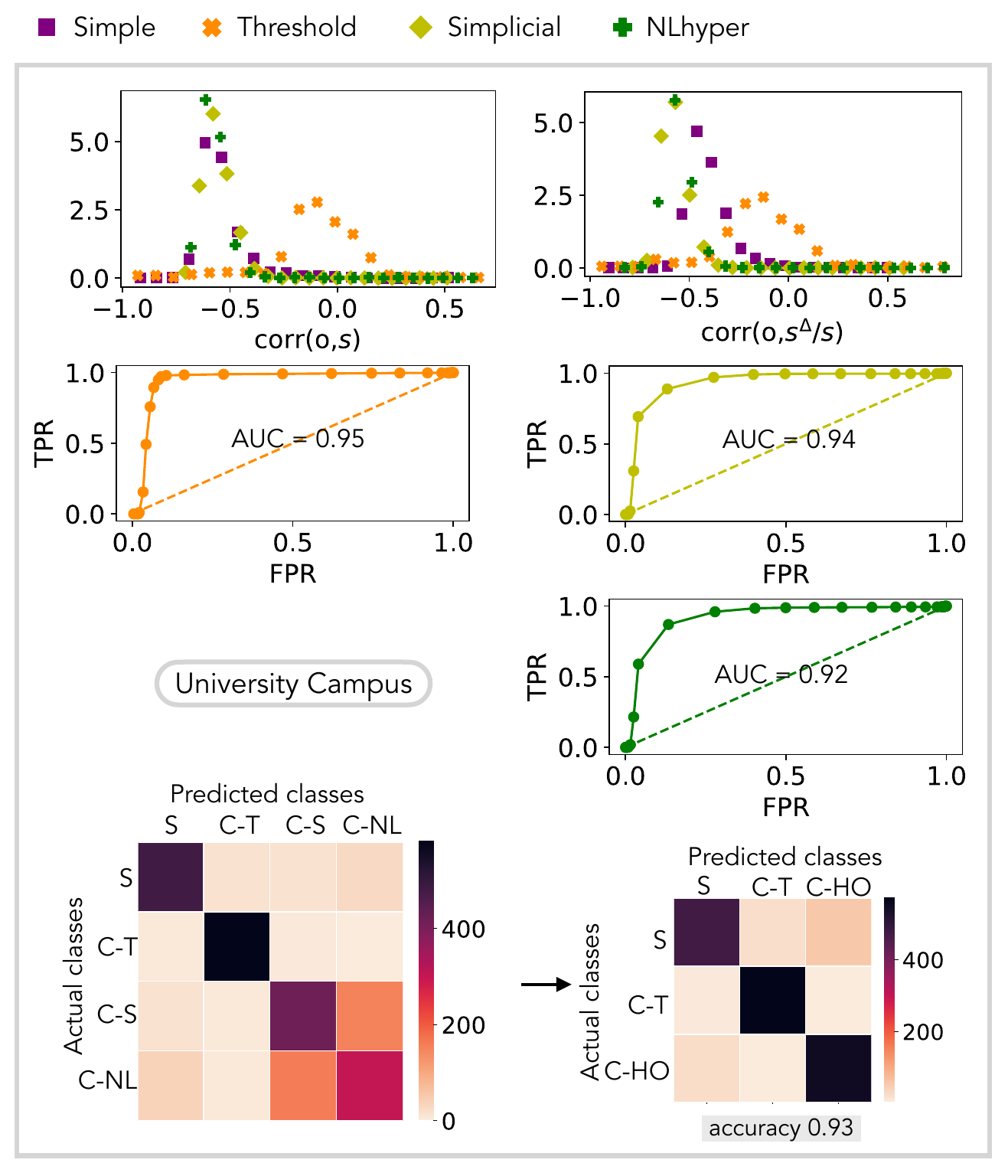}
\includegraphics[width=0.41\textwidth]{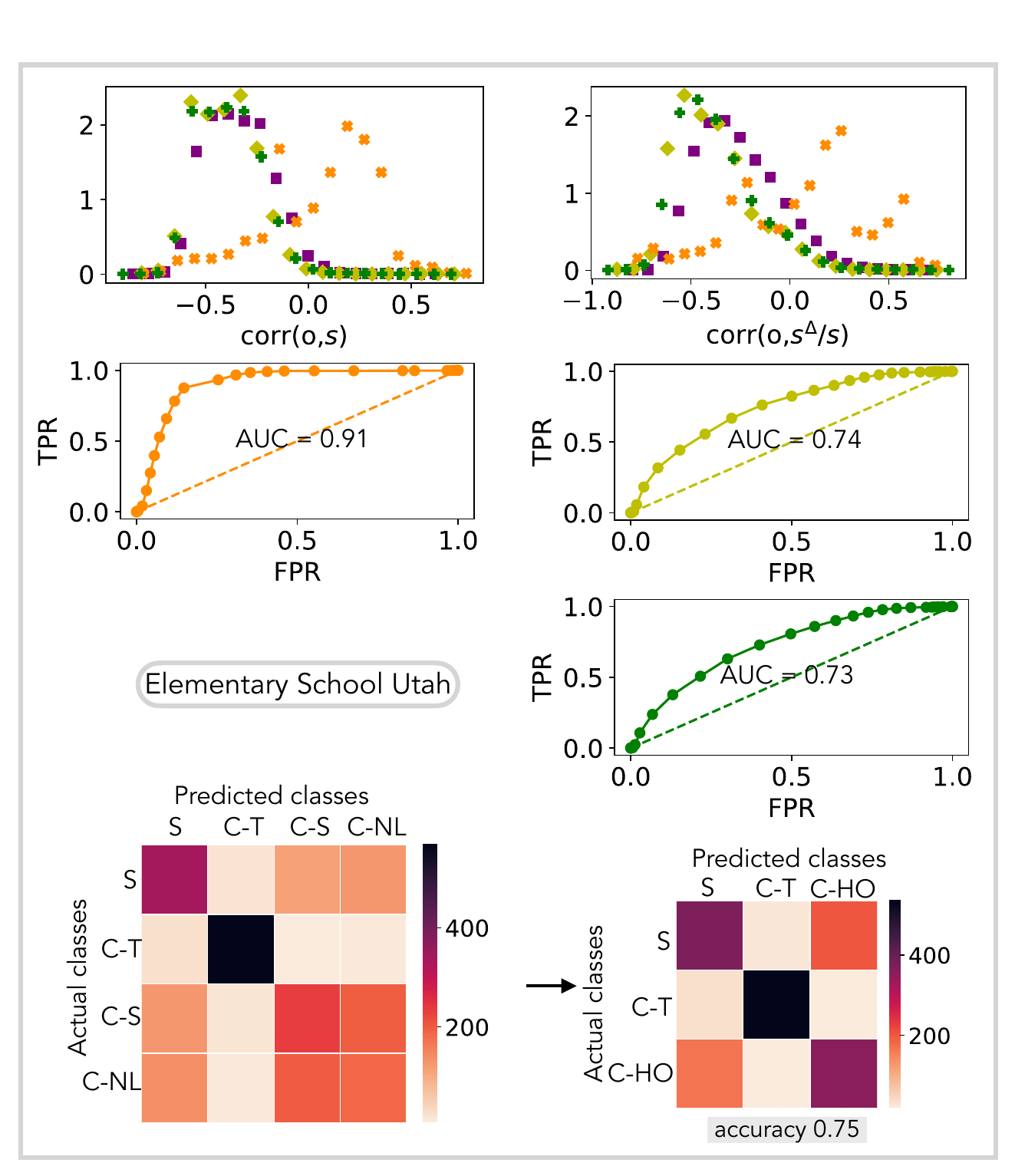}
\includegraphics[width=0.41\textwidth]{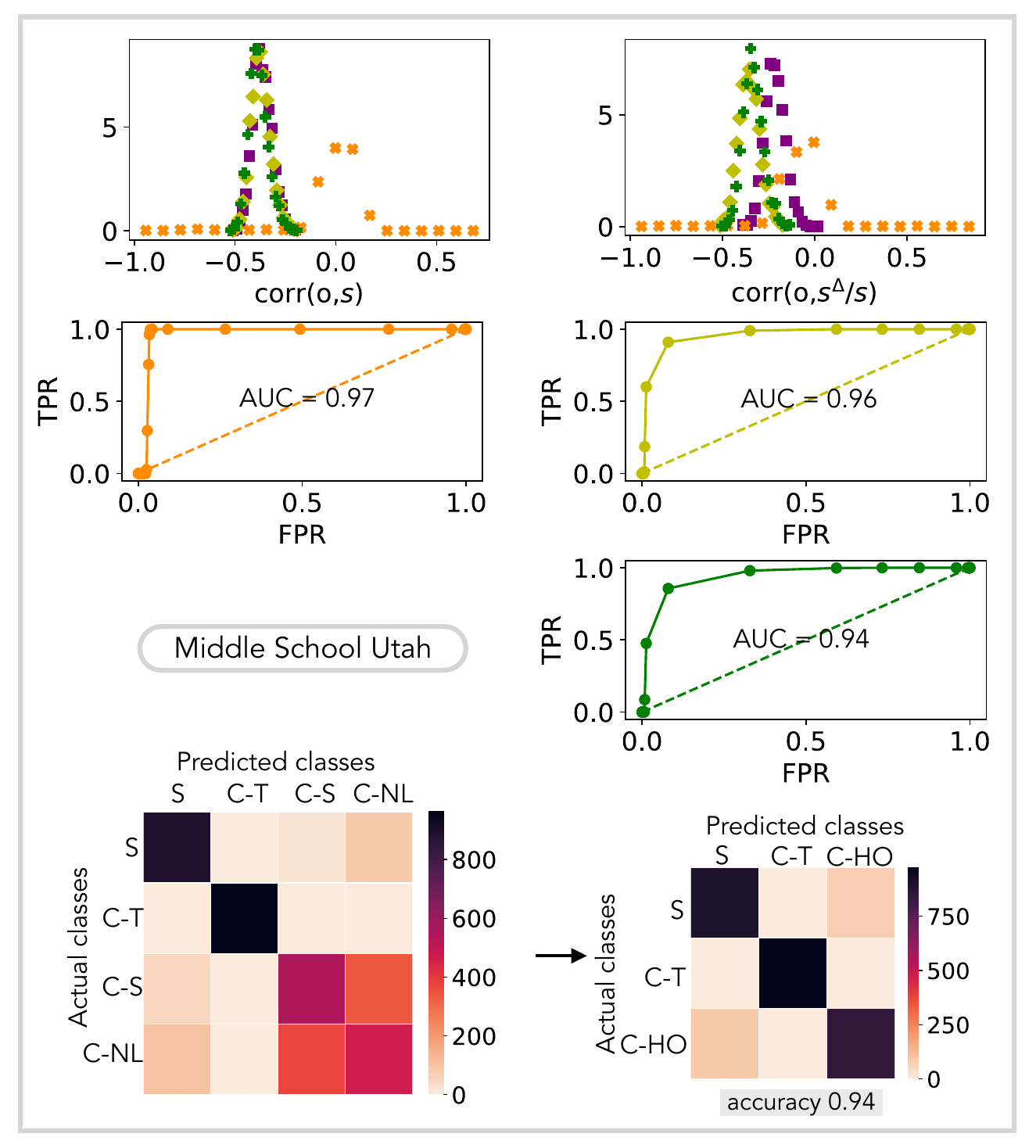}
\includegraphics[width=0.41\textwidth]{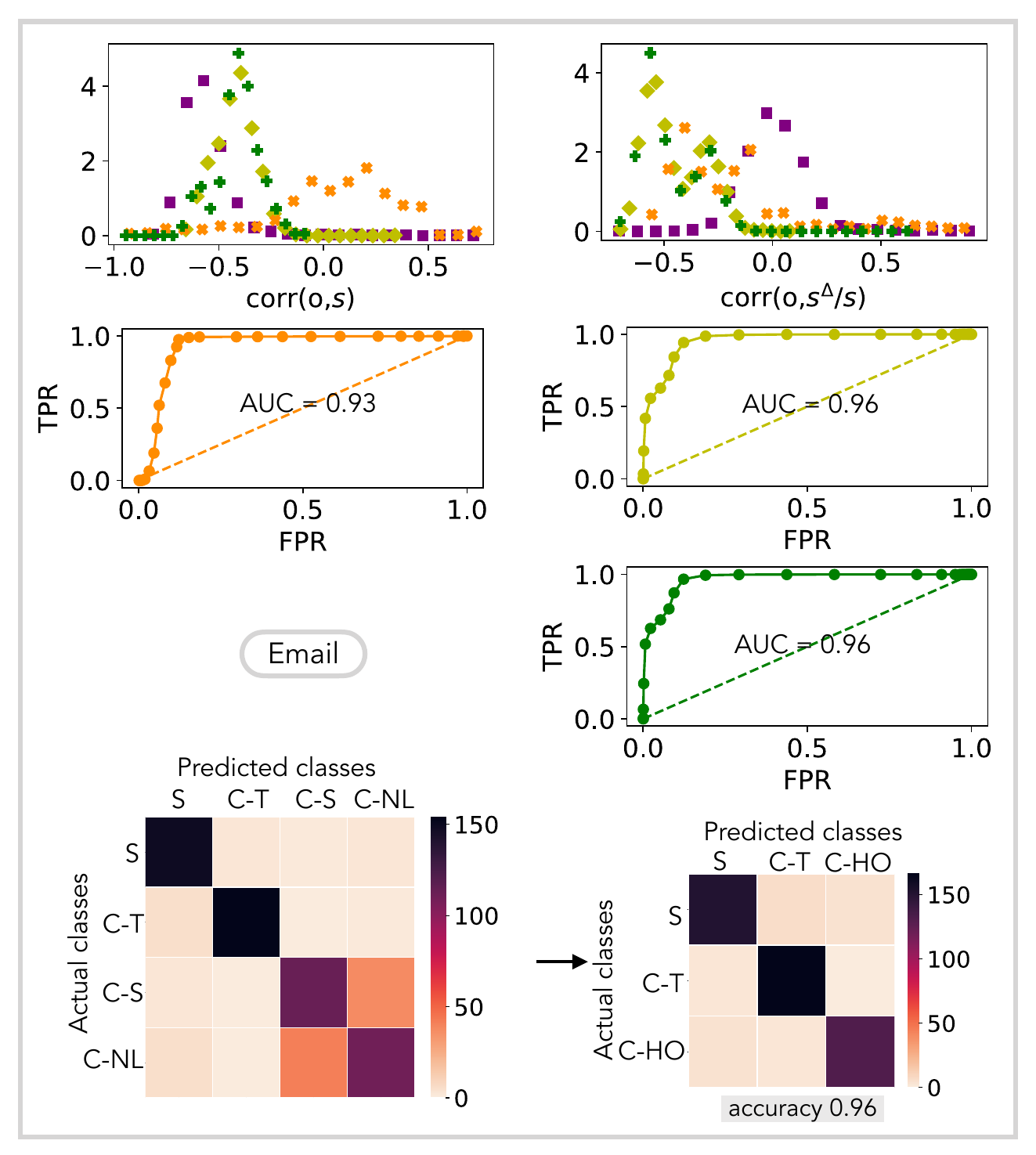}
\caption{\textbf{Weighted SI model.} Same as Fig.~\ref{figSM_distr_weighted1} for the data sets: University campus, elementary school in Utah, middle school in Utah, and email.
\label{figSM_distr_weighted2}}
\end{figure*}

The results are reported in Figs.~\ref{figSM_distr_weighted0}, \ref{figSM_distr_weighted1} and \ref{figSM_distr_weighted2}.
Overall, the classifier performs well in these cases too (accuracy always above 0.7). For some data sets, we observe however a reduced performance with respect to the unweighted case, especially for the ability to distinguish between higher-order complex contagion and simple contagion. In these cases in fact, we have checked that the weights of hyperedges of size larger than 2 are typically much smaller than the weights of the links, so that the spreading processes along links dominate even for higher order processes: as a result, higher-order complex and simple contagion become more similar processes: the decrease in accuracy is thus not a weakness of the classifier but due to the fact that the processes' spreading patterns become closer.

\clearpage
\newpage

\subsection{SIR models on unweighted networks}
\label{secSM_SIR}

The above analyses can be repeated using an SIR instead of an SI model, i.e. including the possibility for nodes to recover (with a fixed probability per unit time $\mu$) after they have been infected. A recovered node cannot infect other nodes or be infected again. The process ends when none of the nodes is infected (they are all susceptible or recovered). 

\begin{figure*}[thb]
\includegraphics[width=0.41\textwidth]{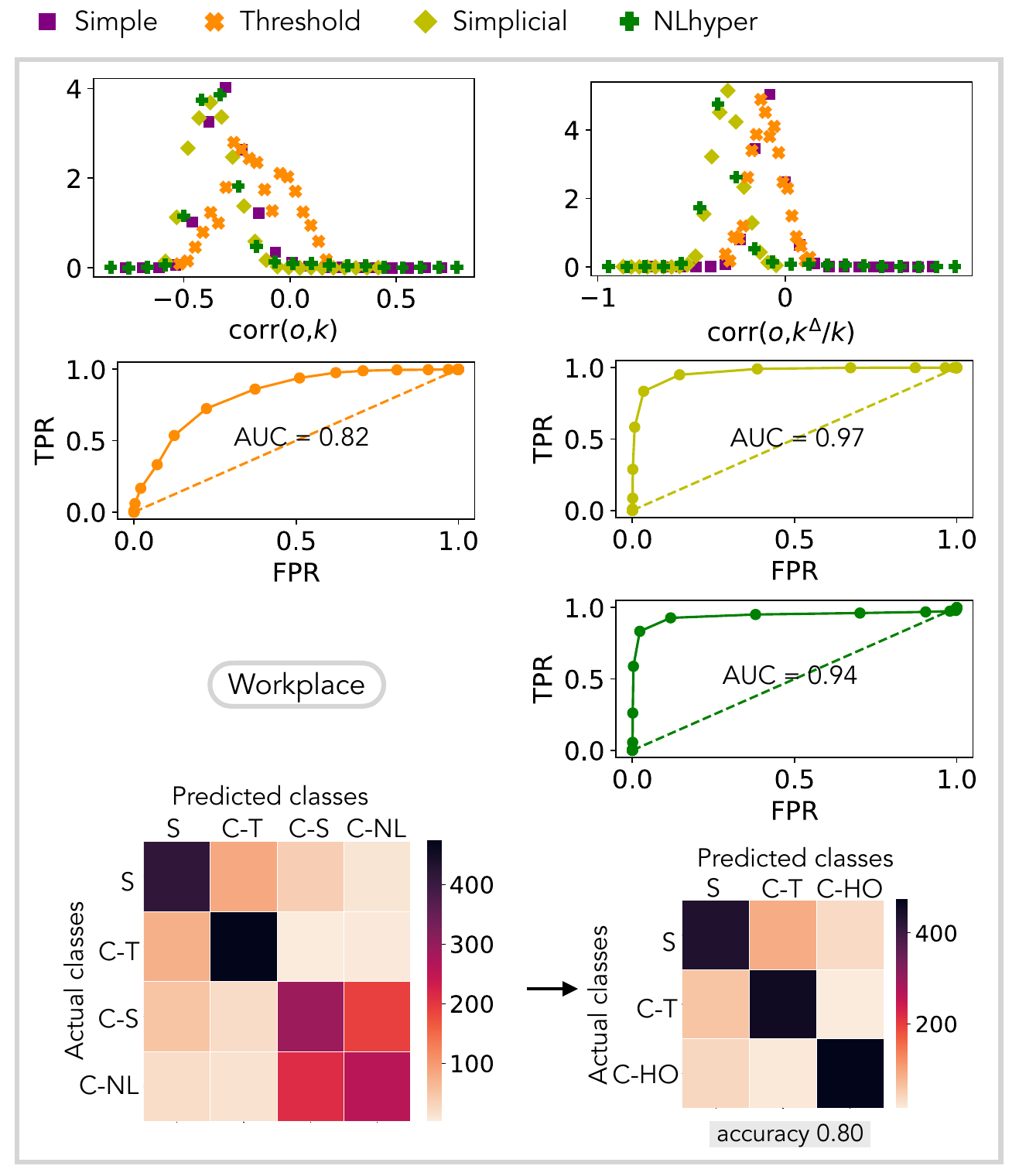}
\caption{\textbf{Unweighted SIR model.} Analogous of Figs.~2 and 3 for the SIR model with $\mu=0.1$ with the workplace network. Additional data sets are shown in Figs.~\ref{figSM_distr_SIR1} and \ref{figSM_distr_SIR2}.
\label{figSM_distr_SIR0}}
\end{figure*}

\begin{figure*}[thb]
\includegraphics[width=0.41\textwidth]{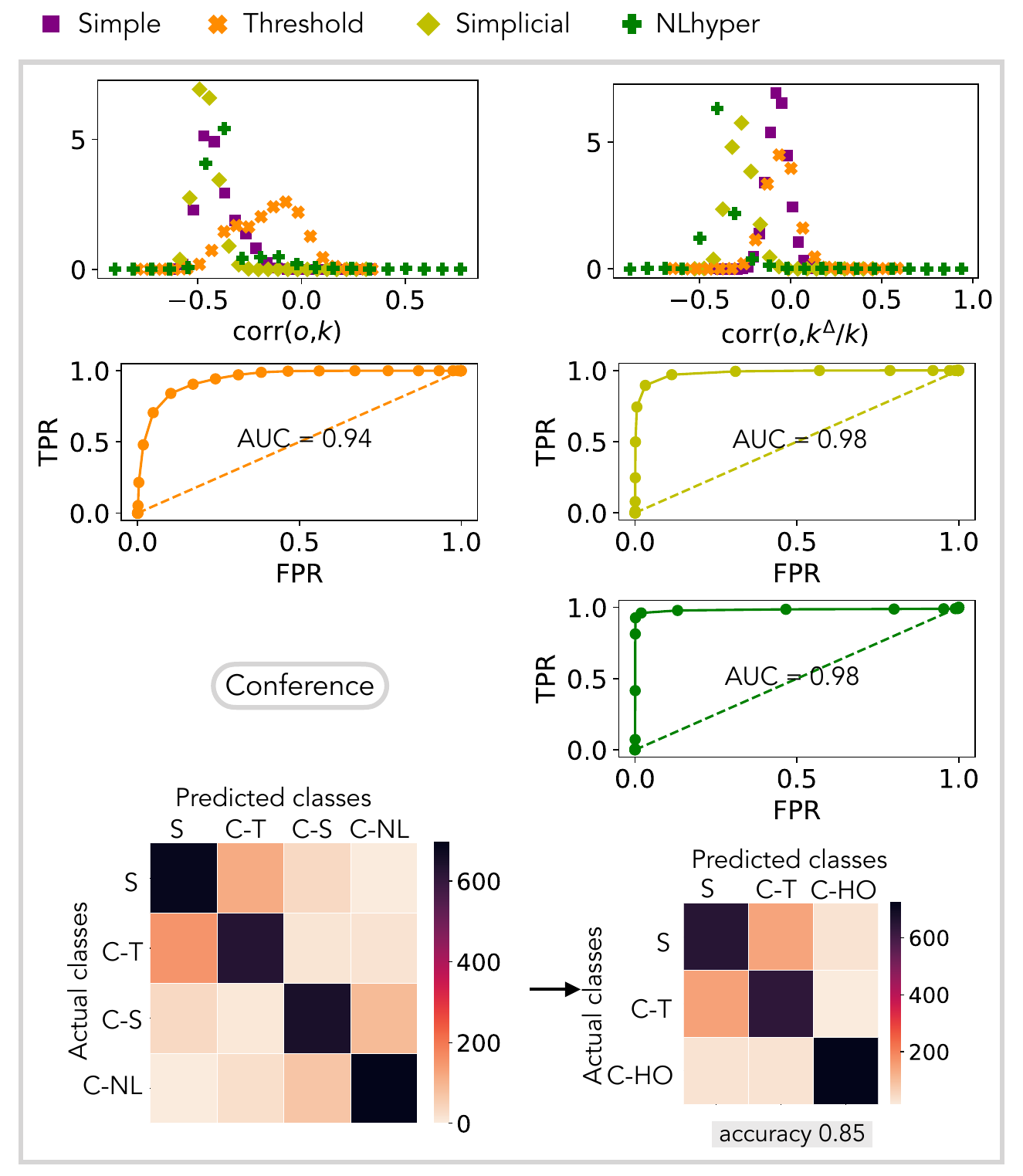}
\includegraphics[width=0.41\textwidth]{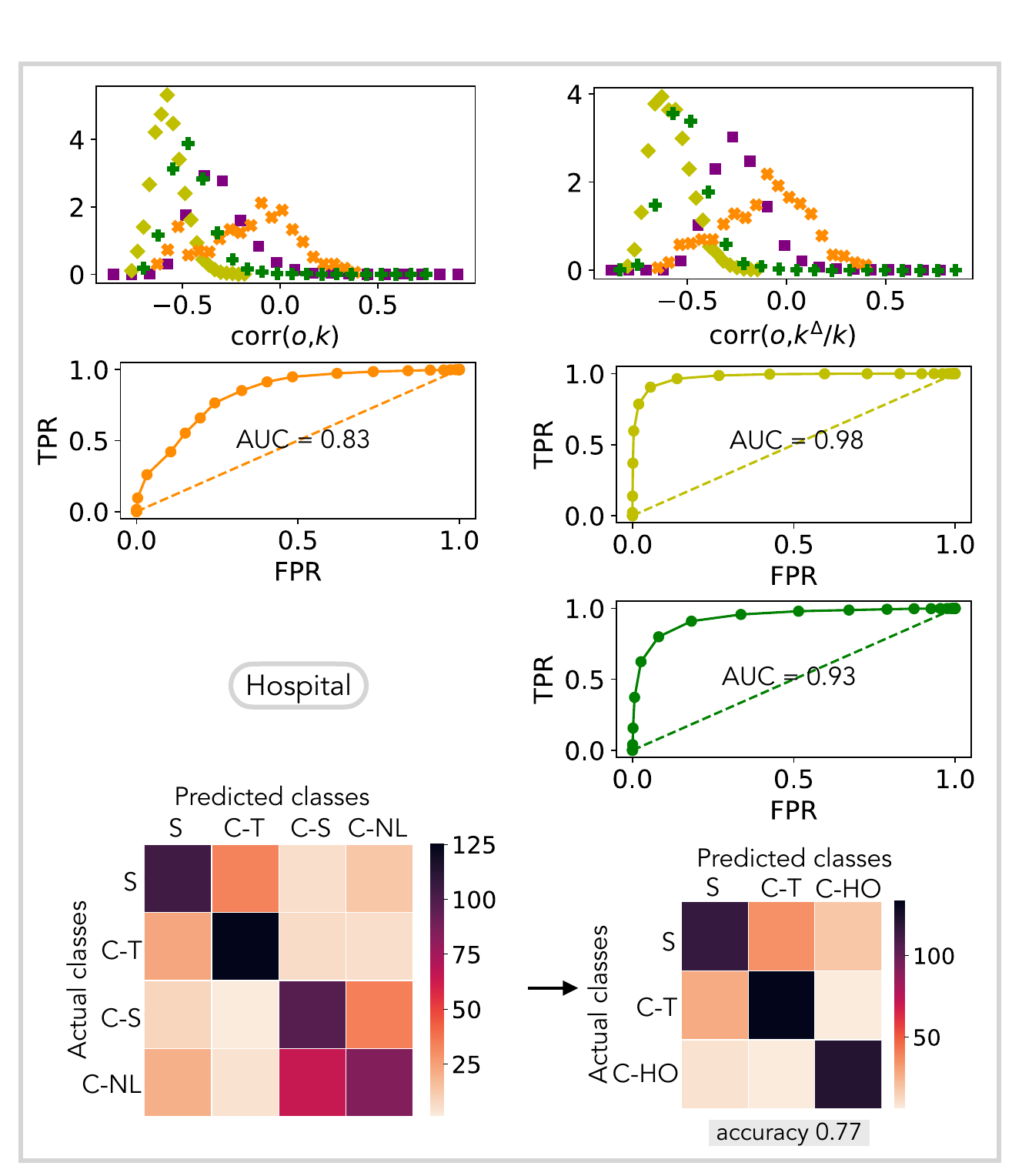}
\includegraphics[width=0.41\textwidth]{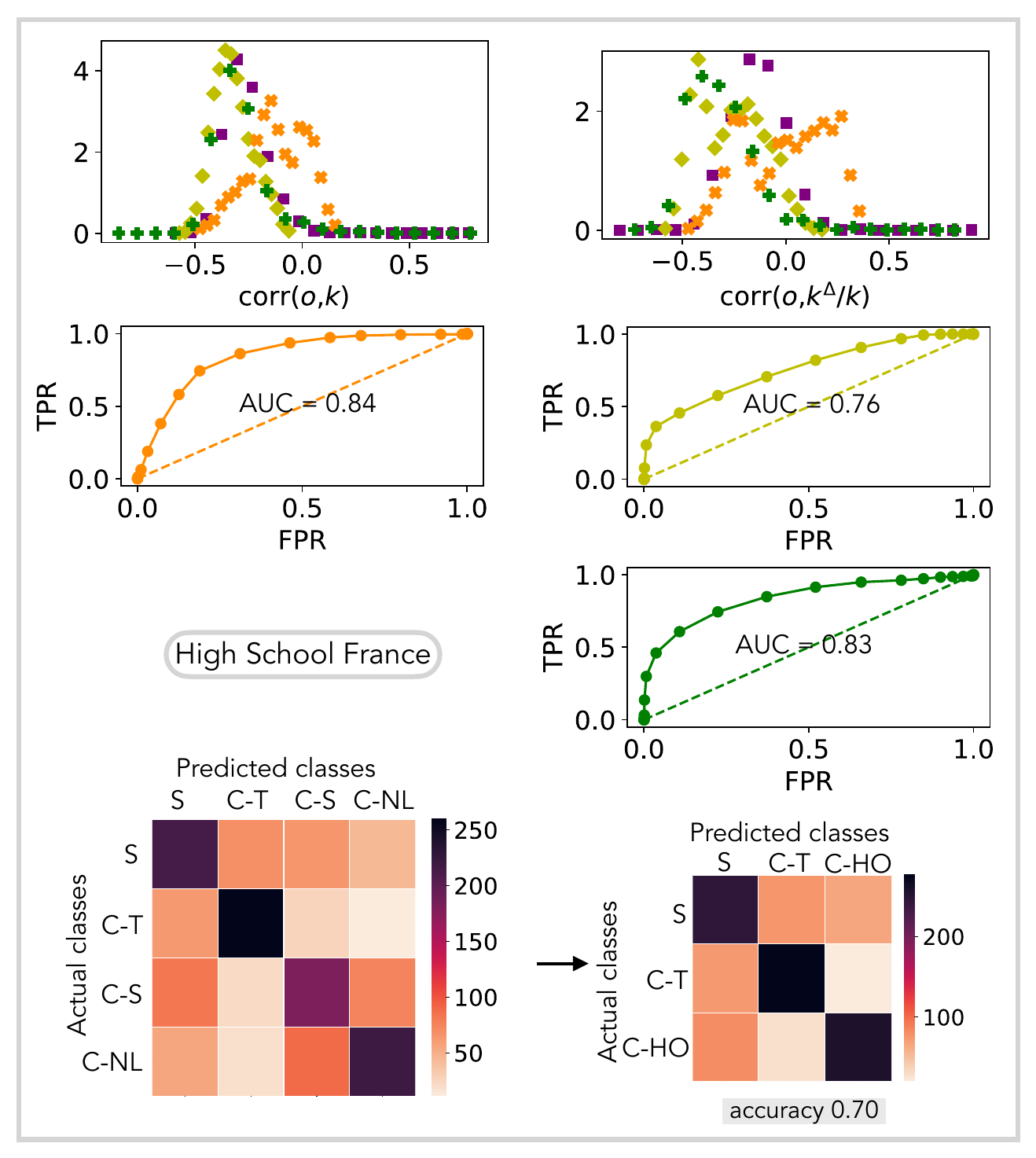}
\includegraphics[width=0.41\textwidth]{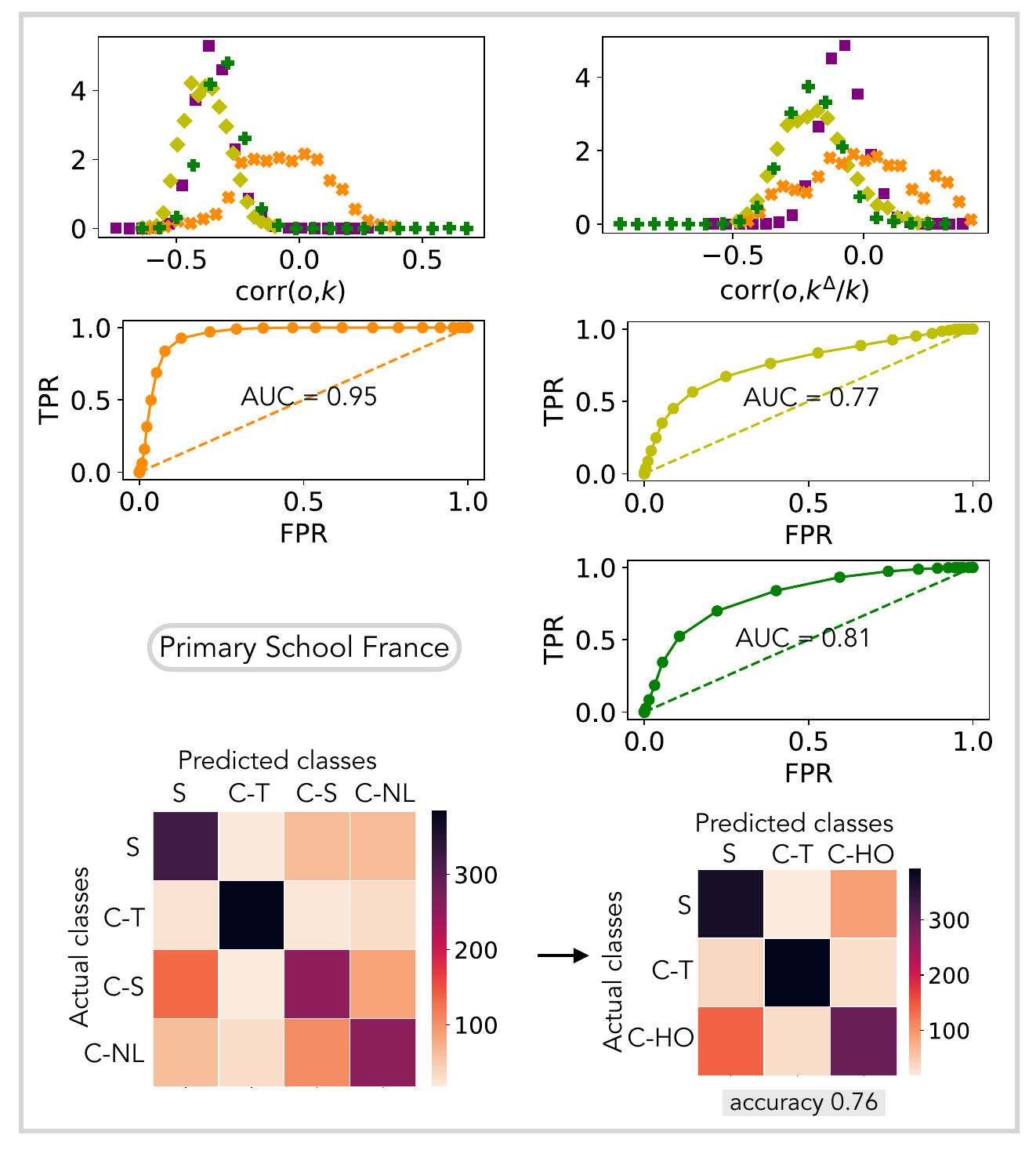}
\caption{\textbf{Unweighted SIR model.} Analogous of Fig.~\ref{figSM_distr_unweighted1} for the SIR model with $\mu=0.1$. Results for additional data sets are shown in Fig.~\ref{figSM_distr_SIR2}.
\label{figSM_distr_SIR1}}
\end{figure*}

\begin{figure*}[thb]
\includegraphics[width=0.41\textwidth]{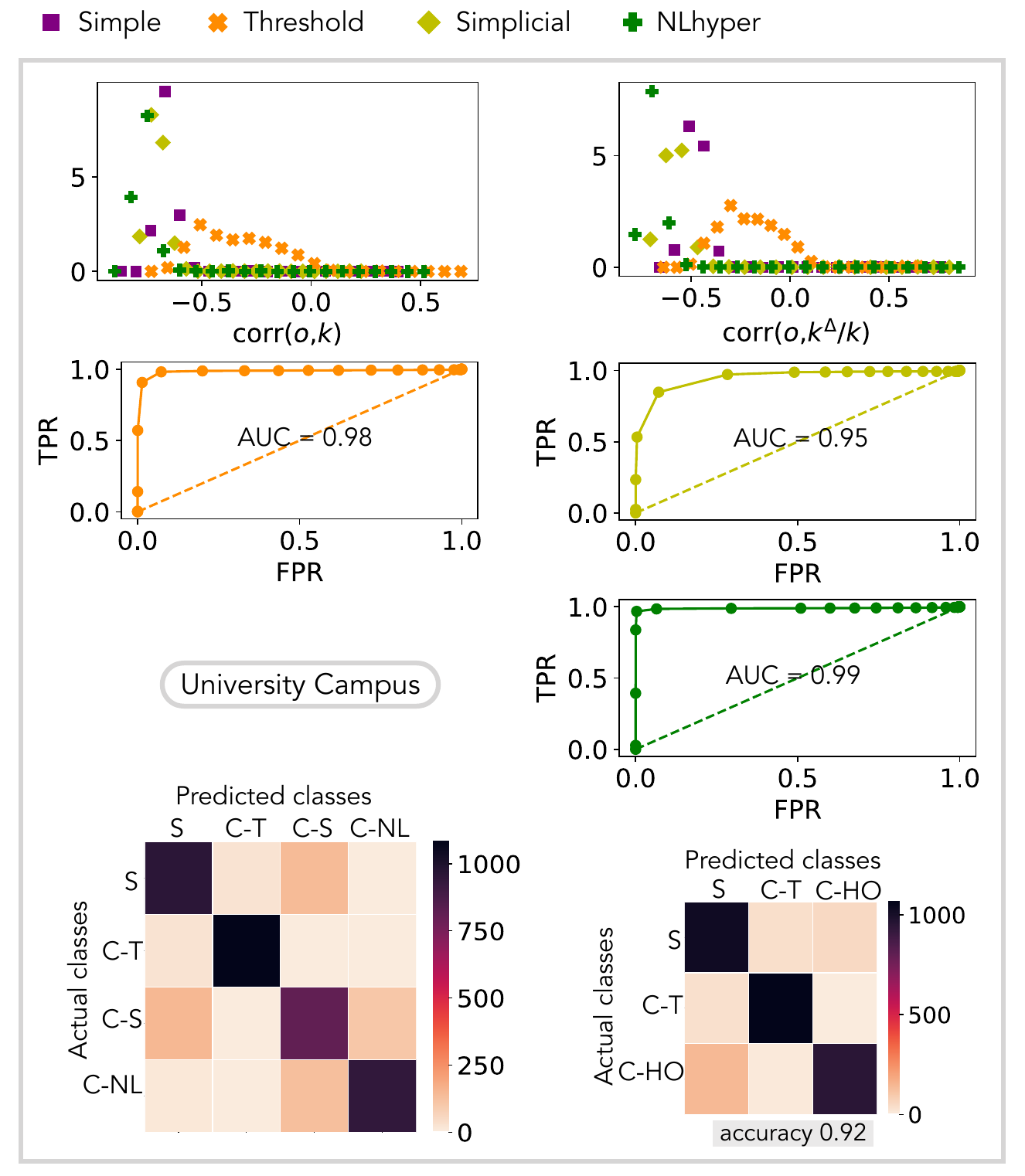}
\includegraphics[width=0.41\textwidth]{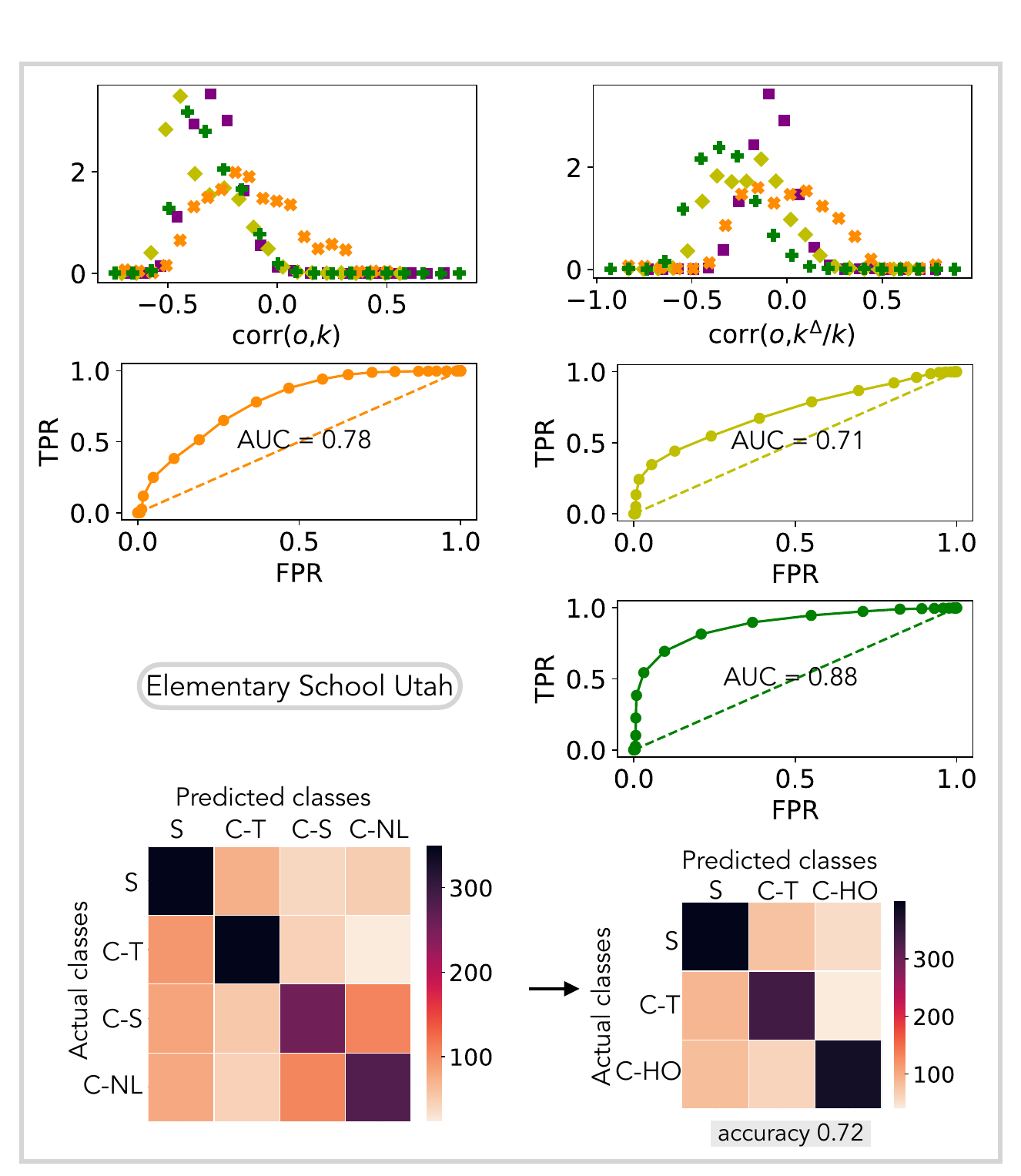}
\includegraphics[width=0.41\textwidth]{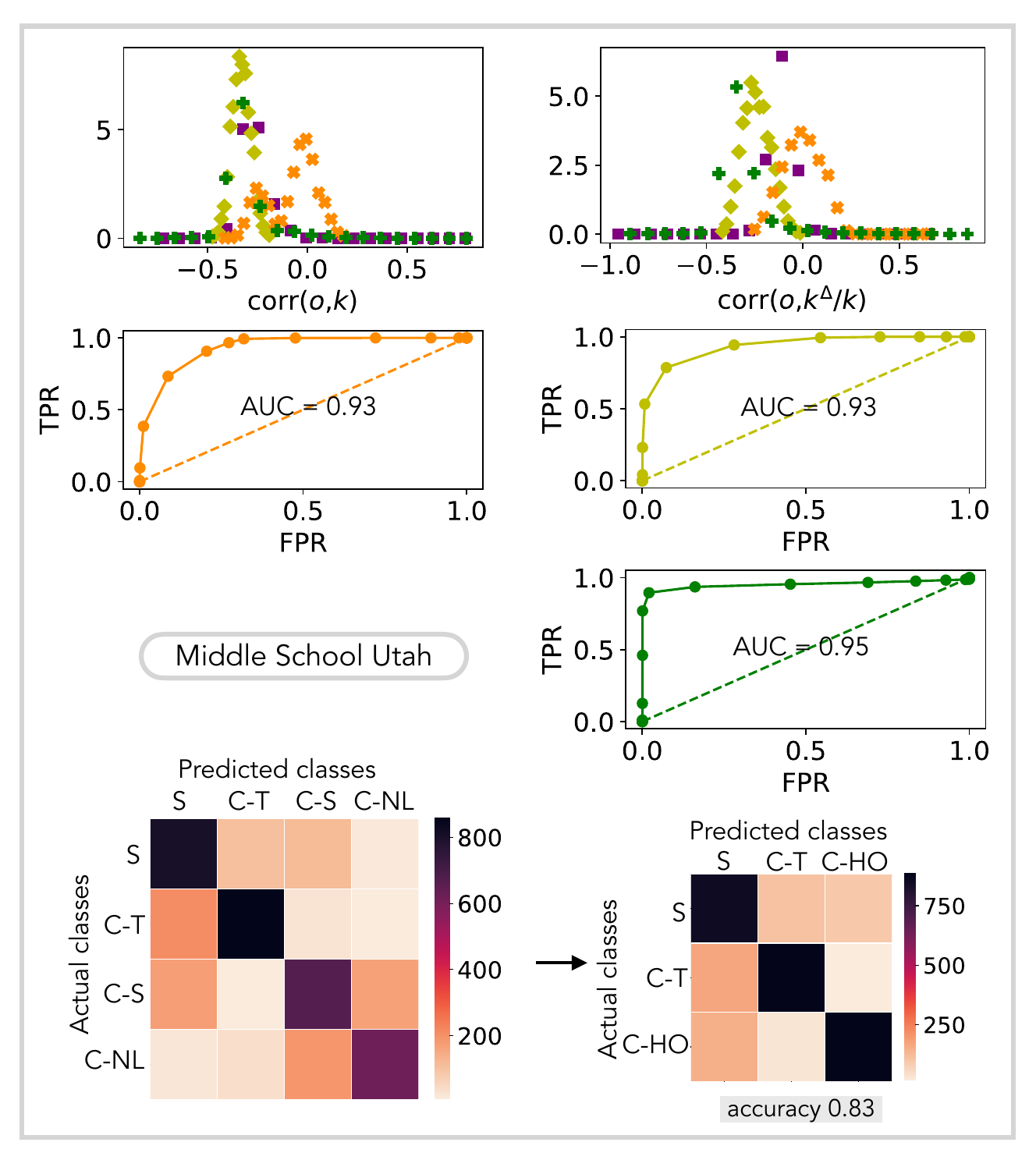}
\includegraphics[width=0.41\textwidth]{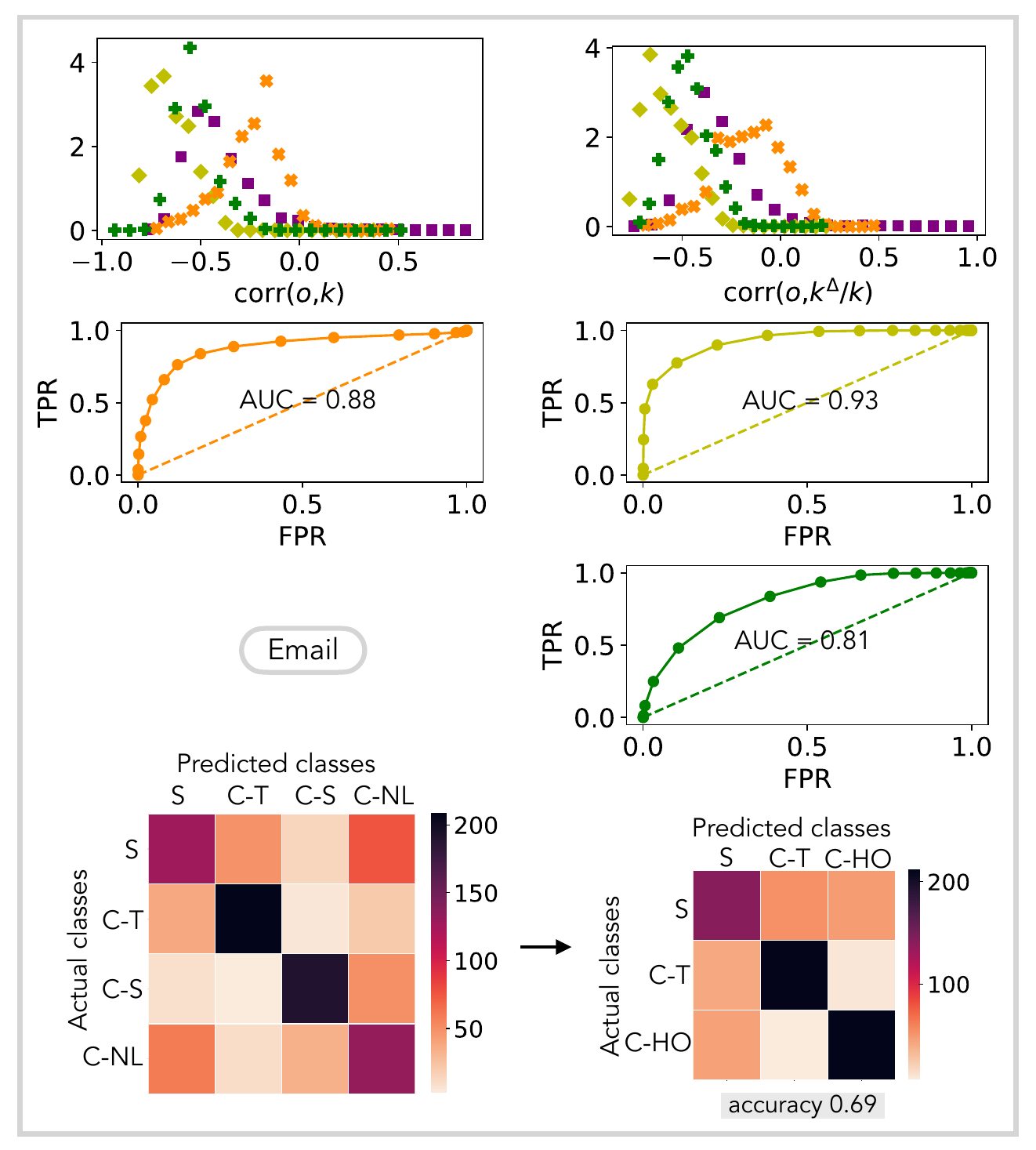}
\caption{\textbf{Unweighted SIR model.} 
Same as Fig.~\ref{figSM_distr_SIR1} for
additional data sets: University campus, elementary school in Utah, middle school in Utah, and email.
\label{figSM_distr_SIR2}}
\end{figure*}

The results of the classifier in this case are reported for unweighted networks in Figs.~\ref{figSM_distr_SIR0}, \ref{figSM_distr_SIR1} and \ref{figSM_distr_SIR2}. 
The classifier performances are slightly reduced with respect to the SI case. This can be understood by the fact that the nodes remain infected only a limited amount of time (independent of their topological properties and of their neighbors' properties). When a node recovers, it does not anymore contribute to the risk of infection of its neighbours. When reinforcement effects are needed (complex contagion), this can thus alter substantially the relationship between the structure of a node's neighborhood and its risk of becoming infected, especially if the recovery is fast. 

Here, as a preliminary analysis, we have considered 
simulations with $\mu=0.1$, and 
the classifier performances remain good, with accuracy values always above 0.7. 
A future deeper analysis will investigate in more details the impact of the recovery parameter on the classifier performance.

\clearpage
\newpage


\section{Robustness with respect to incomplete observations}

The results presented in the main text have been obtained by observing the entire infection chains, i.e. all the infected nodes with their infection order. Here we explore a more realistic case where the access to infection data is limited and only a partial observation of infection chains is possible. Here we show how the proposed classification changes with respect to a reduction in the number of observed infections: first observing just a random percentage of nodes, secondly observing just the first infected.

\subsection{Observing only a fraction of the nodes}

\begin{figure*}[thb]
\includegraphics[width=0.9\textwidth]{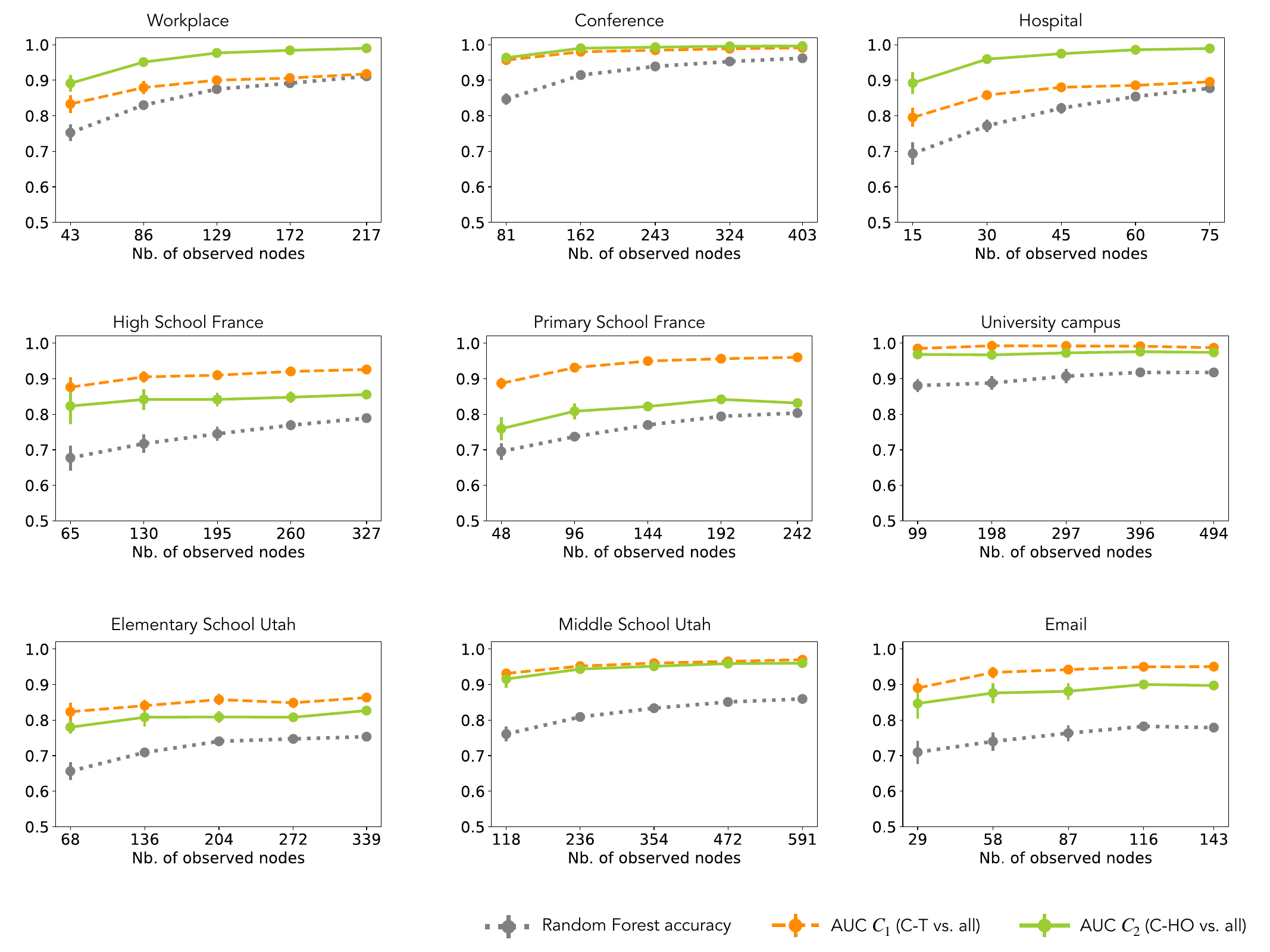}
\caption{Random Forest accuracy and AUC when using $C_1$ and $C_2$, as a function of the number of observed nodes in the different data sets. The $x$-axes report the number of observed nodes, corresponding to $20\%$, $40\%$, $60\%$, $80\%$ and $100\%$ of the total. The dots and bars show average and standard deviation for results collected on 10 realizations of the random subsets for each size.
\label{fig_perc}}
\end{figure*}

Let us assume that only a random subset of nodes can be observed. We rely on the same numerical simulations that are presented in the main text, with the difference that the correlation between infection order and topological features is computed only for an a priori chosen subset of nodes, ignoring all the others. We assume that the local properties of each observed node ($k$, $k_\Delta$,...) are known.
We repeated the experiment for 10 realizations of the random subsets, and for each
of the following subset sizes: $20\%$, $40\%$, $60\%$, $80\%$ and $100\%$ of the network nodes. For each experiment we computed the Random Forest accuracy of the classification in three classes $\{$Simple/Threshold/Higher-order$\}$, the area under the ROC curve when using $C_1$ to single out the threshold model vs. the others and the AUC when using $C_2$ to single out the higher-order models vs.
the simple and threshold ones. The results are reported in Fig.~\ref{fig_perc} for the nine data sets considered: the classification remains very efficient 
in all cases.

\begin{figure*}[thb]
\includegraphics[width=0.9\textwidth]{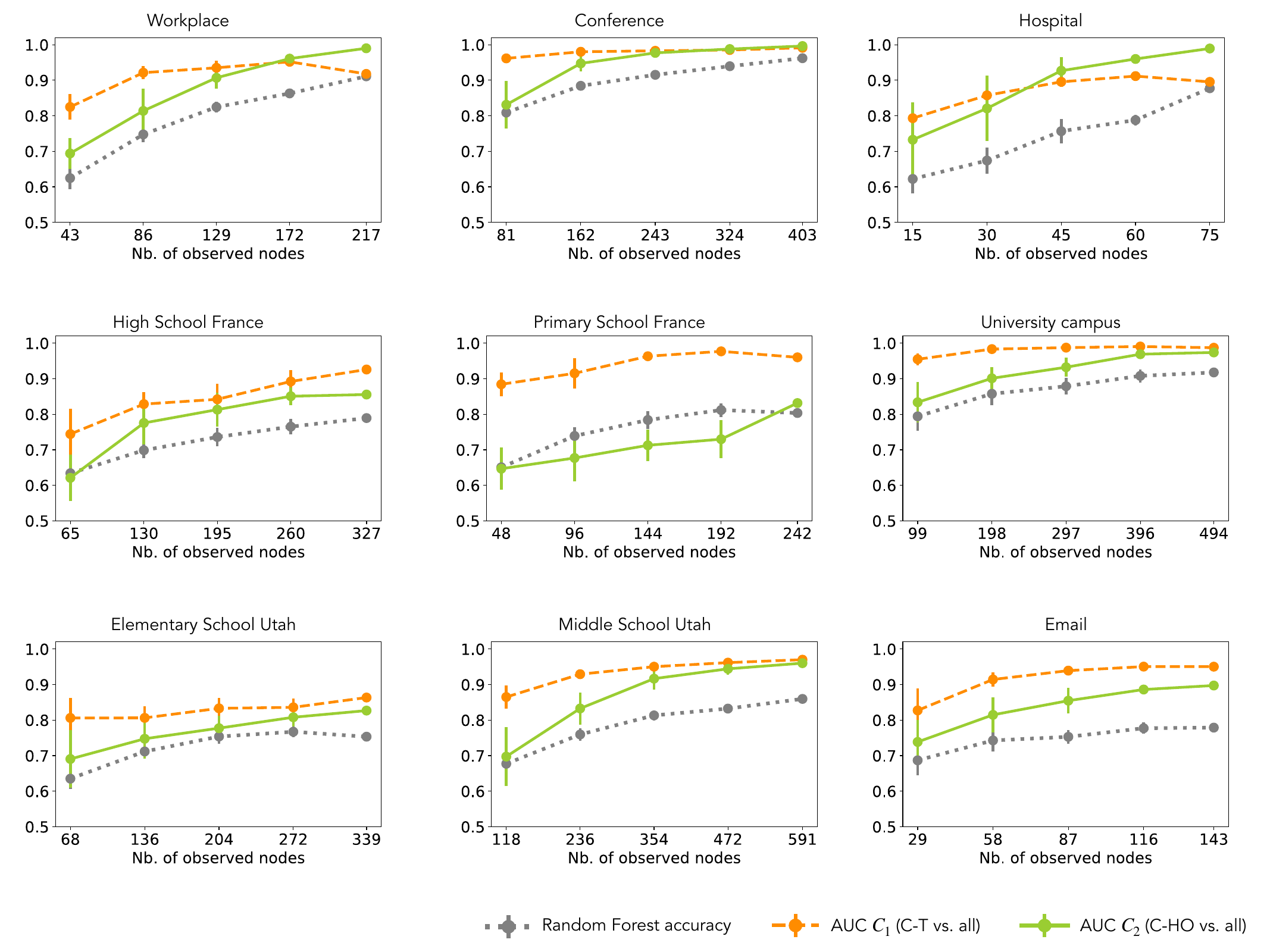}
\caption{Analogous of Fig.~\ref{fig_perc} with recalculated degrees.
\label{fig_perc_recalculated}}
\end{figure*}

We additionally consider the case where, instead of reporting the correlations with the real nodes degrees, the degrees are recalculated by only considering the observed nodes, i.e. excluding the hidden nodes from the computation of the degrees of each observed node. 
Notice that in this case a hyperedge of size $3$, e.g. $(i,j,k)$, can become a link
$(i,j)$ if only nodes $i$ and $j$ are observed. 
Overall, the actual degrees of each observed node are thus not known exactly, which can thus affect the computation of the correlations fed into the classifier. Results are reported in Fig.~\ref{fig_perc_recalculated} where we observe a slight decrease of accuracy with respect to Fig.~\ref{fig_perc}. The accuracy however always remains above 0.7 when at least 50\% of nodes are observed, and in many cases this level of accuracy is reached even with fewer nodes.

\clearpage
\newpage
\subsection{Observing only the first infected nodes}

\begin{figure*}[thb]
\includegraphics[width=1\textwidth]{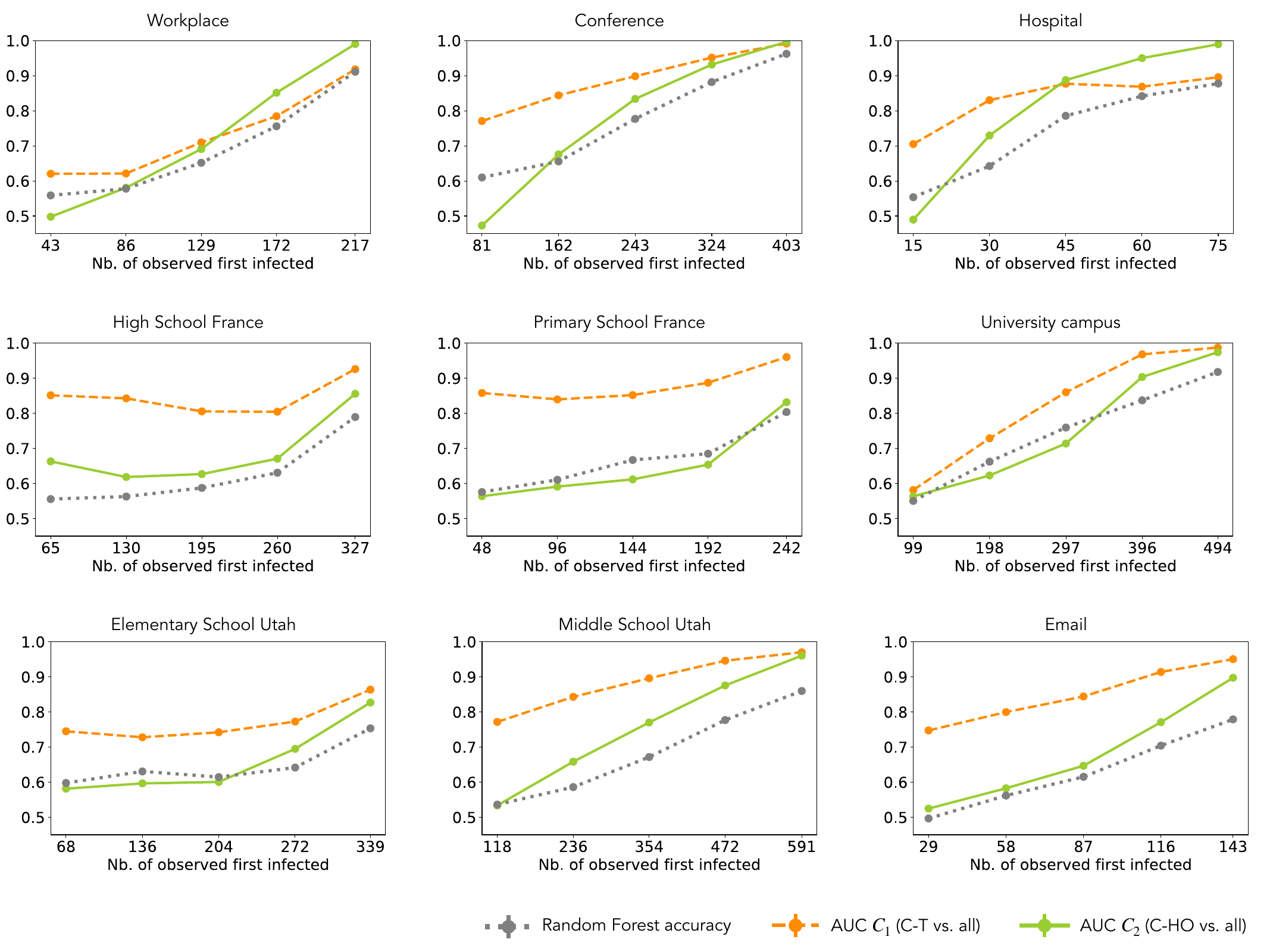}
\caption{Random Forest accuracy and 
AUC when using $C_1$ and $C_2$, obtained by observing only the first infected nodes in each infection chain. The number of observed nodes is reported on the $x$-axes, corresponding to $20\%$, $40\%$, $60\%$, $80\%$ and $100\%$ of the total number of nodes.
}
\label{fig_first_h}
\end{figure*}

As a second experiment, we try to classify contagion models by observing only the first infected nodes. The number of nodes to observe is
chosen as in the previous experiment, corresponding to  $20\%$, $40\%$, $60\%$, $80\%$ and $100\%$ of the total size.
In this case, Figure \ref{fig_first_h} shows that the performance depends more strongly on the fraction of observed nodes,
and is smaller with respect to the case where the same number of nodes have been chosen at random.
Combined with the results of Fig.~\ref{fig_perc}, this suggests that a reduction of the number of observed nodes does not have a drastic
effect if they are distributed along the entire infection chain, but the method is less robust if only the first phase of the infection pattern is observed. 

\clearpage
\newpage

\section{Influence of model parameters}
\label{secSM_param}

The classification experiments shown in the main text take into account the four different models of contagion with several choices of each model's parameters. In some cases the classifier performances depend on these parameters, while in others the parameters choice is less relevant, as discussed in the following. 

\subsection{Parameters of the higher-order models}

\begin{figure*}[thb]
\subfigure[]{\includegraphics[width=0.4\textwidth]{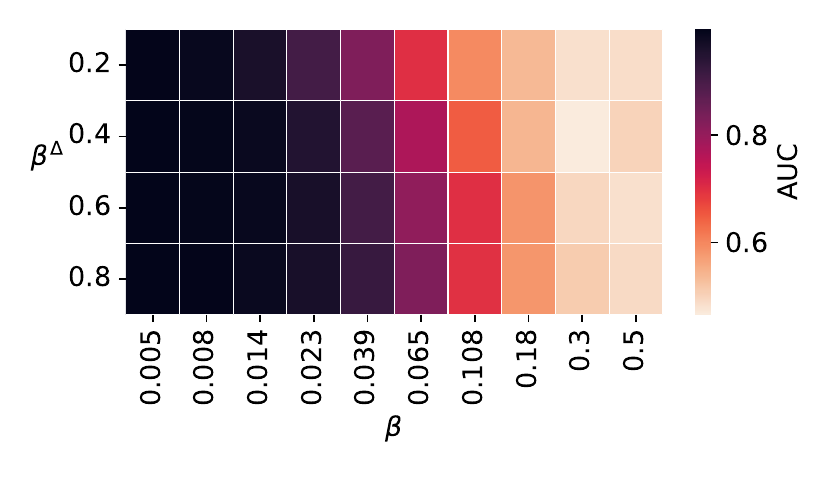}}
\subfigure[]{\includegraphics[width=0.4\textwidth]{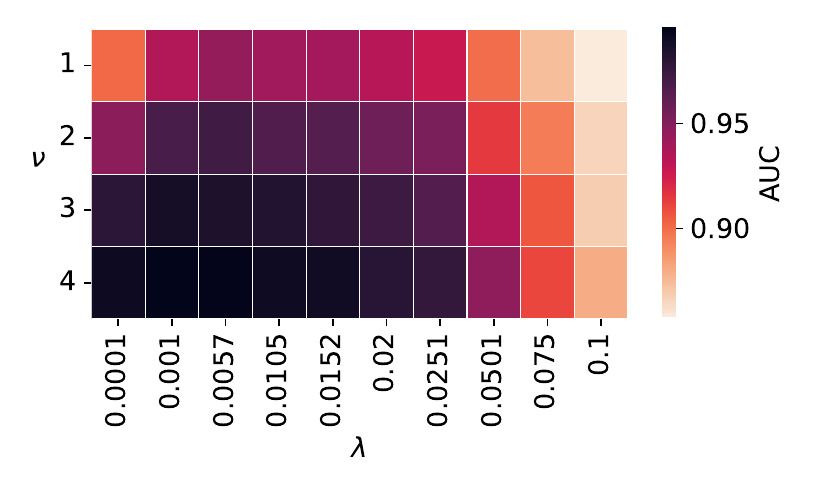}}
\caption{AUC of the ROC curve when using $C_2$ 
to distinguish simplicial vs. simple contagion models
(a) and NL-hyper vs. simple contagion (b), for 
 varying parameters. Each colored pixel is the result of 1000 experiments. The spreading processes are simulated on the workplace data set.}
\label{fig_paramHO}
\end{figure*}

The parameters' choice is particularly relevant when 
the variation of these parameters can make two models' contagion mechanisms more similar. 
For instance the simplicial model is an extension of the simple one and reduces to a simple propagation for $\beta^{\Delta} \to 0$. 
The dynamics -- and hence the order at which nodes are infected -- in the simplicial model appears thus more similar to that of the
simple contagion model for smaller values of $\beta^{\Delta}$ and larger values of $\beta$: in such cases, contagion events take place more via pairwise 
links than via triangles, weakening the importance of the higher-order mechanism.
This naturally affects the classification performances, as shown in Fig.~\ref{fig_paramHO}(a), 
which shows the area under the ROC curve when using $C_2$ to distinguish instances of the simplicial model from instances of the simple propagation, for 
varying values of $\beta$ and $\beta^{\Delta}$
(the instances of the simple model correspond to values of $\beta$ ranging between 0.005 and 0.1). 
The AUC decreases when $\beta$ increases and when $\beta^{\Delta}$ decreases. We emphasize that
this is not a limitation of the classifier but a natural consequence of the fact that
the spreading patterns of the different models actually become similar.

Analogously, in Fig.~\ref{fig_paramHO}(b) we show the AUC obtained when using $C_2$ to classify instances of the NL-hyper model and instances of the simple model, at varying $\nu$ and $\lambda$. 
In this case, the AUC decreases slightly when $\lambda$ increases and $\nu$ decreases, but remains very high for all explored parameter values.

\begin{figure*}[h!]
\includegraphics[width=0.6\textwidth]{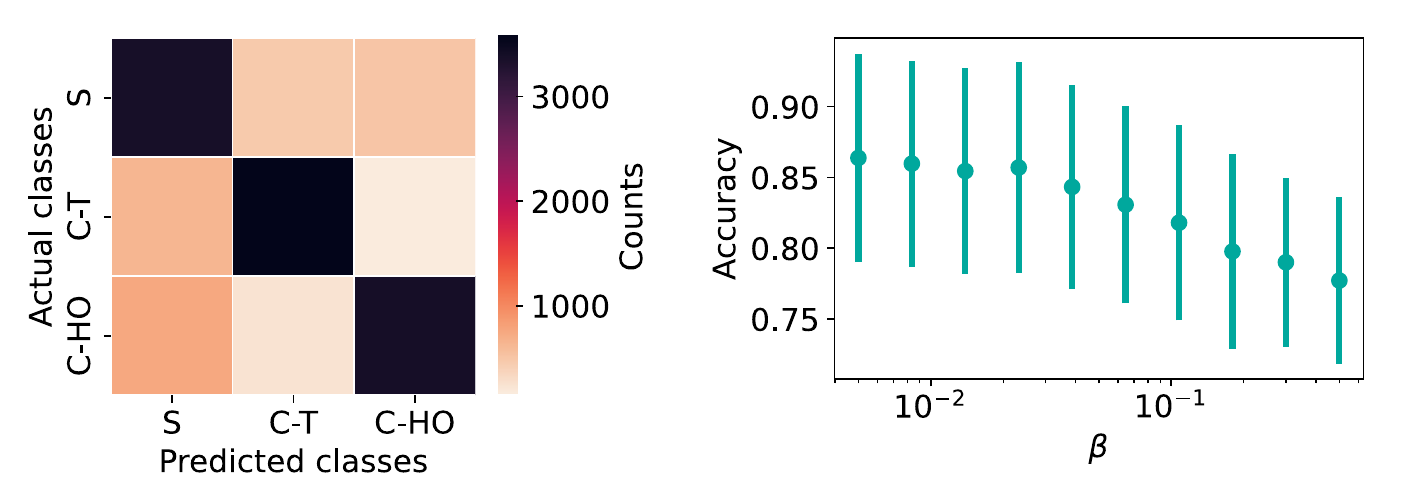}
\caption{Confusion matrix and classification accuracy vs. $\beta$ in simple and simplicial model in the case of classification with all data sets (fig.~3(d) of the main text). Five different splits of the correlation measures in training and testing sets have been considered, the plot reports mean and standard deviation.}
\label{fig_accuracy}
\end{figure*}

In Fig.~3(d) of the main text we showed the confusion matrix obtained when classifying the four contagion models using all the nine considered networks. We depict here the confusion matrix obtained when considering the two higher-order models as a single class, 
and we show the related accuracy obtained at varying the parameter $\beta$ in simple and simplicial contagion (Fig.~\ref{fig_accuracy}).

\begin{figure*}
{\includegraphics[width=1\textwidth]{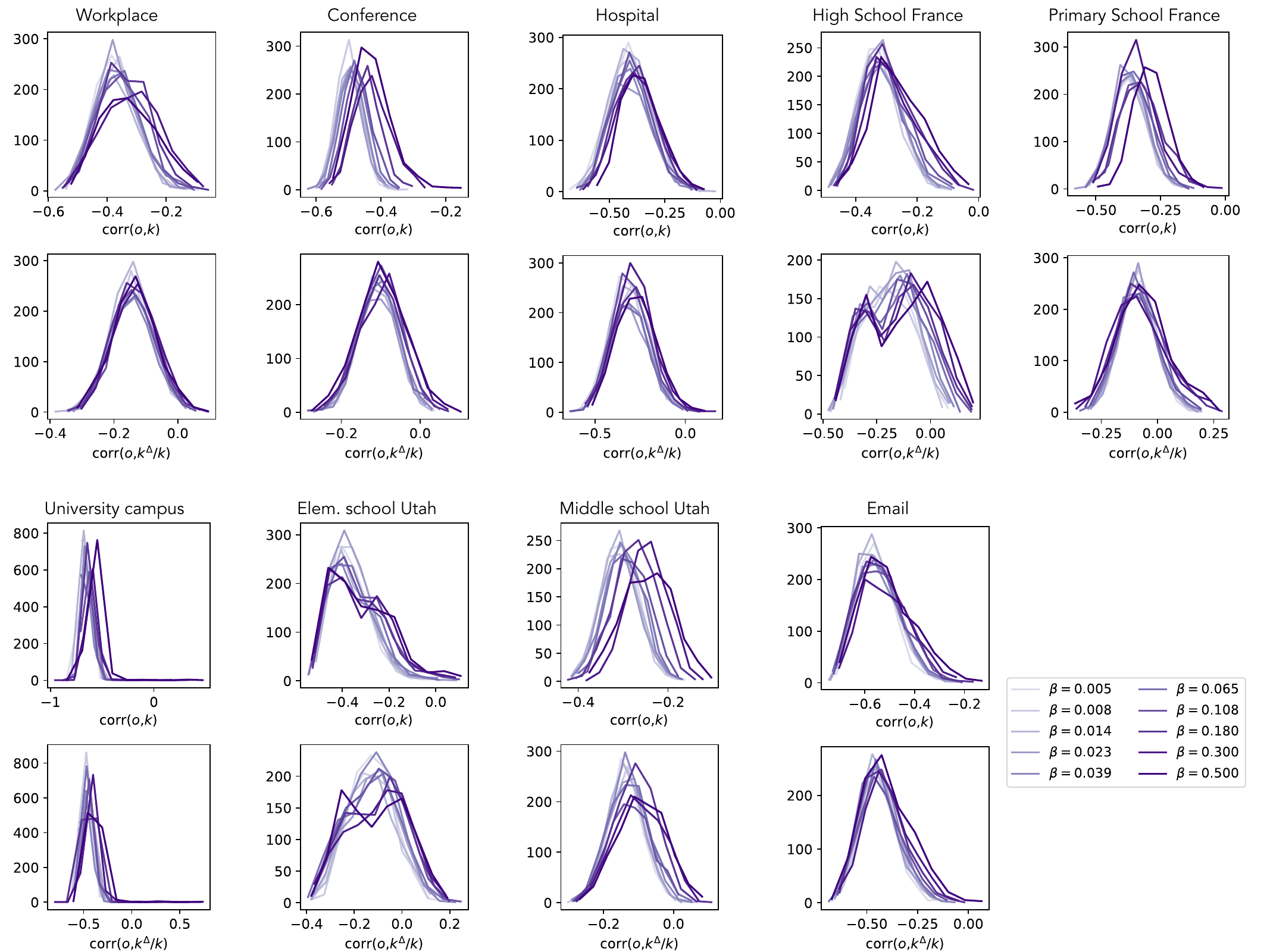}}
\caption{\textbf{Simple contagion model.} Distributions of correlations $C_1$ and $C_2$ at varying parameter $\beta$ in unweighted SI simulations on the nine considered networks.} 
\label{fig_params_simple}
\end{figure*}

\begin{figure*}
{\includegraphics[width=1\textwidth]{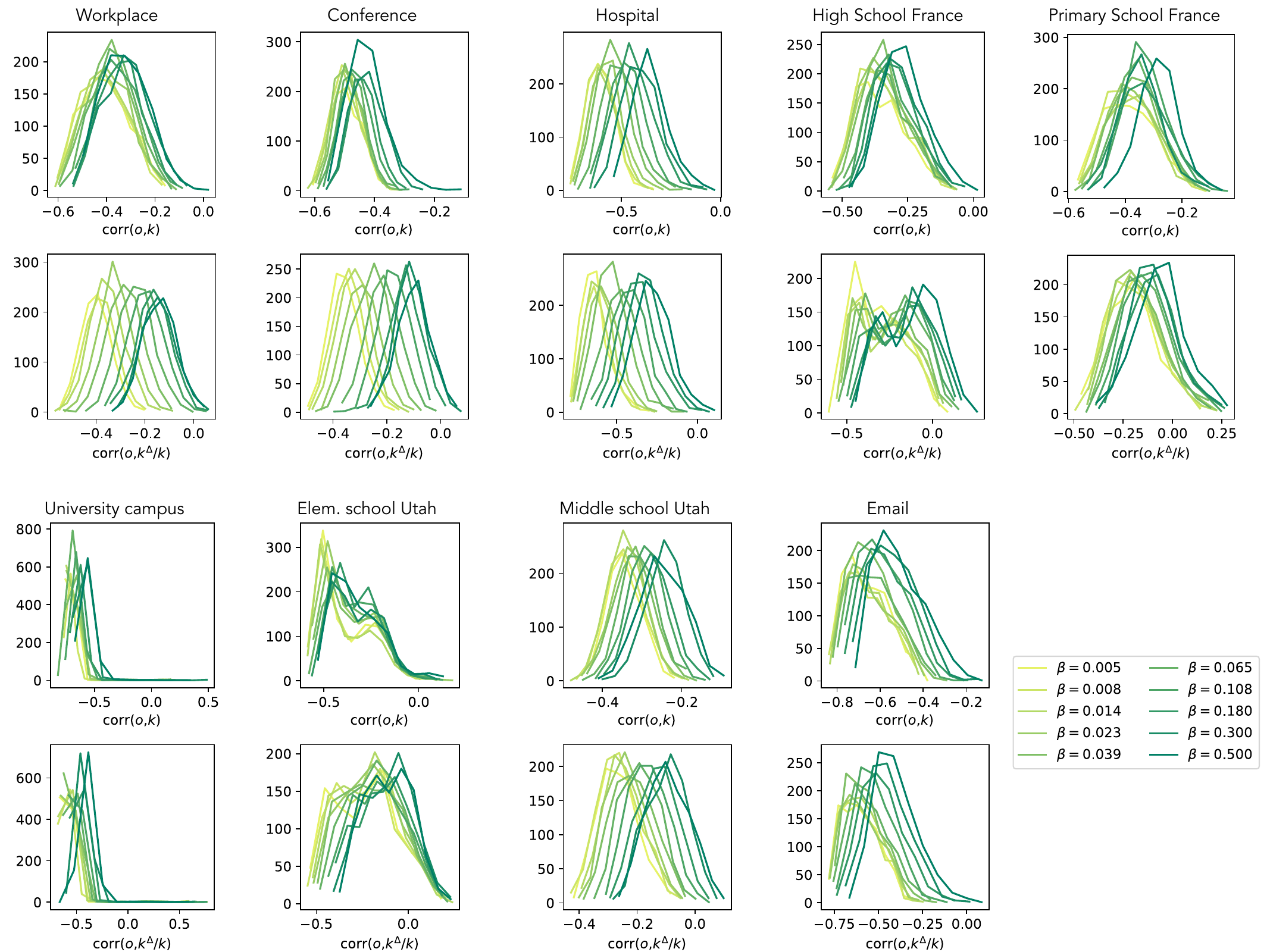}}
\caption{\textbf{Simplicial contagion model.} Distributions of correlations $C_1$ and $C_2$ at varying parameter $\beta$ (while $\beta^{\Delta}$ is fixed at 0.8) in unweighted SI simulations on the nine considered networks.}
\label{fig_params_simplical}
\end{figure*}

\begin{figure*}
{\includegraphics[width=1\textwidth]{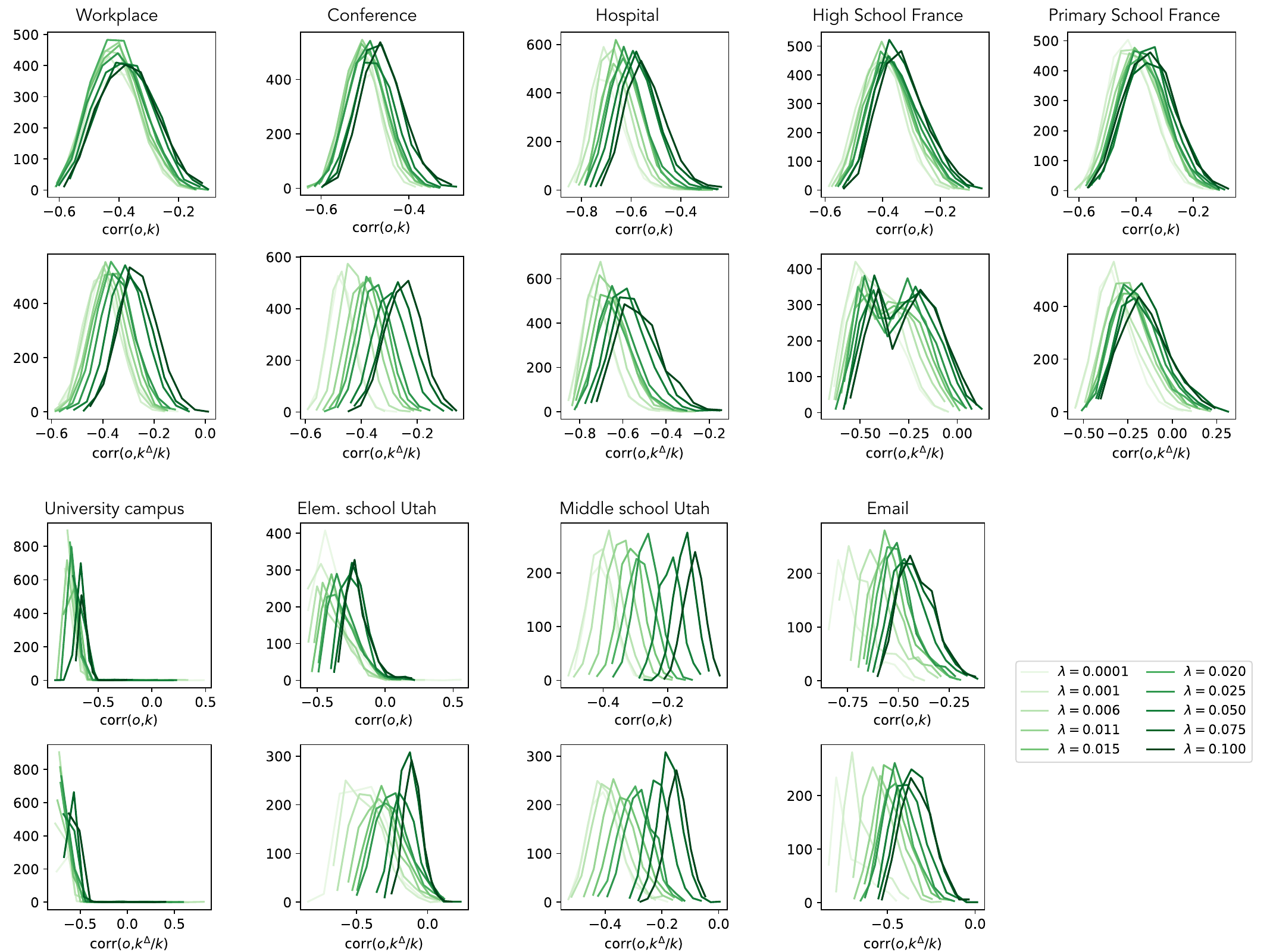}}
\caption{\textbf{NL-hyper model.} Distributions of correlations $C_1$ and $C_2$ at varying parameter $\lambda$ (while $\nu$ is fixed at 4) in unweighted SI simulations on the nine considered networks.}
\label{fig_params_NLhyper}
\end{figure*}

In Figs.~\ref{fig_params_simple}, \ref{fig_params_simplical}, and \ref{fig_params_NLhyper} the distributions of $C_1$ and $C_2$ are shown
at varying parameters $\beta$ and $\lambda$. While the correlation distributions obtained with the simple contagion model 
are not particularly sensitive to the value of its unique parameter $\beta$, 
the effect of $\beta$ is instead important in the simplicial model, where it plays a role in the competition between contagion 
via links and via triangles. Smaller values of $\beta$ (i.e. more contagions on triangles) lead to smaller values of the 
correlation $C_2$ (larger in absolute value), and thus to larger differences with respect to the simple contagion model.
A similar effect is observed in the NL-hyper model at decreasing $\lambda$. 

\clearpage
\newpage

\subsection{Threshold parameter in the threshold model}

\begin{figure*}[th]
{\includegraphics[width=1\textwidth]{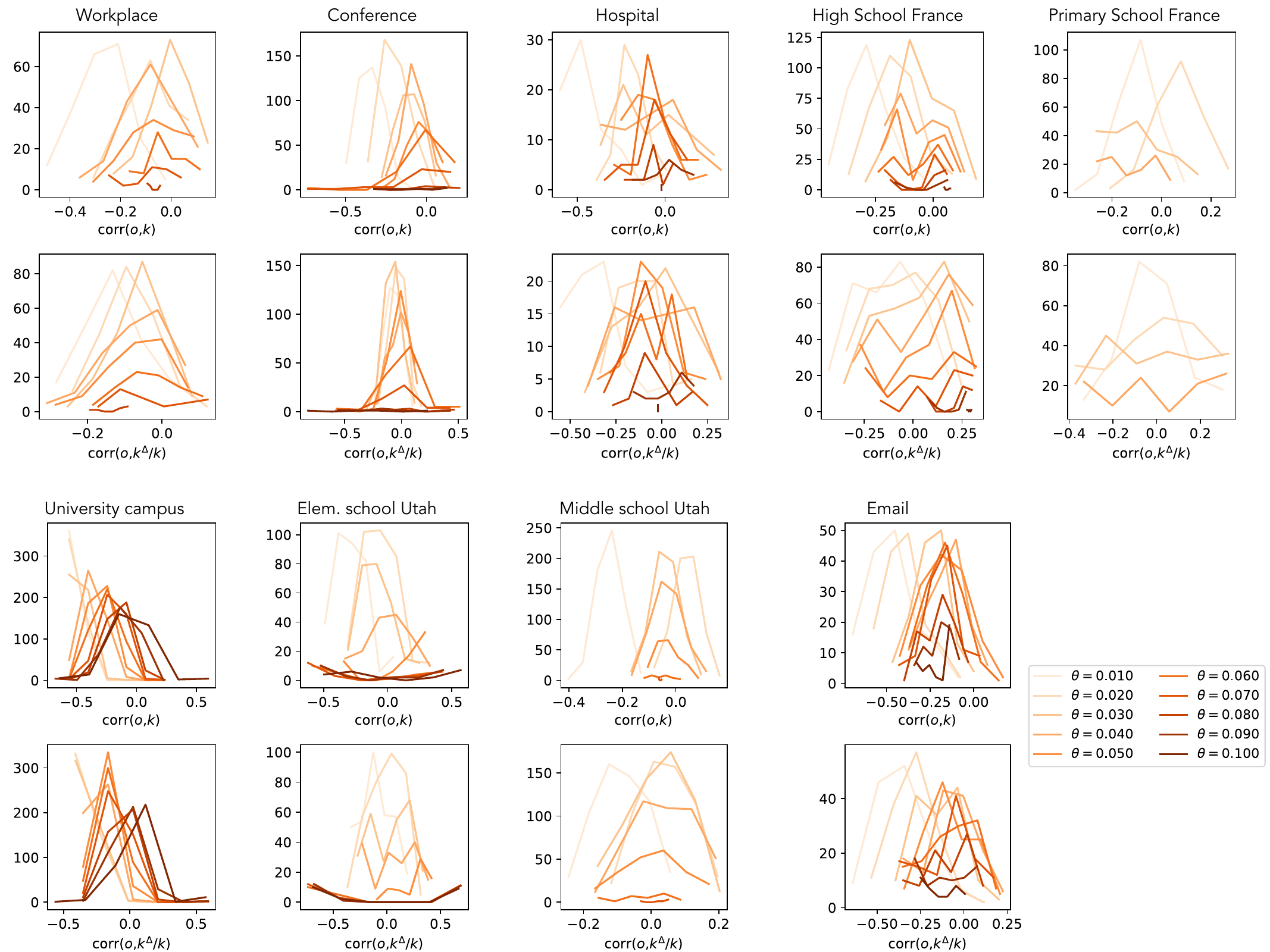}}
\caption{\textbf{Threshold model.} Distributions of correlations $C_1$ and $C_2$ at varying parameter $\theta$ in unweighted SI simulations on the nine considered networks.}
\label{fig_params_thresh}
\end{figure*}

The correlations measured in the threshold model are almost always around zero, since the nodes susceptibility 
usually does not depend on their local neighborhood. This is hence almost independent on the threshold parameter $\theta$, 
as shown in Fig.~\ref{fig_params_thresh}. We however notice that for certain networks smaller values of $\theta$ correspond to 
smaller values of correlations $C_1$ and $C_2$. This may be due to the fact that the smaller is $\theta$ the less important is the 
threshold effect, and the more similar the threshold model becomes to the simple contagion one.
The effect is however small so overall $\theta$ does not affect significantly the classifier performance.

\clearpage
\newpage

\section{Classification of processes occurring on unknown data sets}
\label{sec_train1_test2}

We have until now considered the ideal case in which the process to be classified occurs on the same network as the processes used for training the classifier. To go beyond, we now consider several experiments, in which we train and test the classifier on instances of the spreading processes taking place on distinct networks.

\subsection{Training and testing using different data sets}

\begin{figure*}[thb]
\includegraphics[width=1\textwidth]{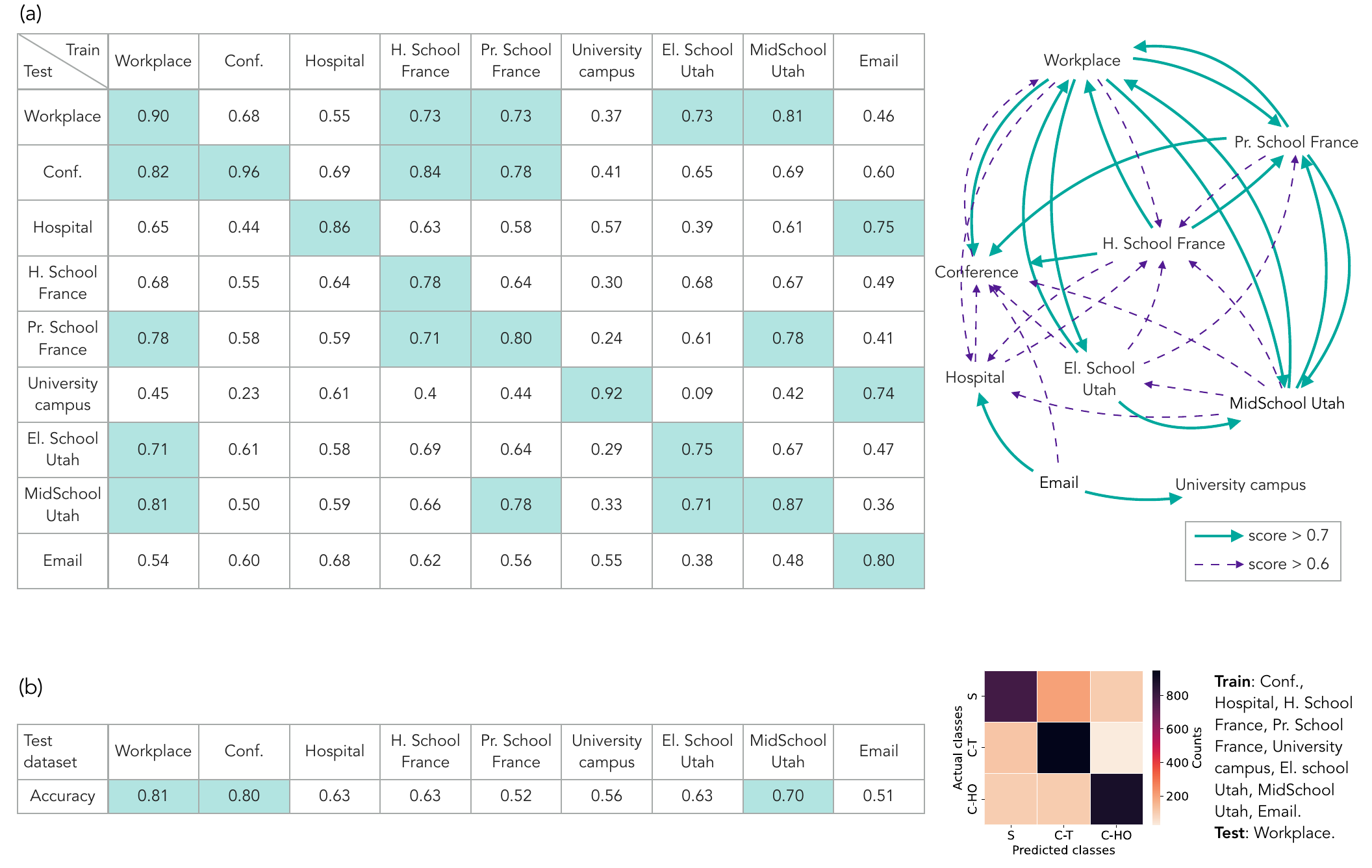}
\caption{(a): Train on one data set, test on another one. The table reports the accuracy for each pair of data sets, while the network only reports the pairs with accuracy above 0.6, where arrows go from the training to the testing data set. (b): Train on eight data sets, test on the ninth. The table reports the accuracies and the confusion matrix is shown for the best case: testing on workplace with training on the other data sets.}
\label{figSM_net_of_net}
\end{figure*}

In a first experiment, we train the classifier using instances of spreading processes run on one data set, and we test on instances simulated on a second data set. The classification accuracy depends on the  networks chosen for training and testing, as reported in the table in Fig.~\ref{figSM_net_of_net}(a), where the results above 0.7 are highlighted (diagonal elements represent the cases where training and testing are implemented using the same data set). 
The same results are also depicted on the right as a network of data sets: arrows start from the data set used for training and end at the data set used for testing. Only arrows associated to an accuracy above 0.6 are drawn, in purple for accuracy between 0.6 and 0.7, in turquoise above 0.7.

The second experiment instead involves training on eight different data sets and testing on the ninth one. The table in Fig.~\ref{figSM_net_of_net}(b) reports the accuracy obtained for testing on each data set. The highest accuracy is obtained when the testing data set is the workplace (we report the corresponding confusion matrix on the right of the figure).

In all these experiments we consider the same model parameter values as in Fig.~2 of the main text:

$\beta \in \{0.005, 0.008, 0.014, 0.023, 0.039\}$ for both simple and simplicial models, 
$\beta^{\Delta} = 0.8$, 

$\lambda \in \{0.0001, 0.001, 0.006, 0.011, 0.015\}$, 
$\nu=4$,

$\theta \in \{0.01, 0.02, 0.03, 0.04, 0.05, 0.06, 0.07, 0.08, 0.09, 0.10\}$.

\clearpage
\newpage

\subsection{Using surrogate networks}
\label{sec_SMsurrogate}

In the previous subsection, we have considered the use of processes unfolding on one or several known data sets for training the classifier, and then applied the classifier on 
processes run on an unknown data set. However, various data sets have very different structural properties (such as different degree distributions, community structures, etc).
As these properties all impact the unfolding of spreading processes, we can not expect to have a "universal" classifier that could be trained using arbitrary data sets and applied on processes run on any other data set.

On the other hand, previous works \cite{machens2013infectious,contreras2022impact} have highlighted how limited information (summary statistics) on a network's structure can be enough information to infer the statistical outcome of a spread, as spreading processes on surrogate data with the same summary statistics as the original one have similar outcomes. Even if these results have been obtained only for simple contagion processes and do not concern single realizations, they hint at a way to explore systematically the generalizability of the classifier and its use to classify processes occurring on a network that has not been used for the classifier training.

To this aim, we assume that the data set ${\cal D}$ on which the process to be classified has taken place is "unknown", in the sense that its precise structure is not known.
However, and coherently with the assumption that we can measure the correlations used in the classifier, we assume that the hyperdegrees in $H^3_{\cal D}$ are known. Note that we do thus not assume to know the hyperedge sizes or hyperdegree sequences beyond the projection on hyperedges of size $3$. 
Moreover, we will also consider the case in which
the structure in groups of the population is known 
(e.g., from metadata about the class structure in a school).
We can then create ensembles of surrogate data with similar statistics, and we simulate spreading processes on these surrogate data: we train the classifier using these simulated processes.

In the following, we consider three different ways of creating the surrogate data sets, which reproduce
the data statistics with different level of precision,
and compute the accuracy of the classifier trained using processes run on surrogate data 
and applied to classify processes run on the
real data set.

\subsubsection{SDk (Size-Degree-k\_sequence) surrogate hypergraphs}

We first assume that the degree sequence is known for each node in $H^3_{\cal D}$:
$\bm{k}(i)=\{k_2,k_3=k^\Delta\}$, 
where $k_m$ indicates the number of hyperedges of size $m$ in which the node $i$ is involved in $H^3_{\cal D}$. 
We then generate an ensemble of surrogate random hypergraphs with the same degree sequence: 
this is the higher-order analogous of the configuration model for networks of pairwise interactions. 
Note that these random hypergraphs have maximal hyperedge size equal to $3$, and have the same  
distribution of hyperedge sizes as $H^3_{\cal D}$
but not as the $H_{\cal D}$. 
The total degree $D=\sum_{m=2}^M k_m$ and its distribution $P(D)$ are also the same as in $H^3_{\cal D}$ and not the same as in the real data. 
Moreover, the surrogate data do not conserve the distributions $P(k)$ of the degree $k$ in the projected network (i.e., the total 
number of nodes to which a node is linked), nor any correlation between the degrees
of connected nodes, nor community structure, of the original data. On the other hand, the distribution
$P(k^{\Delta})$ of the number of hyperedges 
in the hypernetwork restricted to hyperedges of size $3$ at most, is conserved in these
surrogate data sets (see Table \ref{tab_surrogate} and Figs. \ref{fig_P_D}-\ref{fig_P_kDelta}). 

We call this ensemble of surrogate data the 
SDk (Size-Degree-k\_sequence) surrogate hypergraphs. 

In practice, to generate the SDk surrogate hypergraphs for each data set ${\cal D}$, we implement a hypergraph generation method based on a reshuffling approach proposed in Ref. \cite{mancastroppa2023hyper,landry2022assort}
(explained here for an arbitrary hypergraph with hyperedges of any size, and applied in our case
to $H^3_{\cal D}$). 
We start from the data set hypergraph $H_{\cal D}$ and we first set 
$H_{\cal D}'=H_{\cal D}$. We then iterate the following procedure:
 we randomly select two hyperedges $e=\{i_1,i_2,...,i,...,i_m\}$ and $f=\{j_1,j_2,...,j,...,j_m\}$ of same size $m$ in $H'_{\cal D}$, 
  we randomly extract a node within each hyperedge, 
  e.g. $i$ in $e$ and $j$ in $f$, and generate two new hyperedges $e'=\{i_1,i_2,...,j,...,i_m\}$ and $f'=\{j_1,j_2,...,i,...,j_m\}$ 
  by swapping $i$ and $j$. 
  If neither $e'$ nor $f'$ already exist in $H'_{\cal D}$, we replace $e$ and $f$
  by $e'$ and $f'$ in $H'_{\cal D}$.
  The process is iterated $10^5$ times for each size $m \in [2,M]$.

\subsubsection{SCk (Size-Community-k\_sequence) surrogate hypergraphs}

The second possibility we consider, in the spirit of \cite{machens2013infectious,contreras2022impact,genois2015compensating}, is that the overall group structure of the data is known, i.e., the population in each group and the density contact matrix of $H^3_{\cal D}$ giving the number of hyperedges between groups. We also assume that the
within-group degree sequences are known: 
 $\bm{k}^c(i)=\{k_2^c,k_3^c\}$, where $k_m^c$ indicates the number of hyperedges of size $m$ in which node $i$ is involved, and that contain only nodes of the same group.
 We then generate an ensemble of surrogate random hypergraphs with the same degree sequences in each community as in the unknown network and with the same contact matrix, by replicating the same frequency, size and composition of intra- and inter-community interactions.
 
 The surrogate data obtained have the same distribution of hyperedges size as $H^3_{\cal D}$ 
 (but not as the original data), the same group structure (the same contact matrix) and the same average total degree $\langle D \rangle$ as  $H^3_{\cal D}$.
 The correlations potentially existing in the data are however not conserved; moreover,
 the distribution $P(D)$ of the total degree is not preserved either, nor the distributions
 $P(k)$ and $P(k^{\Delta})$ (see Table \ref{tab_surrogate} and Figs. \ref{fig_P_D}-\ref{fig_P_kDelta}). 
 
 We call this ensemble of surrogate data the SCk (Size-Community-k\_sequence) surrogate hypergraphs. 

To generate the SCk surrogate hypergraphs for each data set, we implement a hypergraph generation method based on the reshuffling approach described in the previous section. Let us assume that the data set has $C$ groups, identified by the metadata on the role of the nodes (e.g. workplace, hospital, primary school France, high-school France, elementary school Utah, middle school Utah) or identified by a community detection method (e.g. conference, university campus, email). The reshuffling method described in the SDk case is performed independently within each group, i.e. considering the hyperedges of $H^3_{\cal D}$ only between nodes belonging to the same group. Finally, a random reshuffling is performed on the interactions involving nodes belonging to different groups, preserving the relative fraction of nodes belonging to each: for each hyperedge $e$ of  $H^3_{\cal D}$ of size $m$ in which there are $n_c$ nodes of group $c \in C$, with $\sum_{c \in C} n_c=m$, a hyperedge $e'$ is generated with $n_c$ randomly selected nodes of group $c$, $\forall c \in C$. 

Compared to the SDk surrogate hypergraphs, generating the SCk hypergraphs requires 
information on the group structure and the degree sequence of each node only within their group; however, these data do not conserve any global degree distribution of the original data (see Table \ref{tab_surrogate} and Figs. \ref{fig_P_D}-\ref{fig_P_kDelta}).

\subsubsection{SDC (Size-Degree-Community) surrogate hypergraphs}

The third case we consider is the one in which we assume to know the group structure of the data, as in the SCk case, 
the distribution of the hyperedges size $\Psi(m)$ and the total degree distribution $P(D)$, both in  $H^3_{\cal D}$. 
We then generate an ensemble of surrogate random hypergraphs with the same distributions $P(D)$ and $\Psi(m)$ and with a similar group structure: these hypergraphs thus only partially reproduce the community structure (i.e. the contact matrix) of  $H^3_{\cal D}$,
and, as in the other cases, do not reproduce the correlations present in the data.
The distributions  $P(k)$ and $P(k^{\Delta})$ are also not conserved (see Table \ref{tab_surrogate} and Figs. \ref{fig_P_D}-\ref{fig_P_kDelta}). 

We call this ensemble of surrogate data the SDC (Size-Degree-Community) ensemble of surrogate hypergraphs. 

To generate the SDC surrogate hypergraphs, we use the 
hypergraph generation method proposed in Ref. \cite{ruggeri2022principled}, which consists in a hypergraph sampling algorithm, and apply it to the statistics of  $H^3_{\cal D}$. Let us assume that the original data set has $C$ groups, identified by the data set metadata with the role of the nodes (e.g. workplace, hospital, primary school France, high-school France, elementary school Utah, middle school Utah) or  by a community detection method (e.g. conference, university campus, email). 
The algorithm takes as input the hyperedge-size sequence of  $H^3_{\cal D}$, 
as the total number of hyperedges of each size ($2$ or $3$ in this case), the total degree sequence in  $H^3_{\cal D}$ and its group structure (groups and contact matrix giving for each pair of groups the number of hyperedges of  $H^3_{\cal D}$ 
involving nodes of these two groups).
The method proposed in Ref. \cite{ruggeri2022principled} generates surrogate hypergraphs reproducing these statistics, through a generative and hyperedge sampling approach: we also add a further condition, compared to the original algorithm, so that the process does not generate completely coincident hyperedges.

Compared to the SDk surrogate hypergraphs, the SDC hypergraphs requires knowing the group structure, but only the contact matrix in the projected network, and also only the total degree sequence (instead of the more detailed global or community degree sequence of each node in $H^3_{\cal D}$, as respectively in the SDk and SCk cases). As in the previous case, the surrogate data do not have the same
distributions $P(k)$ and $P(k^{\Delta})$ as the original data. Moreover, in practice 
we observe that the group structure is not very well preserved by the algorithm
(while SCk fully preserves the group structure of $H^3_{\cal D}$). 

\clearpage
\newpage

\subsubsection{Surrogate data properties}

In Table \ref{tab_surrogate} we summarize some properties of the original data ${\cal D}$ and of  $H^3_{\cal D}$ 
that are preserved or not in the surrogate data produced by the various methods proposed. 
Note that in all cases, the degree correlations are not preserved, nor the 
statistics of the overlap between hyperedges.

\begin{figure}[thb]
\includegraphics[width=0.75\textwidth]{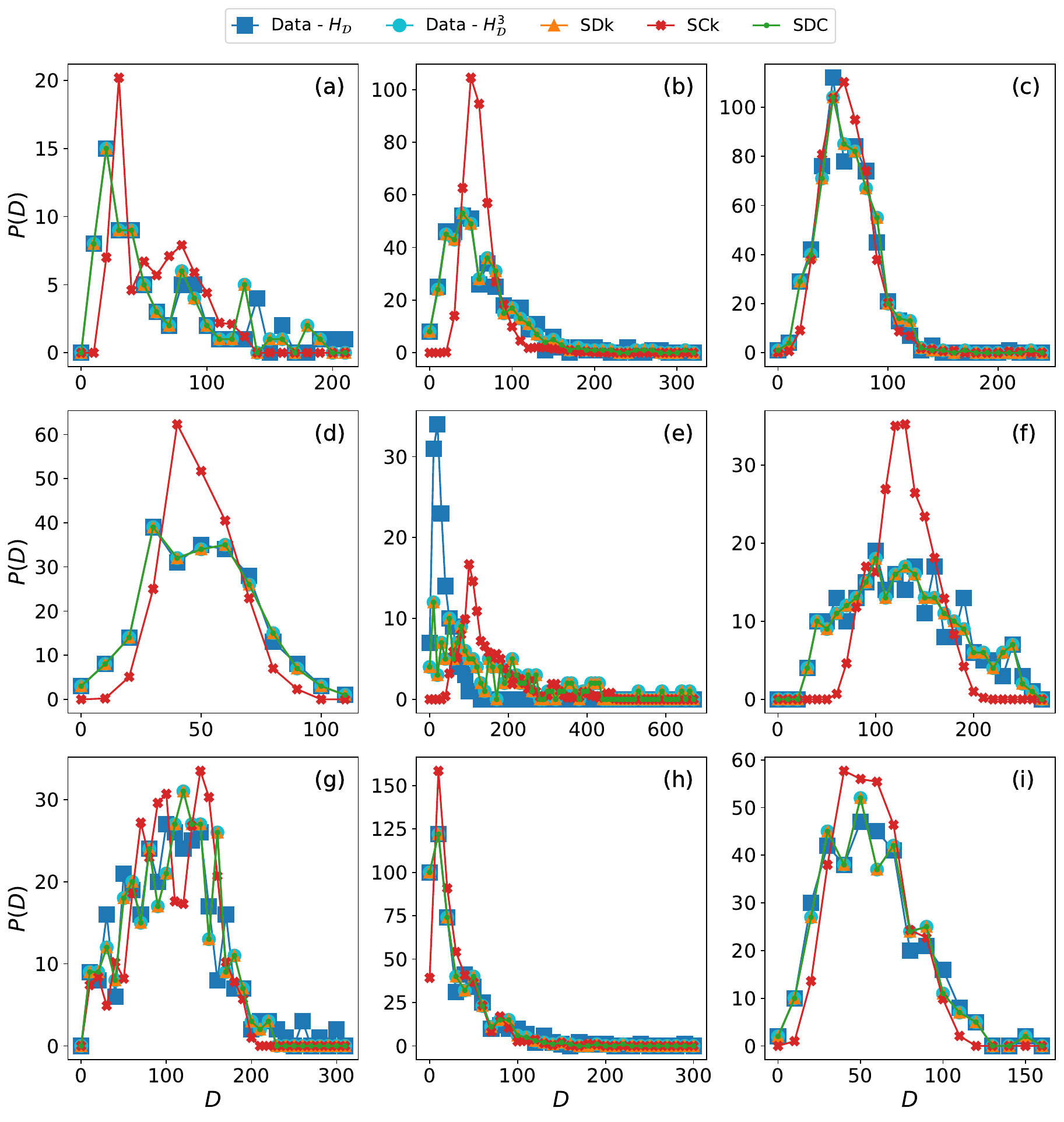}
\caption{\textbf{Total degree distribution.} We show for each data set the distribution $P(D)$ of the total degree, i.e. the number of distinct hyperedges of arbitrary size in which a node is involved, for the original data (both for $H_{\cal D}$ and $H^3_{\cal D}$) and for the 
 corresponding SDk, SCk and SDC surrogate data sets (see legend). 
 The following data sets are considered: (a) hospital, (b) conference, (c) middle school Utah, (d) workplace, (e) email, (f) primary school France, (g) elementary school Utah, (h) university campus and (i) high-school France. 
 The curves corresponding to the SDk, SCk and SDC cases are obtained by averaging over $10$ realizations of surrogate hypergraphs.}
\label{fig_P_D}
\end{figure}

\begin{figure}[thb]
\includegraphics[width=0.75\textwidth]{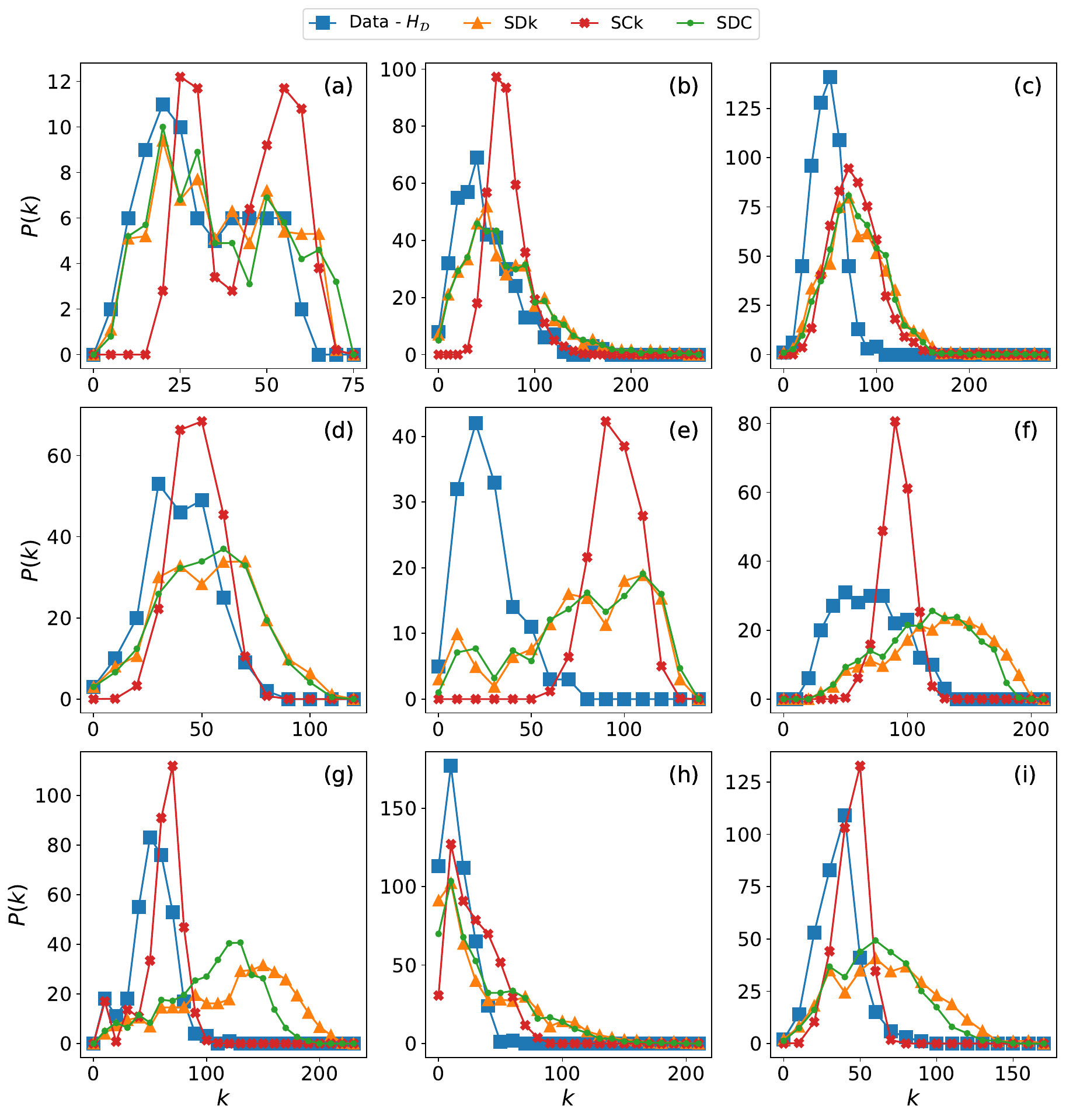}
\caption{\textbf{Pairwise degree distribution.} We show the distribution $P(k)$ of the pairwise degree for all the data sets (obtained by projecting the hypergraphs on a network), considering the original hypergraph and the corresponding SDk, SCk and SDC surrogate hypergraphs (see legend). The following data sets are considered: (a) hospital, (b) conference, (c) middle school Utah, (d) workplace, (e) email, (f) primary school France, (g) elementary school Utah, (h) university campus and (i) high-school France. The curves corresponding to the SDk, SCk and SDC cases are obtained by averaging over $10$ realizations of surrogate hypergraphs.}
\label{fig_P_k}
\end{figure}

\begin{figure}[thb]
\includegraphics[width=0.75\textwidth]{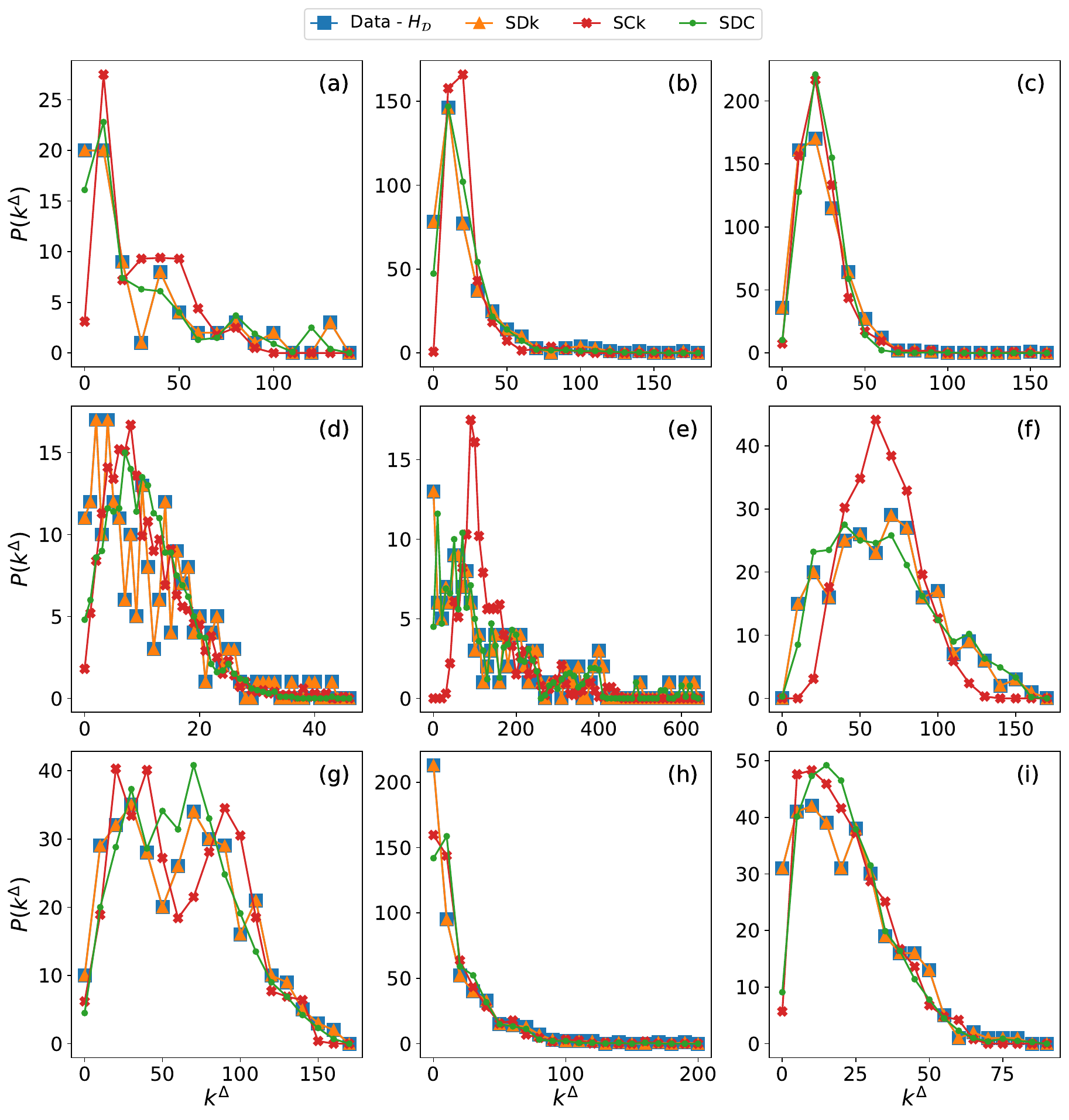}
\caption{\textbf{Distributions $P(k^{\Delta})$.} 
We show the distribution $P(k^{\Delta})$ of the number of hyperedges of size $3$ of each node in the hypergraph obtained by projecting interactions of higher-order on triangles, for each  data set, for the original hypergraph and the corresponding SDk, SCk and SDC surrogate hypergraphs (see legend). The following data sets are considered: (a) hospital, (b) conference, (c) middle school Utah, (d) workplace, (e) email, (f) primary school France, (g) elementary school Utah, (h) university campus and (i) high-school France. The curves corresponding to the SDk, SCk and SDC cases are obtained by averaging over $10$ realizations of surrogate hypergraphs.}
\label{fig_P_kDelta}
\end{figure}

 Figures \ref{fig_P_D}, \ref{fig_P_k} and \ref{fig_P_kDelta}  show respectively the 
 distribution $P(D)$ of the total degree (both for 
 $H_{\cal D}$ and $H^3_{\cal D}$), 
 the distribution $P(k)$ of the degrees in the projected networks, 
 and the distribution $P(k^{\Delta})$ of the number of hyperedges of size 3 in $H^3_{\cal D}$, for all the data sets and surrogate hypergraphs considered. 
 We note that the distributions $P(D)$ are rather well reproduced even in the SCk
 method (which does not explicitly preserve it). On the other hand, the distributions
 $P(k)$ are often shifted to higher values of the degree. This can be attributed to data sets
 where there is an important overlap between hyperedges, which is not preserved in the more
 random surrogate data sets, leading to larger degrees in the projected network.  
 $P(k^{\Delta})$ is well reproduced by SDk (by definition) and SDC, but more deviations are 
 observed with SCk.
 
\begin{table}[h!]
\renewcommand{\arraystretch}{2}
\centering
\begin{tabular}{|c|c|c|c|c|c|c|c|c|>{\centering\arraybackslash}p{10mm}|>{\centering\arraybackslash}p{10mm}|c|}
	\hline
	\multirow{2}{*}{Surrogate hypergraphs} & \multicolumn{2}{c|}{$\Psi(m)$} & \multicolumn{2}{c|}{$P(D)$} & \multirow{2}{*}{$P(k)$} & \multirow{2}{*}{$P(k^{\Delta})$} & \multirow{2}{*}{global $\bm{k}^{H^3_{\cal D}}$} & \multirow{2}{*}{Groups} & \multicolumn{2}{c|}{Contact matrix} & \multirow{2}{*}{within group $\bm{k}_c^{H^3_{\cal D}}$} \\ 
	\cline{2-5}\cline{10-11}
	& $H_{\cal D}$ & $H^3_{\cal D}$ & $H_{\cal D}$ & $H^3_{\cal D}$ & & & & & $H_{\cal D}$ & $H^3_{\cal D}$ & \\ \hline
	SDk & \textcolor{red}{$\bm{\times}$} & \textcolor{green}{$\boldcheckmark$} & \textcolor{red}{$\bm{\times}$} & \textcolor{green}{$\boldcheckmark$} & \textcolor{red}{$\bm{\times}$} & \textcolor{green}{$\boldcheckmark$} & \textcolor{green}{$\boldcheckmark$} & \textcolor{red}{$\bm{\times}$} & \textcolor{red}{$\bm{\times}$} & \textcolor{red}{$\bm{\times}$} & \textcolor{red}{$\bm{\times}$} \\ \hline
	SCk &  \textcolor{red}{$\bm{\times}$} & \textcolor{green}{$\boldcheckmark$}  & \textcolor{red}{$\bm{\times}$} & \textcolor{red}{$\bm{\times}$} & \textcolor{red}{$\bm{\times}$} & \textcolor{red}{$\bm{\times}$} & \textcolor{red}{$\bm{\times}$} & \textcolor{green}{$\boldcheckmark$} &  \textcolor{red}{$\bm{\times}$} & \textcolor{green}{$\boldcheckmark$} & \textcolor{green}{$\boldcheckmark$}  \\ \hline
	SDC &  \textcolor{red}{$\bm{\times}$} & \textcolor{green}{$\boldcheckmark$}  & \textcolor{red}{$\bm{\times}$} & \textcolor{green}{$\boldcheckmark$} & \textcolor{red}{$\bm{\times}$} & \textcolor{red}{$\bm{\times}$} & \textcolor{red}{$\bm{\times}$} & \textcolor{green}{$\boldcheckmark$} &  \textcolor{red}{$\bm{\times}$} & \textcolor{red}{$\bm{\times}$} & \textcolor{red}{$\bm{\times}$}  \\ \hline
\end{tabular}
\caption{\textbf{Properties of the original data conserved 
(\textcolor{green}{$\boldcheckmark$}) or not  (\textcolor{red}{$\bm{\times}$}) in the  surrogate data}. In particular we consider: 
the distribution of the hyperedges size $\Psi(m)$, comparing it with both that of the complete data set, $H_{\cal D}$, and that of the aggregated hypernetwork restricted to hyperedges of size $2$ and $3$, $H^3_{\cal D}$, i.e. the simplicial complex obtained by projecting larger hyperedges on hyperedges of 3 nodes; 
the total degree distribution $P(D)$, comparing it with both that of $H_{\cal D}$ and that of $H^3_{\cal D}$: the total degree $D$ of a node is the total number of hyperedges of arbitrary size in which the node is involved, thus when considering $H_{\cal D}$ the total degree of a node is $D=\sum_{m=2}^M k_m$, with $k_m$ the number of hyperedges of size $m$ in which the node participates, while when considering $H^3_{\cal D}$ the total degree of a node is $D=k^|+k^{\Delta}$, with $k^|$ and $k^{\Delta}$ respectively the number of hyperedges of size $2$ and $3$ in which the node is involved in $H^3_{\cal D}$; 
the pairwise degree distribution $P(k)$, where the pairwise degree $k$ of a node is its degree in the projected network $G_{\cal D}$; 
the distribution $P(k^{\Delta})$, where $k^{\Delta}$ is the number of hyperedges of size $3$ in which the node is involved in $H^3_{\cal D}$;
the global degree sequence of each node in $H^3_{\cal D}$, $\bm{k}^{H^3_{\cal D}}=\{k^|,k^{\Delta}\}$;
the group structure;
the contact matrix in the weighted projected network, comparing it with both that of $H_{\cal D}$ and that of $H^3_{\cal D}$; 
the within-group degree sequence of each node in $H^3_{\cal D}$, $\bm{k}^{H^3_{\cal D}}_c=\{k^|_c,k^{\Delta}_c\}$, where $k^|_c$ and $k^{\Delta}_c$ indicate respectively the number of hyperedges of size $2$ and $3$ in which node is involved in $H^3_{\cal D}$, that are fully contained in its group $c$.
}
\label{tab_surrogate}
\end{table}

\clearpage
\newpage

\subsubsection{Classifier results}

\begin{figure}[thb]
\includegraphics[width=0.71\textwidth]{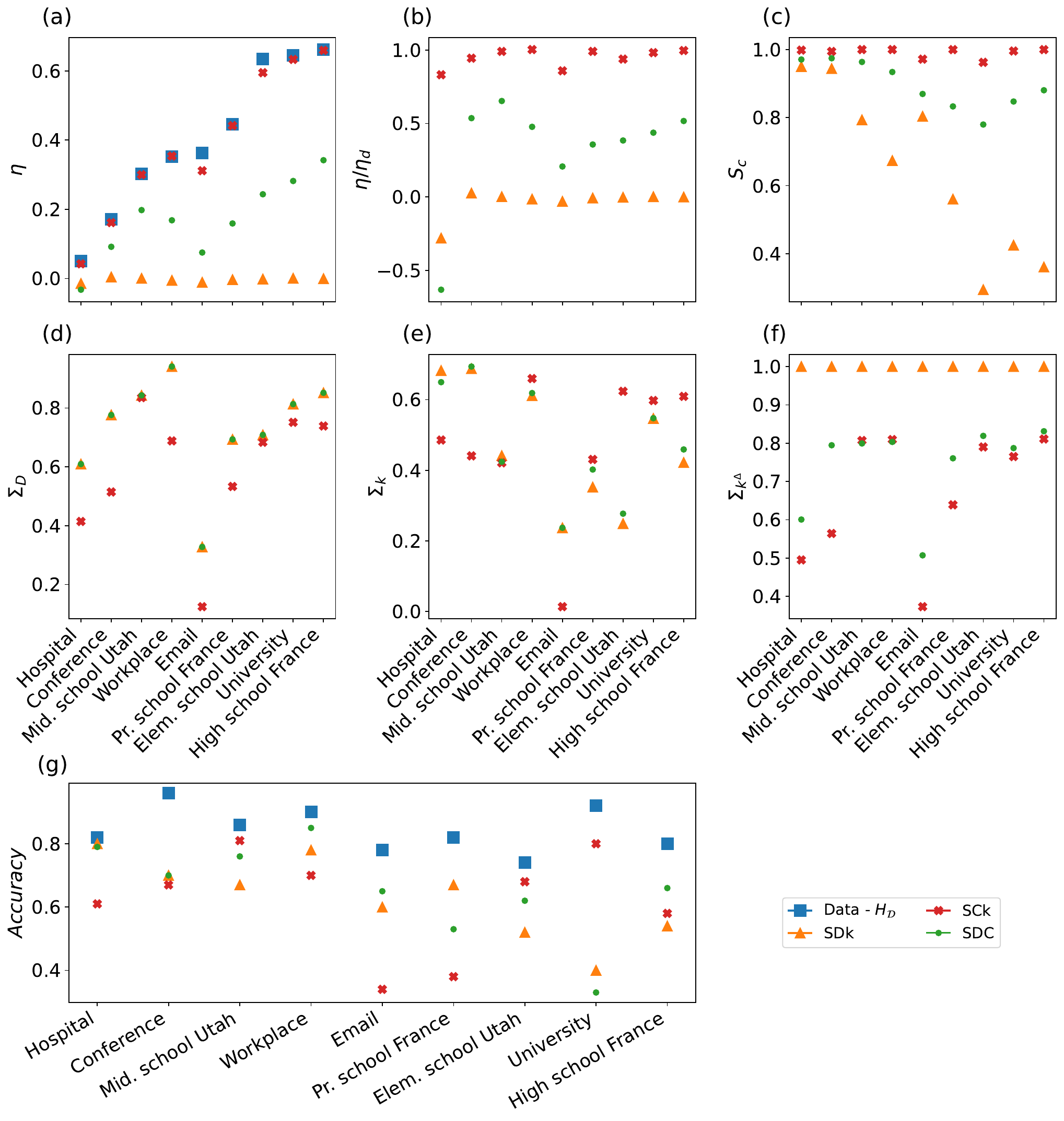}
\caption{\textbf{Data vs. surrogate hypergraphs properties.} Panel (a) shows the value of the modularity $\eta$, evaluated in the weighted projected graph (the weight of each link $i-j$ is the number of hyperedges in which both nodes $i$ and $j$ are involved);  
panel (b) gives the modularity $\eta$ of the surrogate hypergraphs, normalized with the modularity of the corresponding data set $\eta_d$, evaluated for $H_{\cal D}$; 
panel (c) reports the cosine similarity $S_c$ of the contact matrices of the weighted projected graphs of the original data set $H_{\cal D}$ and of the surrogate hypergraphs. 
In panel (d)-(f) we show the similarity $\Sigma$ between the distributions of the total degree $P(D)$ ($\Sigma_D$, panel (d)), of the pairwise degree $P(k)$ ($\Sigma_k$, panel (e)) and of  $P(k^{\Delta})$ ($\Sigma_{k^{\Delta}}$, panel (f)), in the original data sets $H_{\cal D}$ and in the surrogate hypergraphs. 
The similarity is obtained as $1$ minus the Jensen–Shannon divergence between the distributions obtained in the original and surrogate data. 
The values are averaged over $10$ realizations of each type of surrogate hypergraph. 
Panel (g) gives the accuracy of the classifier if trained on processes run on $10$ realizations of the surrogate hypergraphs and then tested on processes run on the original data set; the figure also gives for reference the baseline accuracy 
obtained by the classifier when trained using processes run on the original data set.}
\label{fig_modularity_similarities}
\end{figure}

Figure \ref{fig_modularity_similarities} gives an additional quantitative comparison between several properties of the original data sets and of the corresponding surrogate hypergraphs. The figure also compares the accuracy of the classifier if trained using processes run on $10$ realizations of the surrogate hypergraphs and then tested on processes run on the corresponding original data set. Depending on the original data set, the accuracy obtained when using different surrogate data for training vary, and the optimal surrogate
method changes. In particular, the SDk method, which does not preserve group structure, 
gives very or reasonably good results as long as the modularity of the original data is not
too large, which is somehow expected \cite{machens2013infectious}. The performance of the various methods seem also to depend on a complex interplay of which properties are best preserved by the surrogate data (e.g. for email, $P(k)$ is badly preserved in SCk, so even if the group structure is well preserved, the performance is lower for SCk than for other methods).

\clearpage
\newpage

\begin{figure}[thb]
\includegraphics[width=0.85\textwidth]{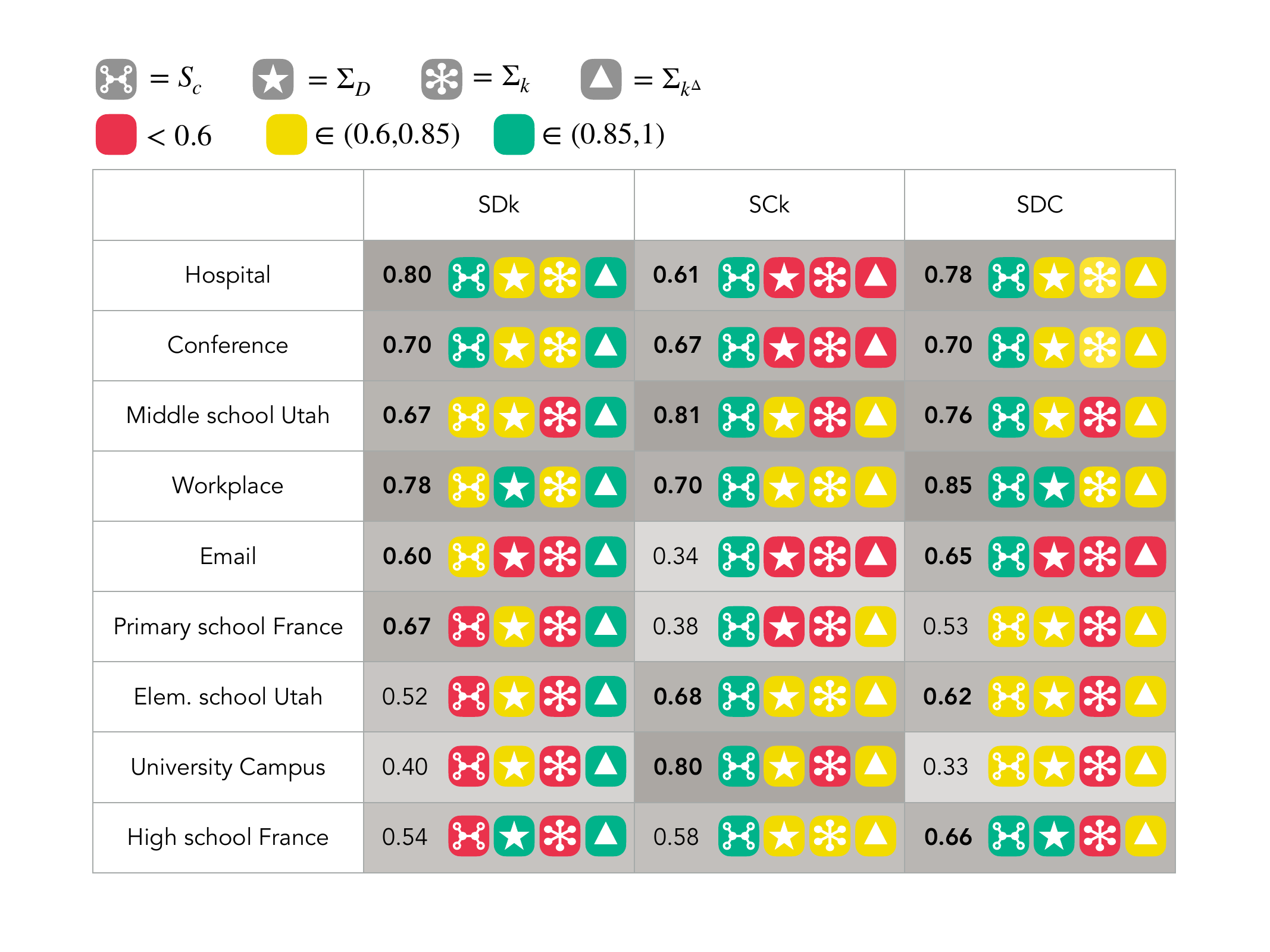}
\vskip -.7cm
\caption{
\textbf{Surrogate networks accuracy.} For each network and surrogate generation method the classification accuracy is reported together with a sketch showing which features are preserved in the surrogate data. The datasets are sorted from the one with lowest modularity to the one with largest modularity in the original network.}
\label{fig_accuracy_table}
\end{figure}

In Fig.~\ref{fig_accuracy_table}, we present a visual summary of these results: 
the data sets are sorted according to the modularity of the corresponding projected network, from the smallest value of $\eta$ (hospital) to the largest (High School France). In each column (SDk, SCk and SDC), we report
the accuracy of the classifier trained using processes run on the corresponding surrogate data (with 
a darker gray for larger accuracies). Moreover, we report with colored symbols the similarity between original and surrogate data with respect to the group structure and to the distributions $P(D)$, $P(k)$ and 
$P(k^{\Delta})$, with green symbols for the highest similarities (above $0.85$), yellow ones for
similarities between $0.6$ and $0.85$, and red ones 
for lower values.

For instance, for the hospital data set we observe a much higher accuracy with SDk and SDC, than with SCk, implying that it is important to replicate the degree distribution $P(D)$, $P(k)$ and $P(k^\Delta)$ for generalization, while the group structure is not a crucial information.
The opposite is true for the University Campus data set, suggesting that the information on the contact matrix plays an important role in this case (note that SDk does not preserve well the group structure in this case). Indeed, the University data set has a high modularity, while the hospital has a low modularity.
Overall, both the group structure and the total degree distribution seem to play an important role. For 
networks with high modularity, 
it is indeed better to use the group structure information when generating surrogates, and so to prefer SCk and SDC methods. 
 
\vskip -.3cm

\vskip -.3cm
Overall, the accuracy of the classifier
trained using surrogate data depends on the original data properties, but seems also limited by the current ability of algorithms to generate surrogate data that correctly preserves the desired statistical properties. 
However, we notice that for each data set it is possible to find a good classification accuracy by using at least one of the proposed surrogate data.
The results presented here provide a first indication of the kind of information that is needed to create useful surrogate data for training the classifier. 
More work using network and hypergraph models with tunable structures will be needed to fully understand this issue. Moreover, it
 would also be highly desirable to devise
an algorithm similar to SCk that would better preserve the degree distributions and group structure of the original data, to obtain better accuracy values.


\end{document}